%% file: ActaMechanica_Manuscript.tex
\documentclass[pdflatex,sn-mathphys-num]{sn-jnl}

\usepackage{graphicx}
\usepackage{multirow}
\usepackage{amsmath,amssymb,amsfonts}
\usepackage{amsthm}
\usepackage{mathrsfs}
\usepackage[title]{appendix}
\usepackage{xcolor}
\usepackage{textcomp}
\usepackage{manyfoot}
\usepackage{booktabs}
\usepackage{algorithm}
\usepackage{algorithmicx}
\usepackage{algpseudocode}
\usepackage{listings}
\usepackage{url}
\usepackage{physics,nicefrac,comment,subcaption}
\graphicspath{{./figures/}}

\theoremstyle{thmstyleone}

\theoremstyle{thmstyletwo}

\theoremstyle{thmstylethree}

\begin{document}

\title{Spatial deformation of a ferromagnetic elastic rod}

\author*[1]{\fnm{G R Krishna Chand} \sur{Avatar}}\email{krishnaagr@iisc.ac.in}

\author*[1]{\fnm{Vivekanand} \sur{Dabade}}\email{dabade@iisc.ac.in}

\affil[1]{\orgdiv{Department of Aerospace Engineering}, \orgname{Indian Institute of Science}, \orgaddress{\street{C V Raman Road}, \city{Bengaluru}, \postcode{560012}, \state{Karnataka}, \country{India}}}

\abstract{

Ferromagnetic elastic slender structures offer the potential for large actuation displacements under modest external magnetic fields, due to the magneto-mechanical coupling. This paper investigates the phase portraits of the Hamiltonian governing the three-dimensional deformation of inextensible ferromagnetic elastic rods subjected to combined terminal tension and twisting moment in the presence of a longitudinal magnetic field. The total energy functional is formulated by combining the Kirchhoff elastic strain energy with micromagnetic energy contributions appropriate to soft and hard ferromagnetic materials: magnetostatic (demagnetization) energy for the former, and exchange and Zeeman energies for the latter. Exploiting the circular cross-sectional symmetry and the integrable structure of the governing equations, conserved Casimir invariants are identified and the Hamiltonian is reduced to a single-degree-of-freedom system in the Euler polar angle. Analysis of the resulting phase portraits reveals that purely elastic and hard ferromagnetic rods undergo a supercritical Hamiltonian Hopf pitchfork bifurcation, whereas soft ferromagnetic rods exhibit this bifurcation only within a restricted range of the magnetoelastic parameter, $0<\tilde{K}_{dM}<1/8$. Both helical and localized post-buckling configurations are analyzed, and the corresponding load-deformation relationships are systematically characterized across a range of loading scenarios. Localized buckling modes, corresponding to homoclinic orbits in the Hamiltonian phase space, are constructed numerically. In contrast to the purely elastic case, the localized configurations of soft ferromagnetic rods exhibit non-collinear extended straight segments, a geometrically distinctive feature arising directly from the magnetoelastic coupling.

}

\keywords{Ferromagnetic rod, Hamiltonian, Spatial buckling, Micromagnetics}

\maketitle

\section{Introduction}

Ferromagnetic elastic rods are slender structures capable of undergoing large deformations in response to external magnetic fields, making them attractive candidates for magnetically actuated robotic systems and adaptive structures. Under combined axial and torsional loading, geometric nonlinearities become the dominant governing mechanisms, causing such slender structures to exhibit post-buckling responses that differ markedly from the classical planar behavior predicted by Euler's theory \cite{timoshenko2012theory}. Earlier work by the present authors \cite{AvatarDabade2024,AvatarDabade2025} addressed the planar deformation of ferromagnetic rods and ribbons. Such planar configurations, characterized by a single nonzero curvature component, admit only in-plane buckling modes. To capture the full spectrum of instabilities in slender ferromagnetic structures, their three-dimensional spatial deformation must be considered. Extending the analysis to three-dimensional configurations introduces coupling between bending and torsion, revealing a variety of instability routes inaccessible in planar models.

\begin{figure}
    \centering
    \includegraphics[width=0.8\linewidth]{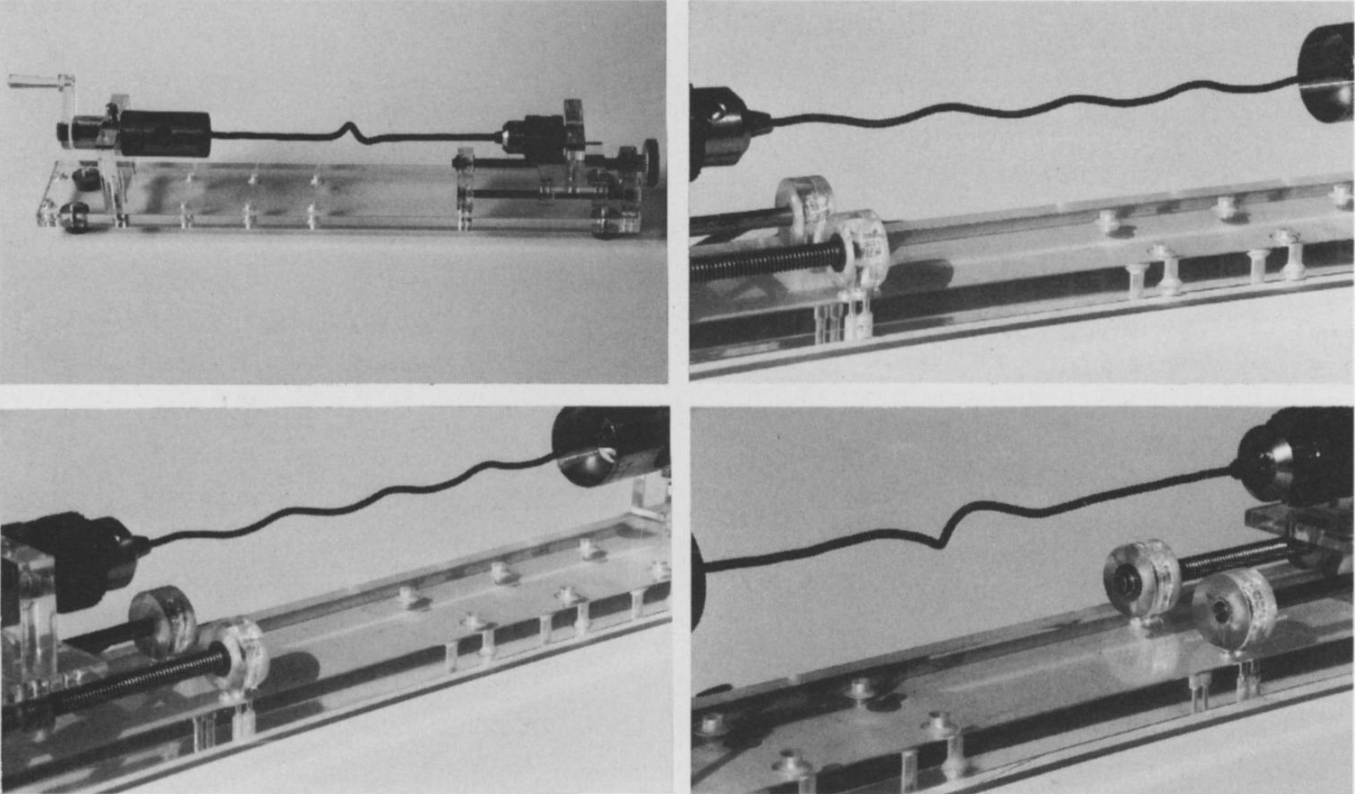}
    \caption{Experiments showing modes of deformation of a circular rubber rod. \cite[Fig. 3]{thompson1996helix}.}
    \label{fig:threeD-thompson-2}
\end{figure}

The coupling between twisting and bending is a key feature underlying deformation phenomena of elastic rods across vastly different length scales: from the supercoiling of DNA and protein filaments at the nanoscale to the coiling and helical perversions observed in plant tendrils and engineered filaments at the macroscale \cite{Gerbode2012,Feng2019}, and the buckling of drilling cables \cite{Wu2024}. Experiments and theory have shown that soft elastic rods subjected to external loading undergo bending and twisting, and beyond critical thresholds, transition through successive buckled configurations. These include helical and solenoidal modes, localized helices, hockles \cite{Ermolaeva2008}, loops, plectonemes, and even physical knots \cite{Coyne1990,Feng2019,Liu2023}. As twist and tension are applied at the rod terminals, this evolution proceeds through several stages: an initial small-amplitude helical bifurcation from the trivial straight configuration, followed by gradual localization, a dynamic transition to a self-contacted state, and eventual writhing to a plectonemic state \cite{thompson1996helix,Hu2024}, a sequence confirmed experimentally by Thompson and Champneys  \cite{thompson1996helix} (see Fig. \ref{fig:threeD-thompson-2}).

\begin{figure}
    \centering
    \includegraphics[width=0.75\linewidth]{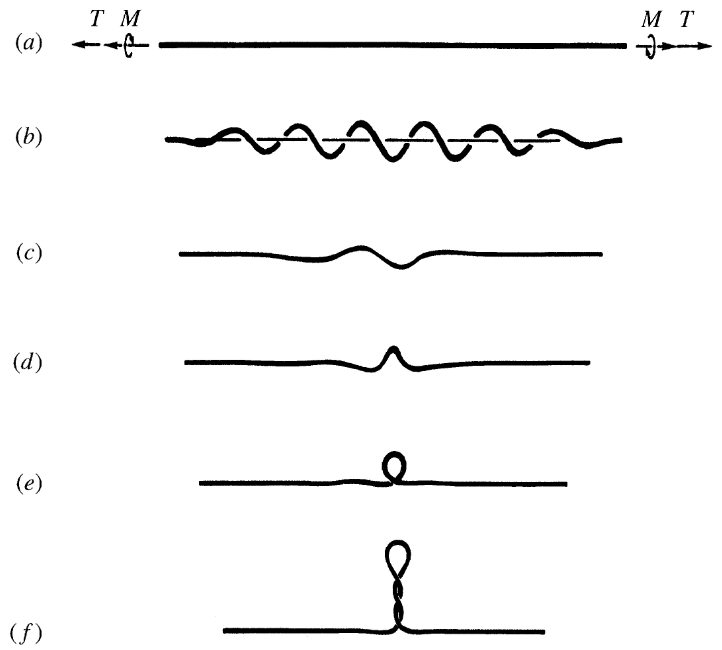}
    \caption{Spatial localization of a stretched and twisted rod: (a) trivial straight configuration, (b) an initial small-amplitude helical bifurcation from the trivial configuration, (c-d) gradual localization, (e) dynamic transition to a self-contacted state, and (f) eventual writhing to a plectonemic state. Reproduced from \cite[Fig. 1]{thompson1996helix}.}
    \label{fig:threeD-thompson-1}
\end{figure}

In localized buckling, deformation is spatially confined, concentrating the strain energy in finite segments and breaking the global symmetry of uniform buckling modes \cite{Champneys1998,champneys1996multiplicity}. Unlike the periodic or helical configurations predicted by linear theory, localised buckling manifests as discrete high-curvature regions separated by nearly straight segments \cite{Audoly2011}. This behaviour arises across diverse physical systems, from tendril perversions \cite{Gerbode2012,Feng2019} to DNA supercoiling and collagen fibril mechanics \cite{Peacock2020}, and is a hallmark of strongly nonlinear rod mechanics, arising when the governing equations admit multiple coexisting equilibria.

To analyze these complex deformations systematically, a dynamical systems formulation is particularly effective. Kirchhoff's kinetic analogy maps the spatial deformation of a rod onto an equivalent dynamical problem in phase space \cite{thompson1996helix,Neukirch2002}, in which the shape of a deformed rod corresponds to a trajectory in Hamiltonian phase space. Localised buckling modes appear as homoclinic or heteroclinic orbits between equilibria \cite{Champneys1998,champneys1996multiplicity}. van der Heijden et al. \cite{vanderHeijden2002} showed that geometric constraints fundamentally reshape the buckling landscape, with spatial localisation arising via heteroclinic connections between equilibrium branches. Champneys and van der Heijden \cite{Champneys1998} demonstrated that twisted rods admit a countably infinite family of localised solutions, each linked to distinct homoclinic orbits of increasing spatial complexity. The shift from globally periodic helices to localised writhing marks a critical breakdown of spatial periodicity while preserving topological properties \cite{thompson1996helix}.

In magnetoelastic rods, applied magnetic fields enrich the buckling landscape by introducing additional instability mechanisms arising from magneto–mechanical coupling. Sinden and van der Heijden \cite{Sinden2008} found that magnetic coupling alters the integrable structure of rod equations, producing chaotic configurations and Hamiltonian–Hopf bifurcations absent in purely mechanical systems. Magneto-mechanical interactions reorganize phase-space topology, opening new pathways to localized and spatially complex buckling.

Furthermore, curvature and torsion also influence the magnetic response of rigid ferromagnetic nanostructures \cite{Sheka2015,sheka2022fundamentals}. At these smaller scales, the exchange energy becomes the dominant contribution to the micromagnetic functional, giving rise to curvature-induced anisotropy and effective Dzyaloshinskii–Moriya–type interactions. The geometry thereby stabilizes topologically nontrivial magnetic configurations, including vortex states on spherical shells and curvature-induced domain walls on Möbius strips \cite{streubel2016magnetism}. Studies on ferromagnetic nanowires \cite{Sheka2015nanowire,Streubel2021} and narrow ribbons \cite{Gaididei2017} have demonstrated that geometric characteristics, such as curvature and torsion, critically determine equilibrium magnetization configurations, thereby establishing geometry as a fundamental mediator of magneto–mechanical coupling. In the present work, we address the complementary perspective by examining how magnetization, in turn, alters the deformation and mechanical response. Specifically, we analyze the deformation of magnetic wires subjected to end loading to elucidate the role of magnetization in governing the onset and character of mechanical instabilities, the emergence of localized deformation, and the overall mechanical response.

Despite the extensive literature on rigid ferromagnetic nanostructures and the spatial buckling of purely elastic rods, a Hamiltonian framework characterizing the three-dimensional magnetoelastic buckling and localization of ferromagnetic elastic rods under combined mechanical and magnetic loading has not previously been developed. The present work addresses this gap. The analysis is restricted to two integrable configurations: a soft ferromagnetic rod with magnetization saturated and aligned with a longitudinal external magnetic field, and a hard ferromagnetic rod with tangential magnetization aligned with the rod axis, also subjected to a longitudinal field, see Section \ref{subsec:note-on-integrability}. In both cases, the rod is loaded by a combination of terminal tension $T$ and twisting moment $M$ applied at its ends. Employing Kirchhoff's phase-space analogy \cite{champneys1996multiplicity}, conserved Hamiltonians are derived for each case and exploited to systematically characterize magnetoelastic buckling and spatial localization. The principal findings are as follows. First, the hard ferromagnetic rod is shown to be structurally equivalent to a purely elastic rod with renormalized bending stiffness and effective tension, and both exhibit a supercritical Hamiltonian Hopf pitchfork bifurcation. Second, the soft ferromagnetic rod exhibits this bifurcation only within the restricted magnetoelastic parameter regime $0<\tilde{K}_{dM}<1/8$, outside this range, the phase portrait topology is qualitatively different and no bifurcation from the straight configuration is observed. Third, the localized buckling configurations of the soft ferromagnetic rod possess a geometrically distinctive feature absent in the purely elastic case: the extended straight segments on either side of the localized region are no longer collinear with each other as a consequence of the magnetoelastic coupling. Fourth, closed-form expressions are derived relating the end displacement of the rod to the composite load parameter, extending the analysis of \cite{Coyne1990} to the magnetoelastic setting. Throughout this paper, the method and style of analysis follow those of the landmark study by \cite{thompson1996helix}, which established the Hamiltonian phase-space framework for the torsional post-buckling of purely elastic rods; the present work extends that framework to the magnetoelastic setting.

\subsection{Organisation of the paper}

The paper is organised as follows. Section~\ref{sec:energy-functional-threeD} formulates the total energy functional of a ferromagnetic elastic rod subjected to combined terminal tension and twisting moment, for both soft and hard ferromagnetic materials. Section~\ref{sec:hamiltonian-threeD} derives the equilibrium equations from the energy functional, identifies the conserved Casimir invariants arising from the circular cross-sectional symmetry and the applied loading, and constructs the Hamiltonian for both material cases via the Legendre transform. The integrable loading configurations, those for which the Hamiltonian admits reduction to a single-degree-of-freedom system in the primary Euler angle are identified and discussed. Section~\ref{sec:threeD-analysis-Hamiltonian} analyses the critical points, bifurcation structure, and phase portraits of the resulting Hamiltonian, contrasting the behaviour of purely elastic, soft ferromagnetic, and hard ferromagnetic rods. Section~\ref{sec:threeD-local-helix} studies the post-buckling response of the rod into a helical configuration, deriving load–deformation relationships under three distinct loading sequences. Section~\ref{sec:threeD-localized-buckling} examines localized buckling modes, which correspond to homoclinic orbits in the Hamiltonian phase space; closed-form expressions for the maximum lateral deflection and the end displacement are derived and compared between the purely elastic and soft ferromagnetic cases. The conclusions are summarized in Section~\ref{sec:conclusions-threeD}.

\section{Energy functional of a space rod} \label{sec:energy-functional-threeD}

In this section, we present the total energy of the ferromagnetic rod subjected to a combination of terminal tension $T$ and twisting moment $M$ \cite{vanderHeijden2000helical,thompson1996helix}, along with its energy
density, represented by the Lagrangian $\mathcal{L}$. The associated Hamiltonian is derived using the
Legendre transform of $\mathcal{L}$.

\subsection{Kinematics}
\begin{figure}
    \centering
    \includegraphics[width=0.5\linewidth]{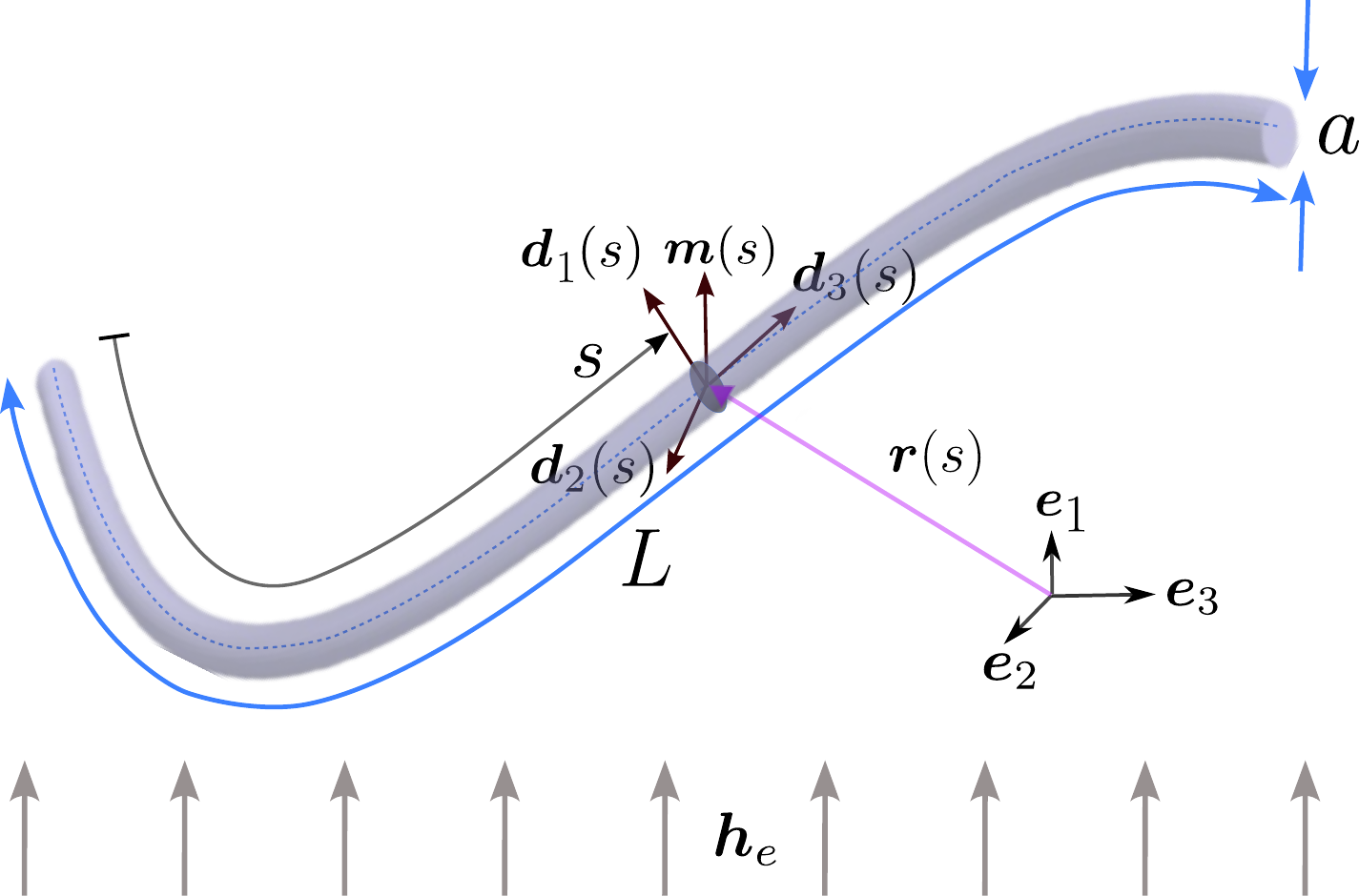}
    \caption{Schematic of a spatial Kirchhoff rod. { Here, $\vb*{m}(s)$ denotes magnetization vector and $\vb*{h}_e$ represents uniform external magnetic field (see Section \ref{sec:magnetic-energy-threeD}).}}
    \label{fig:rod-threeD}
\end{figure}

We briefly review the kinematics of the space Kirchhoff rod as shown in Fig. \ref{fig:rod-threeD}. The rod has a uniform circular cross section of diameter $a$ and it is assumed to be unshearable while its centerline is inextensible. The centerline-based representation of the rod is given as:
\begin{equation}
    \vb*{x}(s,a_1, a_2) = \vb*{r}(s) + a_1\vb*{d}_1(s) + a_2\vb*{d}_2(s),
\end{equation}
where $\vb*{r}(s)$ denotes the deformed centerline, $(\vb*{d}_1(s),\vb*{d}_2(s),\vb*{d}_3(s))$ is the orthonormal director frame, $s$ is the centerline arc length coordinate, and the coordinates $(a_1,a_2)$ parametrize the cross section. Here, $s \in [0,L]$ and $a_1, a_2 \in [0,\nicefrac{a}{2}]$; $L$ denote the characteristic length of the rod. The unshearability condition implies $\vb*{r}'(s)\cdot\vb*{d}_i(s) = 0, i=1,2$, inextensibility constraint enforces $\vb*{r}'(s)\cdot\vb*{d}_3(s) = 1$, and thus $\vb*{r}'(s) = \vb*{d}_3(s)$.

\begin{figure}[h!]
    \centering
    \includegraphics[width=0.9\linewidth]{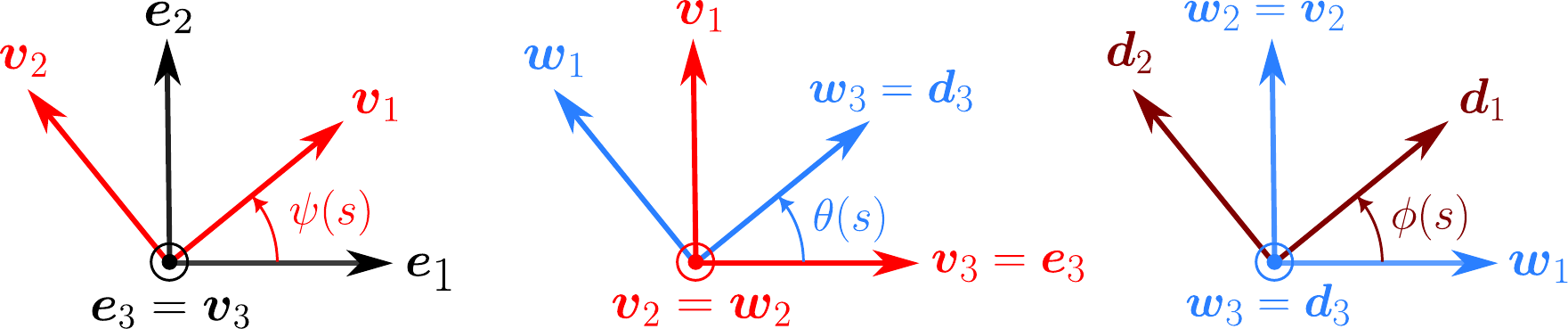}
    \caption{Representation of Euler angles $(\psi,\theta,\psi)$ where $(\psi,\phi) \in [0,2\pi[$ and $\theta \in [0,\pi]$. (Adapted from  \cite{Ameline2017})}
    \label{fig:euler-angles}
\end{figure}

The standard Cartesian basis $(\vb*{e}_1,\vb*{e}_2, \vb*{e}_3)$  is mapped to the  director basis $(\vb*{d}_1(s),\vb*{d}_2(s), \vb*{d}_3(s))$ utilizing Euler angles $(\psi(s), \theta(s), \phi(s))$, such that
\begin{equation}
    \begin{split}
        \vb*{d}_1 &= (-\sin\psi\sin\phi + \cos\psi\cos\phi\cos\theta)\vb*{e}_1 + (\cos\psi\sin\phi + \sin\psi\cos\phi\cos\theta) \vb*{e}_2 - \cos\phi\sin\theta\vb*{e}_3, \\
        \vb*{d}_2 &= (-\sin\psi\cos\phi - \cos\psi\sin\phi\cos\theta)\vb*{e}_1 + (\cos\psi\cos\phi - \sin\psi\sin\phi\cos\theta)\vb*{e}_2 + \sin\phi\sin\theta \vb*{e}_3, \\
        \vb*{d}_3 &= \cos\psi\sin\theta\vb*{e}_1 + \sin\psi\sin\theta\vb*{e}_2 + \cos\theta\vb*{e}_3,
    \end{split}
\end{equation}
where $\theta$ is the angle between $\vb*{d}_3$ and $\vb*{e}_3$, $\psi$ is the azimuthal angle between $\vb*{e}_2$ and the normal to the $\vb*{d}_3-\vb*{e}_3$ plane, that is $\vb*{v}_2$-axis, and $\phi$ is the angle between the normal $\vb*{v}_2$ and $\vb*{d}_2$ axis, see Fig. \ref{fig:euler-angles} for visual representation. The Euler angle convention employed here is in accordance with \cite{Ameline2017,love1944treatise}.

We define $\psi'(s)$ and $\phi'(s)$ as the angular spatial velocities about vertical axis $\vb*{e}_3$ and body axis $\vb*{d}_3$ respectively while $\theta'(s)$ denotes the angular spatial velocity about an intermediate axis $\vb*{v}_2$. The evolution of the director frame is explained by
\begin{equation}
    \vb*{d}'_i(s) = \vb*{\kappa}(s)\cross \vb*{d}_i(s), \qquad i = 1,2,3.
\end{equation}
 where, the components of the curvature vector or rotation gradient $\vb*{\kappa}(s)$ in the director basis are $(\kappa_1, \kappa_2, \kappa_3)$ or $(\kappa_1(s), \kappa_2(s), \tau(s))$. Here, $\kappa_1$ and $\kappa_2$ denote the bending curvatures about $\vb*{d}_1$ and $\vb*{d}_2$ axes respectively, and $\tau$ is called the (local) twist or torsion and measures the rate of rotation of the circular cross section about the body axis of the rod $\vb*{d}_3$.

The strain components $\kappa_i$ are expressed in terms of the Euler angles by using the following identity:
\begin{equation}
    \kappa_i = \frac{1}{2} \varepsilon_{ijk} \vb*{d}_j'\cdot\vb*{d}_k, ~~i,j,k=1,2,3.
\end{equation}
{ where the Levi-Civita permutation tensor $\varepsilon_{ijk}$ is defined as
\begin{equation*}
   \varepsilon_{ijk} = \begin{cases}
     +1, & \text{if $(i,j,k)$ is an even permutation of (1, 2, 3)}, \\
     -1, & \text{if $(i,j,k)$ is an odd permutation of (1, 2, 3)}, \\
     ~~0, & \text{if any index is repeated},
   \end{cases}
\end{equation*}
}
resulting in
\begin{equation}
    \begin{split}
        \kappa_1 &= \theta' \sin\phi - \psi' \sin\theta \cos\phi, \\
        \kappa_2 &= \theta'\cos\phi + \psi'\sin\theta \sin\phi, \\
        \tau &= \phi' + \psi'\cos\theta.
    \end{split} \label{eqn:euler-rates-strains-relations}
\end{equation}
The Cartesian description of the centerline:
\begin{equation}
    \vb*{r}(s) = x(s)\vb*{e}_1 + y(s)\vb*{e}_2 + z(s)\vb*{e}_3 \label{eqn:centerline-coordinates}
\end{equation}
is obtained from the inextensibility condition:
\begin{equation}
    \vb*{r}'(s) = \vb*{d}_3(s) \implies \begin{cases}
       & x'(s) = \cos\psi\sin\theta, \\
       & y'(s) = \sin\psi\sin\theta, \\
       & z'(s) = \cos\theta.
    \end{cases} \label{eqn:centerline-description}
\end{equation}

\subsection{Total energy}
We formulate the total energy functional $\mathcal{E}$ of the ferromagnetic rod as the sum of its elastic strain energy $\mathcal{E}_{\text{elastic}}$, magnetic energy $\mathcal{E}_{\text{magnetic}}$, and work potential $\mathcal{W}$ due to applied end mechanical loads, yielding

\begin{equation}
	\mathcal{E} = \underbrace{\mathcal{E}_{\text{elastic}} - \mathcal{W}}_{\text{Mechanical energy}} + \mathcal{E}_{\text{magnetic}} . \label{eqn:total-energy-functional-threeD-rod}
\end{equation}

\begin{figure}
    \centering
    \includegraphics[width=0.8\linewidth]{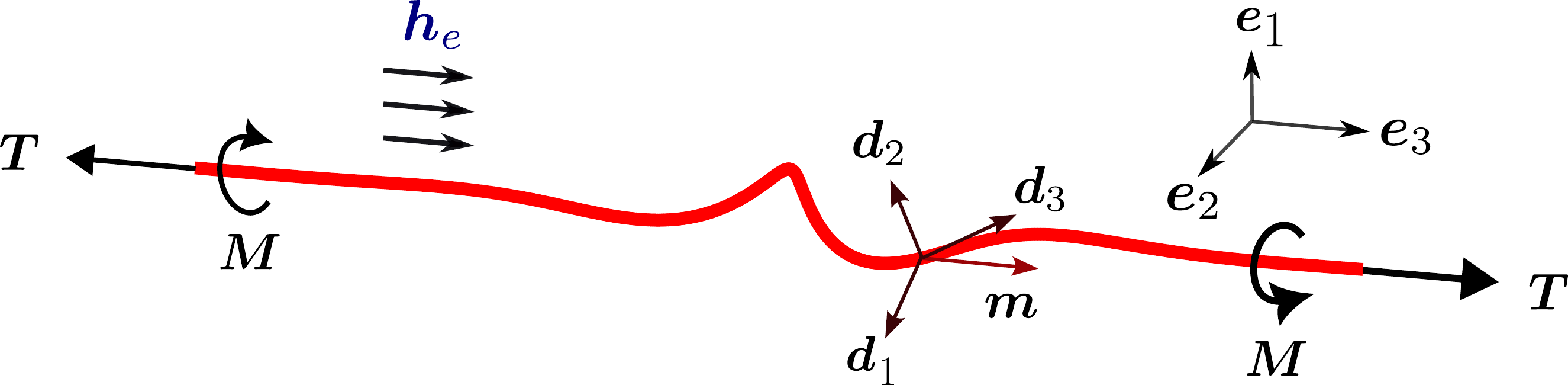}
    \caption{Schematic of a spatial ferromagnetic elastic rod subjected to terminal load $\vb*{T}$ and twisting moment $\vb*{M}$ placed in a uniform external magnetic field $\vb*{h}_e$.}
    \label{fig:threeD-problem-setup}
\end{figure}

The mechanical contribution consists of the elastic strain energy and work potential terms alone. Fig. \ref{fig:threeD-problem-setup} illustrates the setup: a spatial ferromagnetic elastic rod subjected to combined tension $T$, twisting moment $M$, and immersed in a uniform external magnetic field $\vb*{h}_e$.

\subsubsection{Mechanical energy}
We consider a uniform, initially straight Kirchhoff rod with circular cross-section, so that the principal bending stiffnesses satisfy $EI_1 = EI_2$. In this setting, the circular symmetry of the cross-section ensures that the corresponding Hamiltonian formulation of the rod equations is completely integrable, as shown in \cite{Mielke1988,champneys1996multiplicity}. The elastic energy of a linearly elastic, uniform, and isotropic rod of circular cross-section can be written as \cite{Audoly2010elasticity,AvatarDabade2025}
\begin{equation}
    \mathcal{E}_{\text{elastic}} = \int_{0}^{L} \left[\frac{1}{2}EI_1\kappa_1^2 + \frac{1}{2}EI_2\kappa_2^2 + \frac{1}{2}GJ\tau^2\right] ds, \label{eqn:elastic-enery-threeD-rod}
\end{equation}
where $E$ denotes Young's modulus, $G$ the shear modulus, $I_1 = I_2 = I = \nicefrac{\pi a^4}{64}$ are the principal area moments of inertia, $J = \nicefrac{\pi a^4}{32}$ is polar area moment of inertia, and $GJ$ is the torsional stiffness of the cross section.
We now introduce an axial dead load by applying a constant axial tension $T\vb*{e}_3$ at the terminal of the rod, where $\vb*{e}_3$ is the fixed unit vector along the undeformed rod axis. We also apply a twisting moment $M \vb*{e}_3$ at the end terminal which generates a twist rotation $R\vb*{e}_3$ of the end cross section. The corresponding work potential is expressed as
\begin{multline}
 \mathcal{W} = T\vb*{e}_3 \cdot \left(\int_{0}^{L} \vb*{d}_3~ds - L\vb*{e}_3\right) + M\vb*{e}_3\cdot R\vb*{e}_3 = T \left(\int_{0}^{L} \vb*{d}_3\cdot \vb*{e}_3 ~ds - L \right) + MR \\
  = T \int_{0}^{L} \left(\vb*{d}_3\cdot \vb*{e}_3 - 1 \right)ds + MR. \label{eqn:work-potential-threeD-rod}
\end{multline}

{

}

\subsubsection{Magnetic energy} \label{sec:magnetic-energy-threeD}
The magnetic energy of a ferromagnetic rod is derived using the theory of micromagnetics. Micromagnetics is based on a variational principle of the micromagnetic functional that successfully predicts the domain structures in a ferromagnet \cite{choksi1999domain}. The total micromagnetic energy is a sum of four contributions: the exchange energy, which penalizes spatial variations in the magnetization direction; the magnetocrystalline anisotropy energy, which reflects the preference of the magnetization to align along crystallographic easy axes; the magnetostatic energy, which accounts for the long-range dipolar interactions arising from the divergence of the magnetization field; and the Zeeman energy, which captures the interaction of the magnetization with an externally applied magnetic field. For a detailed exposition of each of these energy contributions in the context of a ferromagnetic rod, we refer the reader to Avatar and Dabade \cite{AvatarDabade2024} and to Appendices \ref{sec:app-slastikov-derivation} and \ref{sec:app-exchange-energy-hard-magnetic} of the present paper, where these terms are derived and discussed systematically. Here, we will use this micromagnetic functional to find the magnetic energy of a ferromagnetic rod. The magnetic energy expressions for both soft and hard ferromagnetic rods are derived in \autoref{sec:app-slastikov-derivation} and \autoref{sec:app-exchange-energy-hard-magnetic}.

In the present work, for a soft ferromagnetic rod, we examine the influence of magnetization on the mechanical deformation of the rod, and not the converse; that is, the magnetization distribution is prescribed and held fixed. The magnetization is assumed to be aligned with the applied external magnetic field, which is taken to be sufficiently large to saturate the sample irrespective of its deformed configuration. With the magnetization thus fixed, and the anisotropy energy being negligibly small for a soft ferromagnetic material, the magnetic energy contribution in the soft ferromagnetic case arises solely from the magnetostatic energy. For a soft ferromagnetic rod, the magnetostatic energy is given as:
\begin{equation}
    \mathcal{E}_{\text{mag,soft}} =  \frac{K_d\pi a^2}{4} \int_{0}^{L}\left[(\vb*{m}\cdot\vb*{d}_1)^2 + (\vb*{m}\cdot\vb*{d}_2)^2\right] ds. \label{eqn:magnetostatic-energy-threeD-rod}
\end{equation}

For a hard ferromagnetic rod, the anisotropy energy is very large compared to the magnetostatic energy, and consequently dominates the magnetic response. We assume that the magnetization vector $\vb*{m}(s)$ remains aligned with the easy axis of the magnet $\vb*{p}(s)$, which is taken to coincide with the axial direction of the rod $\vb*{d}(s)$, so that the magnetization is everywhere tangential to the centerline; $\vb*{m}(s) = \vb*{p}(s) = \vb*{d}_3(s)$. For a hard ferromagnetic rod with tangential magnetization, the magnetostatic energy reduces to a constant, which in fact vanishes identically, as shown in Eqn. \ref{eqn:magnetostatic-energy-threeD-rod}. Consequently, the total magnetic energy of a hard ferromagnetic rod is given by the sum of its exchange energy and Zeeman energy, the former penalizing spatial variations in the magnetization direction induced by the deformation of the rod, and the latter capturing the interaction of the magnetization with the applied external magnetic field. For hard ferromagnetic rods, the magnetic energy takes the following general form:
\begin{equation}
    \mathcal{E}_{\text{mag,hard}} = \underbrace{\frac{A\pi a^2}{4}\int_{0}^{L} \mathbb{E}_{\text{ex}} ds}_{\text{Exchange energy}} + \underbrace{\frac{K_a\pi a^2}{4}\int_{0}^{L}(\vb*{m}\cdot\vb*{p})^2ds}_{\text{Anisotropy energy}} - \underbrace{\frac{K_d\pi a^2}{2}\int_{0}^{L}\vb*{m}\cdot\vb*{h}_e~ds}_{\text{Zeeman energy}},
\end{equation}
where $\mathbb{E}_{\text{ex}}$ is the exchange energy density and $\vb*{p}$ is the anisotropy easy axis. The exchange energies for special magnetization distributions $\vb*{m} = \vb*{d}_i$ ($i=1,2,3$) are derived in \autoref{sec:app-exchange-energy-hard-magnetic} and reproduced as
\begin{equation}
    \begin{split}
        &\vb*{m} = \vb*{d}_1(s): \mathbb{E}_{\text{ex}} = \kappa_2^2 + \kappa_3^2 \\
        &\vb*{m} = \vb*{d}_2(s): \mathbb{E}_{\text{ex}} = \kappa_1^2 + \kappa_3^2 \\
        &\vb*{m} = \vb*{d}_3(s): \mathbb{E}_{\text{ex}} = \kappa_1^2 + \kappa_2^2
    \end{split}
\end{equation}
As discussed previously, for a uniaxially hard ferromagnetic circular rod with tangential magnetization $\vb*{m}(s) = \vb*{d}_3(s)$ and easy axis $\vb*{p} = \vb*{d}_3(s)$, the exchange and Zeeman energies constitute the total magnetic energy, which has been derived in Eqn. \ref{app:hard-magnetic-energy-rod} as
\begin{equation}
    \mathcal{E}_{\text{mag,hard}}(\vb*{m}) = \underbrace{\frac{A\pi a^2}{4}\int_{0}^{L} (\kappa_1^2 + \kappa_2^2) ds}_{\text{Exchange energy}} - \underbrace{\frac{K_d\pi a^2}{2}\int_{0}^{L}\vb*{d}_3\cdot\vb*{h}_e ds}_{\text{Zeeman energy}}. \label{eqn:hard-magnetic-energy-rod}
\end{equation}
Next, we construct the total energy functionals for both soft and hard ferromagnetic rods.

\subsubsection{Total energy functional}

{
\color{blue}
}

\paragraph{Soft ferromagnetic rod:}
Combining Eqns. \ref{eqn:elastic-enery-threeD-rod}, \ref{eqn:work-potential-threeD-rod} and \ref{eqn:magnetostatic-energy-threeD-rod}, we obtain the total energy functional for the soft ferromagnetic rod:
\begin{multline}
        \mathcal{E} = \int_{0}^{L} \Bigg[\underbrace{\frac{1}{2}EI\kappa_1^2 + \frac{1}{2}EI\kappa_2^2 + \frac{1}{2}GJ\tau^2}_{\text{Elastic energy}} + \underbrace{\frac{K_d\pi a^2}{4} \left[(\vb*{m}\cdot\vb*{d}_1)^2  + (\vb*{m}\cdot\vb*{d}_2)^2\right]}_\text{Magnetostatic energy} \\ - \underbrace{T\left(\vb*{d}_3\cdot \vb*{e}_3 - 1 \right)}_{\text{Work due to $T$}} \Bigg] ds - \underbrace{MR}_{\text{Work due to $M$}}. \label{eqn:total-energy-threeD-soft-ferromagnetic-rod}
\end{multline}

\paragraph{Hard ferromagnetic rod:}
The total energy of the hard ferromagnetic rod with tangential magnetization distribution ($\vb*{m} = \vb*{d}_3$) is obtained by summing Eqns. \ref{eqn:elastic-enery-threeD-rod}, \ref{eqn:work-potential-threeD-rod} and \ref{eqn:hard-magnetic-energy-rod}. Dropping the constant term, the energy expression is
\begin{multline}
        \mathcal{E} = \int_{0}^{L} \Bigg[\underbrace{\frac{1}{2} \left(EI + \frac{A\pi a^2}{4}\right) \kappa_1^2 + \frac{1}{2} \left(EI + \frac{A\pi a^2}{4}\right)\kappa_2^2 + \frac{1}{2}GJ\tau^2}_\text{Elastic + Exchange energies} \\- \underbrace{\frac{K_d\pi a^2}{2} \vb*{d}_3 \cdot\vb*{h}_e}_\text{Zeeman energy} - \underbrace{T\left(\vb*{d}_3\cdot \vb*{e}_3 - 1 \right)}_\text{Work due to $T$}\Bigg] ds - \underbrace{MR}_{\text{Work due to $M$}}. \label{eqn:total-energy-threeD-hard-ferromagnetic-rod}
\end{multline}
We next derive the associated Euler-Lagrange (equilibrium) equations for both soft and hard ferromagnetic rods.

\subsection{Equilibrium equations}
{
\color{blue}
}
\paragraph{Soft ferromagnetic rod:} Following the variational procedure of \cite[Section 3.6]{Audoly2010elasticity}, we set the first variation of the total energy functional $\mathcal{E}$ from Eqn. \ref{eqn:total-energy-threeD-soft-ferromagnetic-rod} to zero. This yields the Euler-Lagrange equations expressing the balance of forces and moments along the rod respectively:
\begin{equation}
    \begin{split}
        &\vb*{F}'(s) = \vb*{0} \implies \vb*{F}(s) = \vb*{F}(0) ~(\text{constant}), \\
        &\vb*{M}'(s) + \vb*{d}_3(s)\cross \vb*{F}(s) + \frac{K_d\pi a^2}{2}\left(\vb*{d}_3(s)\cdot \vb*{m}(s)\right)\left(\vb*{d}_3(s)\cross \vb*{m}(s)\right)  = \vb*{0}
    \end{split}
\end{equation}
where $\vb*{F}(s)$ is the internal contact force and $\vb*{M}(s)$ is the internal contact moment, expressed in the director basis as $\vb*{M}(s) = M_1(s)\vb*{d}_1(s) + M_2(s)\vb*{d}_2(s) + M_3(s)\vb*{d}_3(s)$.
From Kirchhoff constitutive relations, these internal moment components relate linearly to the kinematic strains:
\begin{equation}
    M_1 = EI\kappa_1, ~M_2 = EI\kappa_2, ~M_3 = GJ\tau.
\end{equation}

{
From the variational procedure, the following boundary conditions are obtained, as discussed in \cite[Section 3.6.4]{Audoly2010elasticity}:
\begin{equation*}
    \begin{split}
       & (\vb*{F}(0) + \vb*{F}_0)\cdot \delta\vb*{r}(0) = 0, \\
       & (\vb*{M}(0) + \vb*{M}_0)\cdot \delta\vb*{\phi}(0) = 0, \\
       & (-\vb*{M}(L) + \vb*{M}_L)\cdot \delta\vb*{\phi}(L) = 0. \\
    \end{split}
\end{equation*}
Here, $\vb*{F}_0 = -T\vb*{e}_3$ denotes the external total force applied to the rod cross section at $s=0$, while $\vb*{M}_0 = -M\vb{e}_3$ and $\vb*{M}_L = M\vb*{e}_3$ represent the prescribed moments at $s=0$ and $s=L$,  respectively. Also, $\delta\vb*{r}(0)$ is the variation in the position of the rod end point at $s=0$, while $\delta\vb*{\phi}(0)$ and $\delta\vb*{\phi}(L)$ characterizes the variation in the orientation of the rod sections at $s=0$ and $s=L$ respectively.

In our configuration (see Fig. \ref{fig:threeD-problem-setup}), we consider a free-standing elastica in which both endpoints of the rod are unconstrained. Specifically, the end point of the rod at $s=0$ is free to move, making the variation $\delta \vb*{r}(0)$ arbitrary. As a consequence, the boundary condition reduces to $\vb*{F}(0) + \vb*{F}_0 = \vb*{0}$, which implies $\vb*{F}(0) = T\vb*{e}_3$. Similarly, since the orientations at both ends are unconstrained, the variations $\delta\vb*{\phi}(0)$ and $\delta \vb*{\phi}(L)$ are arbitrary. This enforces that the total torques $\vb*{M}(0) + \vb*{M}_0 = \vb*{0}$ and $-\vb*{M}(L) + \vb*{M}_L = \vb*{0}$, yielding $\vb*{M}(0) = M\vb*{e}_3$ and $\vb*{M}(L) = M\vb*{e}_3$.
}

Now, the constant internal contact force assumes the value
\begin{equation}
    \vb*{F} = T \vb*{e}_3.
\end{equation}
Substituting into the moment balance equation thus yields
\begin{equation}
    \vb*{M}' + T\vb*{d}_3\cross\vb*{e}_3 + \frac{K_d\pi a^2}{2}(\vb*{d}_3 \cdot \vb*{m})(\vb*{d}_3\cross \vb*{m}) = \vb*{0}. \label{eqn:moment-equation-soft-ferromagnetic-rod}
\end{equation}

\paragraph{Hard ferromagnetic rod:} Repeating the variational procedure for Eqn. \ref{eqn:total-energy-threeD-hard-ferromagnetic-rod}, the balance of forces and moments for a hard ferromagnetic circular rod with $\vb*{m}(s) = \vb*{d}_3(s)$ are obtained as
\begin{equation}
    \begin{split}
        &\vb*{F}' = \vb*{0} \implies \vb*{F} = T\vb*{e}_3, \\
        &\vb*{M}' + \vb*{d}_3\cross \vb*{F} + \vb*{q}_{\text{mag,hard}}  = \vb*{0}
    \end{split}
\end{equation}
where, the distributed magnetic couple derived in Eqn. \ref{app:hard-magnetic-rod-moment} is
\begin{equation}
    \vb*{q}_{\text{mag,hard}}(s) = \frac{A\pi a^2}{2}\left(\kappa_1(s)\vb*{d}_1(s) + \kappa_2(s)\vb*{d}_2(s)\right)' + \frac{K_d\pi a^2}{2}\vb*{d}_3(s)\cross\vb*{h}_e.  \label{hard-magnetic-rod-moment}
\end{equation}
Substituting $\vb*{M} = EI\kappa_1\vb*{d}_1 + EI\kappa_2\vb*{d}_2 + GJ\tau\vb*{d}_3$, the moment balance equation for the rod subjected to a longitudinal field $\vb*{h}_e = h_e\vb*{e}_3$ reads
\begin{equation}
       \vb*{M}_h' + \left(T + \frac{K_d\pi a^2}{2}h_e\right)\vb*{d}_3\cross\vb*{e}_3 = \vb*{0}, \label{eqn:moment-equation-hard-ferromagnetic-rod}
\end{equation}
where $\vb*{M}_h$ denotes the augmented internal moment and is expressed as
\begin{equation}
    \vb*{M}_h = \left(EI + \frac{A\pi a^2}{2}\right)\kappa_1 \vb*{d}_1 + \left(EI + \frac{A\pi a^2}{2}\right)\kappa_2\vb*{d}_2 + GJ\tau\vb*{d}_3.
\end{equation}
In the next section, the Hamiltonian formulations are constructed for soft and hard ferromagnetic cases.

\section{Hamiltonian formulation} \label{sec:hamiltonian-threeD}

{
\color{blue}
}

In this section, the static spatial deformation of the rod is placed in correspondence with the dynamics of a spinning top through Kirchhoff's static–dynamic analogy \cite{vanderHeijden2000helical}. The Euler–Lagrange equations are derived from the total energy functionals (Eqns.\ \ref{eqn:total-energy-threeD-soft-ferromagnetic-rod} and \ref{eqn:total-energy-threeD-hard-ferromagnetic-rod}), and the associated conserved quantities, the Casimir invariants, are identified. The Hamiltonian formulation is subsequently obtained via the Legendre transform of the energy density (Lagrangian). Completely integrable forms of the Hamiltonian are identified for both soft and hard ferromagnetic rods under the loading configurations specified in Section \ref{subsec:note-on-integrability}, and the Hamiltonian is reduced to a one-degree-of-freedom system expressed solely in terms of the primary Euler angle $\theta$. The resulting Hamiltonian structure is then employed to systematically analyze the spatial buckling and localized deformation of the ferromagnetic rod.

The Lagrangian or energy density $\mathcal{L}$ is defined as the integrand of the total energy functionals $\mathcal{E}$ in Eqns. \ref{eqn:total-energy-threeD-soft-ferromagnetic-rod} and \ref{eqn:total-energy-threeD-hard-ferromagnetic-rod}. Following the standard Legendre transform as prescribed in \cite{Dichmann1996,neukirch2025noetherian,AvatarDabade2025}, the Hamiltonian $\mathcal{H}$ is given by
\begin{equation}
  \mathcal{H} = \pdv{\mathcal{L}}{\kappa_1}\kappa_1 + \pdv{\mathcal{L}}{\kappa_2}\kappa_2 + \pdv{\mathcal{L}}{\tau}\tau - \mathcal{L} = \text{constant}.
\end{equation}

Since $\mathcal{L}$ does not depend explicitly on the arc-length coordinate $s$, i.e.\ $\pdv{\mathcal{L}}{s} = 0$, it follows that $\dv{\mathcal{H}}{s} = 0$ along any solution of the Euler-Lagrange equations, see \cite[Equations 18 and 19]{AvatarDabade2025}. Consequently, $\mathcal{H}$ is a first integral of the equilibrium equations and is conserved along the rod centerline. The Hamiltonian formulations for the soft and hard ferromagnetic rods are derived in the following sections.

\subsection{Soft ferromagnetic rod} \label{sec:soft-ferromagnetic-threeD}
From the total energy functional $\mathcal{E}$ in Eqn. \ref{eqn:total-energy-threeD-soft-ferromagnetic-rod}, the Lagrangian or energy density is
\begin{equation}
    \mathcal{L} = \frac{1}{2}EI\kappa_1^2 + \frac{1}{2}EI\kappa_2^2 + \frac{1}{2}GJ\tau^2 +  \frac{K_d\pi a^2}{4} \left[(\vb*{m}\cdot\vb*{d}_1)^2  + (\vb*{m}\cdot\vb*{d}_2)^2\right] - T\vb*{d}_3\cdot \vb*{e}_3
\end{equation}
The Hamiltonian can be derived and written as follows:
\begin{multline}
  \mathcal{H} = \pdv{\mathcal{L}}{\kappa_1}\kappa_1 + \pdv{\mathcal{L}}{\kappa_2}\kappa_2 + \pdv{\mathcal{L}}{\tau}\tau - \mathcal{L} \\
      = \frac{1}{2}EI\kappa_1^2 + \frac{1}{2}EI\kappa_2^2 + \frac{1}{2}GJ\tau^2 -  \frac{K_d\pi a^2}{4} \left[(\vb*{m}\cdot\vb*{d}_1)^2  + (\vb*{m}\cdot\vb*{d}_2)^2\right] + T\vb*{d}_3\cdot \vb*{e}_3.
\end{multline}
Using the moment balance equation, Eqn.~\ref{eqn:moment-equation-soft-ferromagnetic-rod}, together with the circular symmetry and specific loading conditions, we now derive additional conserved quantities, also known as Casimir invariants in classical mechanics literature \cite{goldstein2011classical}, that enable explicit reduction of the Hamiltonian. For the soft ferromagnetic rod, the moment balance equation reads:
\begin{equation}
    \vb*{M}' + T\vb*{d}_3\cross\vb*{e}_3 + K_d\frac{\pi a^2}{2}(\vb*{d}_3 \cdot \vb*{m})(\vb*{d}_3\cross \vb*{m}) = \vb*{0}.
\end{equation}
Taking dot product with $\vb*{d}_3$ produces $\vb*{M}' \cdot \vb*{d}_3 = 0$, or equivalently,
\begin{equation}
    \dv{(\vb*{M}\cdot\vb*{d}_3)}{s} - \vb*{M}\cdot\vb*{d}_3' = 0.
\end{equation}
Now, utilizing $EI_1 = EI_2 = EI$ for the circular cross-section, we have
\begin{multline}
    \vb*{M}\cdot\vb*{d}_3' = \vb*{M}\cdot(\vb*{\kappa}\cross\vb*{d}_3) = \vb*{M}\cdot(-\kappa_1\vb*{d}_2 + \kappa_2\vb*{d}_1) = -M_2\kappa_1 + M_1\kappa_2 \\ = -EI_2\kappa_2\kappa_1 + EI_1\kappa_1\kappa_2 = 0,
\end{multline}
resulting in the first Casimir invariant \cite{Sinden2008}:
\begin{equation}
   \vb*{M}(s)\cdot\vb*{d}(s) = M_3(s) = GJ\tau = \beta ~~(\text{constant}). \label{eqn:casimir-invariant-1}
\end{equation}
The analysis is henceforth restricted to the case in which the applied longitudinal magnetic field $\vb*{h}_e = h_e\vb*{e}_3$ is sufficiently large to saturate the magnetization, so that $\vb*{m} = \vb*{e}_3$ uniformly along the rod, irrespective of the deformed configuration of the centerline. The moment balance equation simplifies to:
\begin{equation}
    \vb*{M}' + \left(\frac{K_d\pi a^2}{2}(\vb*{d}_3 \cdot \vb*{e}_3) + T\right)(\vb*{d}_3\cross \vb*{e}_3) = \vb*{0}.
\end{equation}
Projecting this onto $\vb*{e}_3$ gives the second Casimir invariant:
\begin{equation}
    \begin{split}
        \dv{(\vb*{M}(s)\cdot\vb*{e}_3)}{s} &= 0 \implies \vb*{M}(s)\cdot\vb*{e}_3 = \alpha = M, \label{eqn:casimir-invariant-2}
    \end{split}
\end{equation}
where $M$ prescribes the applied end twisting moment.
We now expand the conserved quantity $\alpha$ in terms of the Euler angles as follows:
\begin{equation}
\begin{split}
    \alpha = \vb*{M}(s)\cdot\vb*{e}_3 &= EI\kappa_1\vb*{d}_1\cdot\vb*{e}_3 + EI\kappa_2\vb*{d}_2\cdot\vb*{e}_3 + GJ\tau \vb*{d}_3\cdot\vb*{e}_3 \\
    &= -EI\kappa_1 \cos\phi\sin\theta + EI\kappa_2\sin\phi\sin\theta + \beta\cos\theta.
\end{split}
\end{equation}
Substituting the strain--angular rate relations from Eqn. \ref{eqn:euler-rates-strains-relations} and simplifying yields
\begin{equation}
\begin{split}
    \alpha &= EI\psi'\sin^2\theta + \beta\cos\theta \implies \psi' = \frac{\alpha - \beta\cos\theta}{EI\sin^2\theta},
\end{split}
\end{equation}
and
\begin{equation}
    \phi' = \frac{\beta}{GJ} - \psi' \cos\theta.
\end{equation}
The curvatures and twist then follow as
\begin{equation}
    \begin{split}
        \kappa_1 &= \theta'\sin\phi - \frac{(\alpha - \beta\cos\theta)}{EI\sin\theta}\cos\phi, \\
        \kappa_2 &= \theta'\cos\phi + \frac{\alpha - \beta\cos\theta}{EI\sin\theta}\sin\phi, \\
        \tau &= \frac{\beta}{GJ}.
    \end{split}
    \label{eqn:threeD-curvatures-in-terms-of-euler-angles}
\end{equation}
With appropriate substitutions, the Hamiltonian for soft ferromagnetic rod subjected to longitudinal magnetic field reduces to
\begin{equation}
    \mathcal{H} = \frac{1}{2}EI\theta'^2 + \frac{1}{2}\frac{(\alpha - \beta\cos\theta)^2}{EI\sin^2\theta} + \frac{1}{2}\frac{\beta^2}{GJ} - \frac{K_d\pi a^2}{4}\sin^2\theta + T\cos\theta.
\end{equation}
The Hamiltonian is thus expressed solely in terms of the Euler angle $\theta$ and its arc-length derivative $\theta^{'}$, rendering the system amenable to systematic phase-space analysis. For the localized buckling case where $\vb*{d}_3$ is parallel to $\vb*{e}_3$ at certain points along the rod, we set
\begin{equation}
    \alpha = \beta = M = GJ\tau. \label{eqn:threeD-m-tau-relation}
\end{equation}
Dropping the constant term $\frac{1}{2}\frac{\beta^2}{GJ}$, the Hamiltonian simplifies to
\begin{equation}
       \mathcal{H}_\text{soft} = \frac{1}{2}EI\theta'^2 + \frac{1}{2}\frac{M^2}{EI}\frac{(1 - \cos\theta)}{(1 + \cos\theta)} - \frac{K_d\pi a^2}{4}\sin^2\theta + T\cos\theta. \label{eqn:threeD-soft-ferromagnetic-longitudinal-Hamiltonian}
\end{equation}
The Euler angle rates $\psi'$ and $\phi'$ correspondingly simplify as
\begin{equation}
    \begin{split}
        \psi' &= \frac{M(1 - \cos\theta)}{EI\sin^2\theta} = \frac{M}{EI}\frac{1}{1 + \cos\theta}, \\
        \phi' &= \frac{M}{GJ} - \psi' \cos\theta = \frac{M}{EI}\frac{EI}{GJ} - \frac{M}{EI}\frac{1}{1 + \cos\theta} = \frac{M}{EI}\left(\nu + \frac{1}{1 + \cos\theta}\right),
    \end{split} \label{}
\end{equation}
where $\frac{EI}{GJ} = 1 + \nu$, with shear modulus $G = \frac{E}{2(1+\nu)}$, polar area moment $J = 2I$, and Poisson's ratio $\nu$.

\noindent The Hamiltonian for the purely elastic case is recovered by setting $K_d = 0$ in Eqn. \ref{eqn:threeD-soft-ferromagnetic-longitudinal-Hamiltonian}, and is consistent with the expression reported in \cite{vanderHeijden2000helical}.

\begin{equation}
\mathcal{H}_{\text{elastic}} = \frac{1}{2}EI\theta'^2 + \frac{1}{2}\frac{M^2}{EI}\frac{(1 - \cos\theta)}{(1 + \cos\theta)} + T\cos\theta.    \label{eqn:threeD-purely-elastic-Hamiltonian}
\end{equation}

\subsubsection*{Non-dimensionalization of the Hamiltonian}
The Hamiltonian in Eqn.~\ref{eqn:threeD-soft-ferromagnetic-longitudinal-Hamiltonian} admits two natural non-dimensionalizations, depending on whether the twisting moment $M$ or tension $T$ is held fixed as the primary control parameter.

\paragraph{Fixed-$M$ scaling:} We define the following non-dimensional quantities
\begin{equation}
    \begin{split}
        &\tilde{s} = s\frac{M}{EI},~ \dot{\theta} = \frac{EI}{M}\theta',~\dot{\psi} = \frac{EI}{M}\psi',~\dot{\phi} = \frac{EI}{M}\phi', \\
        &\tilde{K}_d = \frac{K_d\pi a^2}{4M^2/EI}, ~\tilde{M} = \frac{M}{\sqrt{EIT}}, ~\tilde{\mathcal{H}}_\text{soft} = \frac{\mathcal{H}_\text{soft}}{M^2/EI}.
    \end{split}
\end{equation}
For notational simplicity, we denote $\dot{()}$ by $()'$, where $()$ denotes $\theta$, $\psi$ and $\phi$. Under this convention, the non-dimensional Hamiltonian becomes
\begin{equation}
    \tilde{\mathcal{H}}_\text{soft} = \frac{1}{2}\theta'^2 + \frac{1}{2}\frac{(1 - \cos\theta)}{(1 + \cos\theta)} - \tilde{K}_{dM}\sin^2\theta + \frac{\cos\theta}{\tilde{M}^2}.
    \label{eqn:threeD-soft-ferromagnetic-longitudinal-Hamiltonian-non-dimensionalized-M}
\end{equation}
with the non-dimensional Euler angle rates
\begin{equation}
    \begin{split}
        \psi' = \frac{1}{1 + \cos\theta}, \qquad
        \phi' = \nu + \frac{1}{1 + \cos\theta}.
    \end{split} \label{eqn:euler-angle-rates-non-dimensionalized-M}
\end{equation}
\paragraph{Fixed-$T$ scaling:} With the introduction of the following non-dimensional quantities
\begin{equation}
\begin{split}
    &\tilde{s} = s\sqrt{\frac{T}{EI}},~ \dot{\theta} = \sqrt{\frac{EI}{T}}\theta',~~ \dot{\psi} = \sqrt{\frac{EI}{T}}\psi',~\dot{\phi} = \sqrt{\frac{EI}{T}}\phi', \\
    &\tilde{K}_{dT} = \frac{K_d\pi a^2}{4T}, ~\tilde{\mathcal{H}}_\text{soft} = \frac{\mathcal{H}_\text{soft}}{T},
\end{split}
\end{equation}
the non-dimensional Hamiltonian (where, by a slight abuse of notation, $()'$ is used in place of $\dot{()}$ where $() \equiv  \theta,\psi,\phi$) takes the form
\begin{equation}
    \tilde{\mathcal{H}}_\text{soft} = \frac{1}{2}\theta'^2 + \frac{\tilde{M}^2}{2}\frac{(1 - \cos\theta)}{(1 + \cos\theta)} - \tilde{K}_{dT}\sin^2\theta + \cos\theta,
    \label{eqn:threeD-soft-ferromagnetic-longitudinal-Hamiltonian-non-dimensionalized-T}
\end{equation}
and the Euler angle rates are
\begin{equation}
    \begin{split}
        \psi' &= \frac{\tilde{M}}{1 + \cos\theta}, \qquad \phi' = \tilde{M}\left(\nu + \frac{1}{1 + \cos\theta}\right).
    \end{split} \label{eqn:euler-angle-rates-non-dimensionalized-T}
\end{equation}
Both $\tilde{K}_{dM}$ and $\tilde{K}_{dT}$ represent magnetoelastic parameters, while $\tilde{M} = \frac{M}{\sqrt{EIT}}$ is referred to as the composite mechanical load parameter. As remarked by Thompson and Champneys \cite{thompson1996helix}, controlling $M$ or $\frac{1}{\sqrt{T}}$ is equivalent and its dynamic similarities has been previously established by Champneys and Thompson \cite{champneys1996multiplicity}.

\subsection{Hard ferromagnetic rod}
We now analyze the Hamiltonian of a uniaxial hard ferromagnetic circular rod with tangential magnetization $\vb*{m}(s) = \vb*{d}_3(s)$. The Lagrangian $\mathcal{L}$ is derived from the total energy expression in Eqn. \ref{eqn:total-energy-threeD-hard-ferromagnetic-rod} as
\begin{equation}
    \mathcal{L} = \frac{1}{2} \left(EI + \frac{A\pi a^2}{4}\right) \kappa_1^2 + \frac{1}{2} \left(EI + \frac{A\pi a^2}{4}\right)\kappa_2^2 + \frac{1}{2}GJ\tau^2 \\- \frac{K_d\pi a^2}{2} \vb*{d}_3 \cdot\vb*{h}_e - T\left(\vb*{d}_3\cdot \vb*{e}_3 - 1 \right)
\end{equation}
From Eqn. \ref{eqn:moment-equation-hard-ferromagnetic-rod}, the moment balance equation is obtained as
\begin{equation}
       \vb*{M}_h' + \left(T + \frac{K_d\pi a^2}{2}h_e\right)\vb*{d}_3\cross\vb*{e}_3 = \vb*{0},
\end{equation}
where $\vb*{M}_h = \bigl( EI + \frac{A \pi a^2}{2} \bigr) (\kappa_1 \vb*{d}_1 + \kappa_2 \vb*{d}_2) + GJ \tau \vb*{d}_3$.

Similar to the procedure for the soft ferromagnetic rod, projecting  the above equation onto $\vb*{d}_3$ and $\vb*{e}_3$ produces two Casimir invariants:
\begin{equation}
    \begin{split}
        \vb*{M}_h(s)\cdot \vb*{e}_3 &= \alpha_h, \\
        \vb*{M}_h(s)\cdot \vb*{d}_3 &= \beta_h.
    \end{split}
\end{equation}
The Hamiltonian is derived as
\begin{equation}
    \mathcal{H} = \frac{1}{2}\left(EI + \frac{A\pi a^2}{2}\right)\kappa_1^2 + \frac{1}{2}\left(EI + \frac{A\pi a^2}{2}\right)\kappa_2^2 + \frac{1}{2}GJ\tau^2 + \frac{K_d\pi a^2}{2} \vb*{m}\cdot\vb*{h}_e + T\vb*{d}_3\cdot \vb*{e}_3.
\end{equation}
Substituting the above Casimir invariants $(\alpha_h,\beta_h)$ in the expressions for curvatures $(\kappa_1,\kappa_2,\tau)$ (Eqns. \ref{eqn:threeD-curvatures-in-terms-of-euler-angles}), the Hamiltonian is expressed in terms of the primary Euler angle $\theta$ and its derivative $\theta^{'}$ as
\begin{equation}
    \mathcal{H} = \frac{1}{2}\left(EI + \frac{A\pi a^2}{2}\right)\theta'^2 + \frac{1}{2}\frac{(\alpha_h - \beta_h\cos\theta)^2}{\left(EI + \frac{A\pi a^2}{2}\right)\sin^2\theta} + \frac{1}{2}\frac{\beta^2}{C} + \frac{K_d\pi a^2}{2}\cos\theta + T\cos\theta.
\end{equation}
As discussed previously in the localized buckling of a soft ferromagnetic rod, $\vb*{d}_3$ is parallel to $\vb*{e}_3$ at certain points along the rod. This leads us to the condition: $\alpha_h = \beta_h = M$. Dropping the constant $\nicefrac{\beta_h^2}{2 GJ}$, the Hamiltonian reduces to
\begin{equation}
       \mathcal{H}_\text{hard} = \frac{1}{2}\left(EI + \frac{A\pi a^2}{2}\right)\theta'^2 + \frac{1}{2}\frac{M^2}{\left(EI + \frac{A\pi a^2}{2}\right)}\frac{(1 - \cos\theta)}{(1 + \cos\theta)} + \left(T + \frac{K_d\pi a^2}{2}h_e\right)\cos\theta. \label{eqn:threeD-hard-ferromagnetic-longitudinal-Hamiltonian}
\end{equation}
\subsubsection{Structural equivalence to the purely elastic Hamiltonian} \label{subsec:structural-equivalence-Hamiltonian}
Although the Hamiltonian for the hard ferromagnetic case preserves the structural form of its purely elastic counterpart (Eqn. \ref{eqn:threeD-purely-elastic-Hamiltonian}), magnetic effects induce two key modifications: (i) the exchange energy augments the effective bending stiffness to $\left(EI + \frac{A\pi a^2}{2}\right)$, and (ii) the Zeeman energy enhances the external terminal load to $\left(T + \frac{K_d\pi a^2}{2}h_e\right)$. Accordingly, the bifurcation characteristics of the hard ferromagnetic elastic rod mirror those of a purely elastic rod, provided the parameters in Eqn. \ref{eqn:threeD-purely-elastic-Hamiltonian} are replaced by their renormalized counterparts.

\subsection{A short note on integrability} \label{subsec:note-on-integrability}
The subsequent sections present a Hamiltonian analysis for the purely elastic and soft ferromagnetic rods.

These reductions highlight the integrable structure preserved under the stated
assumptions, as summarized next. It is well established
\cite{vanderHeijden2000helical,vanderHeijden2001,vanderHeijden2002} that
complete integrable Kirchhoff rod systems admit phase space analysis via
conserved quantities. Integrability requires two conditions: (i) an isotropic
cross-section with equal principal bending stiffnesses $EI_1 = EI_2$
\cite{vanderHeijden2001}, and (ii) mechanical/magnetic loading that preserves
the two Casimir invariants of the bending moment
(Eqns.~\ref{eqn:casimir-invariant-1} and \ref{eqn:casimir-invariant-2}).
The significance of these two conditions is as follows. When both are satisfied,
the two Casimir invariants permit the elimination of the Euler angles $\psi$
and $\phi$ and their derivatives from the Hamiltonian, reducing it to a
function of the single primary Euler angle $\theta$ and its arc-length
derivative $\theta'$ alone, that is, $H = H(\theta, \theta') = \text{constant}$.
This reduction to a one-degree-of-freedom system is the essential consequence
of integrability: the phase space becomes two-dimensional with coordinates
$(\theta, \theta')$, and since $H$ is conserved along every solution trajectory,
each trajectory lies on a level curve of $H(\theta, \theta')$. The complete
solution structure: equilibria, periodic helical orbits, and homoclinic
orbits corresponding to localized buckling modes, is therefore encoded
in the topology of these level curves and is directly accessible through
phase portrait analysis. Without integrability, the Hamiltonian would depend
on all three Euler angles and their derivatives, yielding a higher-dimensional
phase space in which level sets of $H$ alone do not determine solution
trajectories, and the phase portrait analysis presented in Section
\ref{sec:threeD-analysis-Hamiltonian} would not be possible. The analyzed scenarios
satisfying both conditions are:
\begin{itemize}
    \item Case (i): Soft ferromagnetic rod immersed in longitudinal magnetic
    field $\vb*{h}_e = h_e\vb*{e}_3$, with uniform magnetization
    $\vb*{m}(s) = \vb*{e}_3$;
    \item Case (ii): Hard ferromagnetic rod with tangential magnetization
    $\vb*{m}_3(s) = \vb*{d}_3(s)$ subjected to longitudinal field
    $\vb*{h}_e = h_e\vb*{e}_3$.
\end{itemize}
The subsequent sections present a Hamiltonian analysis for the purely elastic
and soft ferromagnetic rods.

\section{Analysis of the Hamiltonian} \label{sec:threeD-analysis-Hamiltonian}
The analysis that follows is conducted primarily in terms of the non-dimensionalized Hamiltonian under fixed-$M$ scaling, where $M$ denotes the applied end twisting moment. We begin by evaluating the critical points of the Hamiltonian. We then employ a linear local analysis of the equilibrium equations to locate the bifurcation points where the sole trivial configuration of the rod gives rise to additional equilibrium solutions. We then examine the corresponding phase portraits.

\subsection{Critical points}
Following the procedure in \cite[Section 3]{AvatarDabade2025}, we determine the critical points of the Hamiltonian of a soft ferromagnetic rod immersed in longitudinal magnetic field $\tilde{\mathcal{H}} = \tilde{\mathcal{H}}_\text{soft}$ as given in Eqn.~\ref{eqn:threeD-soft-ferromagnetic-longitudinal-Hamiltonian-non-dimensionalized-M} under fixed-$M$ condition and classify their nature. These critical points satisfy $\grad \tilde{\mathcal{H}}  = \vb*{0}$, or
\begin{equation}
	\grad \tilde{\mathcal{H}}  = \begin{pmatrix}
		\pdv{\tilde{\mathcal{H}}}{\theta} \\
		\pdv{\tilde{\mathcal{H}}}{\theta'}
	\end{pmatrix} = \vb*{0}.
\end{equation}
For the soft ferromagnetic rod in longitudinal field $\vb*{h}_e$, Eqn. \ref{eqn:threeD-soft-ferromagnetic-longitudinal-Hamiltonian-non-dimensionalized-M} gives
\begin{equation}
	\begin{pmatrix}
		\pdv{\tilde{\mathcal{H}}}{\theta} \\
		\pdv{\tilde{\mathcal{H}}}{\theta'}
	\end{pmatrix} = \vb*{0} \implies \begin{cases}
		\sin\theta \bigg[\frac{1}{(1+\cos\theta)^2} - 2\tilde{K}_{dM}\cos\theta - \frac{1}{\tilde{M}^2}\bigg] &= 0, \\  \theta'  &= 0,
	\end{cases} \label{eqn:threeD-critical-point-equations}
\end{equation}
and $\theta_c' = 0$ for all critical points. The physical constraint $\sin\theta\big\rvert_{\theta=\theta_c} = 0$ (analogous to the spinning top) admits only the trivial solution $\theta_c = 0$. Nontrivial solutions $\theta_c \neq 0$, if they exist, solve
\begin{equation}
    \frac{1}{(1+\cos\theta)^2} - 2\tilde{K}_{dM}\cos\theta - \frac{1}{\tilde{M}^2}\bigg\rvert_{\theta=\theta_c} = 0. \label{eqn:threeD-critical-points}
\end{equation}

The trivial critical point $p_1 = (0,0)$ exists for all values of $\tilde{M}$ and $\tilde{K}_{dM} > 0$, and corresponds to the straight, undeformed equilibrium configuration of the rod. Non-trivial critical points, when they exist, appear as the symmetric pair $p_{2,3} = (\mp \theta_c, 0)$, each corresponding to a uniformly helically deformed equilibrium state. In the purely elastic limit $\tilde{K}_{dM}=0$, the condition in Eqn.~\ref{eqn:threeD-critical-points} yields no real solution for $\theta_c$ when $\tilde{M} > 2$, and hence the straight configuration is the unique equilibrium of the rod in this regime.

The nature of each critical point is determined from the Hessian matrix of $\tilde{\mathcal{H}}$:
\begin{equation}
	\grad^2\tilde{\mathcal{H}} = \begin{bmatrix}
    \frac{\partial^2\tilde{\mathcal{H}}}{\partial\theta^2}  & \pdv[2]{\tilde{\mathcal{H}}}{\theta}{\theta'} \\
		\pdv[2]{\tilde{\mathcal{H}}}{\theta'}{\theta} & \frac{\partial^2\tilde{\mathcal{H}}}{\partial\theta'^2}
	\end{bmatrix} = \begin{bmatrix}
	\bigg(\frac{1 + \sin^2\theta + \cos\theta}{(1+\cos\theta)^3} - 2\tilde{K}_{dM}\cos 2\theta - \frac{\cos\theta}{\tilde{M}^2}\bigg) & 0 \\
	0 & 1
\end{bmatrix}.
\end{equation}

The Hessian is diagonal, and its two eigenvalues are therefore
$\lambda_1 = \pdv[2]{\tilde{\mathcal{H}}}{\theta}$ and $\lambda_2 = 1$. Since $\lambda_2$ is strictly positive, the character of each critical point is governed entirely by the sign of $\lambda_1$. A critical point is a
center when $\lambda_1 > 0$, in which case both eigenvalues are positive and the surrounding phase-plane trajectories are closed orbits corresponding to periodic, helical buckling modes. The critical point is a saddle when $\lambda_1 <0 $, in which case the eigenvalues are of opposite sign and the associated phase-plane trajectories are homoclinic orbits corresponding to localized buckling modes.

\subsection{Bifurcation points of the Hamiltonian}\label{sec:bifurcation-points-Hamiltonian}

Having classified the critical points, we analyze the bifurcation characteristics
of the Hamiltonian for the soft ferromagnetic rod given in Case (i) (Eqn. \ref{eqn:threeD-soft-ferromagnetic-longitudinal-Hamiltonian}), given as follows:
\begin{equation}
    \frac{1}{2}EI\theta'^2 + \frac{1}{2}\frac{M^2}{EI}\frac{(1 - \cos\theta)}{(1 + \cos\theta)} - \frac{K_d\pi a^2}{4}\sin^2\theta + T\cos\theta = \mathcal{H}_\text{soft}.
\end{equation}

Since $\mathcal{H}_\text{soft}$ is a first integral of the Euler--Lagrange equations \cite{vanderHeijden2000helical}, differentiating it with respect to $s$
and simplifying yields:
\begin{equation}
    EI\theta'\theta'' + \left(\frac{M^2}{EI}\frac{\sin\theta}{(1 + \cos\theta)^2} - \frac{K_d\pi a^2}{2}\sin\theta\cos\theta - T\sin\theta\right)\theta' = 0,
\end{equation}

The above equation admits the trivial solution $\theta' = 0$, which corresponds to the straight, undeformed equilibrium configuration of the rod. Factoring out $\theta'$ and retaining the non-trivial factor recovers the Euler--Lagrange equation:
\begin{equation}
    EI\theta'' + \left(\frac{M^2}{EI}\frac{\sin\theta}{(1 + \cos\theta)^2} - \frac{K_d\pi a^2}{2}\sin\theta\cos\theta - T\sin\theta\right) = 0.
\end{equation}

The Euler--Lagrange equation is linearized about the trivial equilibrium
$\theta = 0$ by substituting $\theta = \epsilon\bar{\theta}$, where
$|\epsilon| \ll 1$, yielding
\begin{equation}
    EI\bar{\theta}'' + \left(\frac{M^2}{4EI} - \frac{K_d\pi a^2}{2}
    - T\right)\bar{\theta} = 0.
\end{equation}
Substituting the ansatz $\bar{\theta} \sim e^{\lambda s}$ into the above
yields the characteristic equation
\begin{equation}
    EI\lambda^2 + \left(\frac{M^2}{4EI} - \frac{K_d\pi a^2}{2}
    - T\right) = 0,
\end{equation}
with eigenvalues
\begin{equation}
    \lambda_{1,2} = \pm\, i \sqrt{\frac{1}{EI}\left(\frac{M^2}{4EI}
    - \frac{K_d\pi a^2}{2} - T\right)}.
\end{equation}
The trivial equilibrium is stable, and the rod remains straight, when the
eigenvalues $\lambda_{1,2}$ are purely imaginary, which requires
\begin{equation}
    \frac{M^2}{4EI} - \frac{K_d\pi a^2}{2} - T \geq 0 \qquad
    \Longleftrightarrow \qquad \frac{M^2}{4EI} - T \geq
    \frac{K_d\pi a^2}{2}.
\end{equation}
Bifurcation from the trivial equilibrium occurs at the transition point
where the eigenvalues become zero, i.e.\ when
\begin{equation}
    \frac{M^2}{4EI} - T = \frac{K_d\pi a^2}{2}.
\end{equation}
Expressing this bifurcation condition in non-dimensional form under each
scaling yields the following results. \\

\noindent \textit{Fixed-$M$ scaling}: The bifurcation condition becomes
   \begin{equation}
    \frac{1}{4} - \frac{1}{\tilde{M}^2} = 2\tilde{K}_{dM} \implies \tilde{M}^2 = \frac{1}{\frac{1}{4} - 2\tilde{K}_{dM}}.
    \end{equation}
For $\tilde{M}^2$ to be real and finite, it is necessary that
$\frac{1}{4} - 2\tilde{K}_{dM} > 0$, which imposes the restriction
\begin{equation}
    \tilde{K}_{dM} < \frac{1}{8}.
\end{equation}
Consequently, the soft ferromagnetic rod exhibits bifurcation only within
the magnetoelastic parameter regime $0 < \tilde{K}_{dM} < \frac{1}{8}$.
For $\tilde{K}_{dM} \geq \frac{1}{8}$, the bifurcation is suppressed
entirely, as discussed further in Section \ref{subsec:hamiltonian-phase-portrait-soft-ferromagnetic}.
\\

\noindent \textit{Fixed-$T$ scaling}: The bifurcation condition reduces to
\begin{equation}
        \frac{\tilde{M}^2}{4} - 1 = 2\tilde{K}_{dT} \implies \tilde{M}^2 = 4(1 + 2\tilde{K}_{dT}),
\end{equation}
which admits a real finite bifurcation point for all $\tilde{K}_{dT} >
-\frac{1}{2}$. Since $\tilde{K}_{dT} = K_d\pi a^2/4T > 0$ for all
physically admissible parameter values, this condition is always satisfied,
and the supercritical Hamiltonian Hopf pitchfork bifurcation is exhibited
by the soft ferromagnetic rod for all physically admissible values of the
magnetoelastic parameter under fixed-$T$ scaling.

\subsection{Hamiltonian phase portraits}

We analyze the Hamiltonian phase portraits for purely elastic, soft ferromagnetic, and hard ferromagnetic rods as the composite mechanical load parameter $\tilde{M}$ varies.

\subsubsection{Purely elastic rod}
\begin{figure}[ht!]
    \centering
    \includegraphics[width=0.8\linewidth]{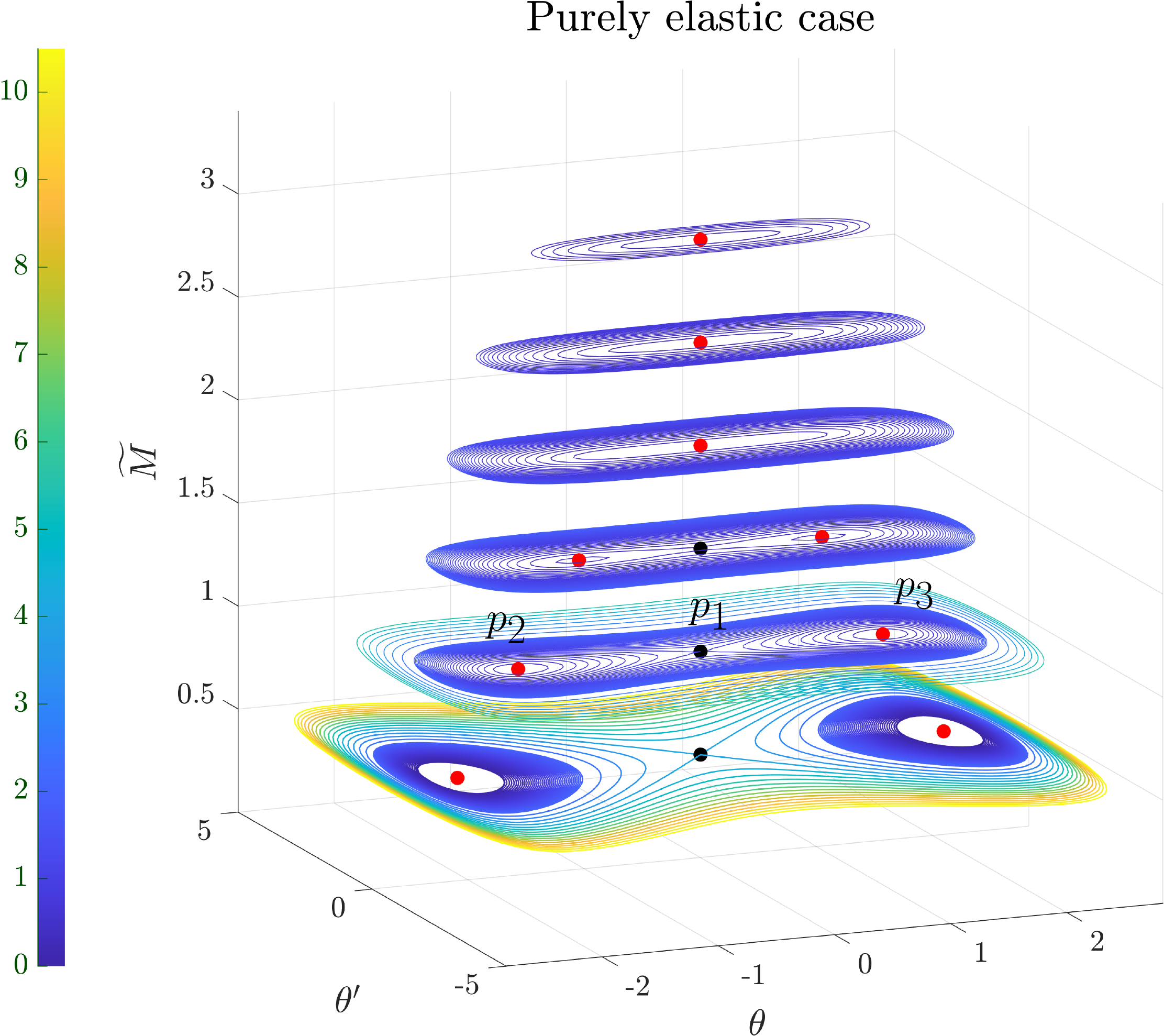}
    \caption{Purely elastic case: Evolution of phase portraits as $\tilde{M}$ varies.}
    \label{fig:phase-portrait-evolution-purely-elastic}
\end{figure}

We track the evolution of the Hamiltonian phase portrait for the purely elastic case versus the composite mechanical load parameter $\tilde{M}$ in Fig.~\ref{fig:phase-portrait-evolution-purely-elastic}. For $\tilde{M} > 2$, a single center exists at $p_1$. As $\tilde{M}$ decreases through the bifurcation point $\tilde{M} = 2$, $p_1$ undergoes a supercritical Hamiltonian Hopf pitchfork bifurcation \cite{vanderHeijden2000helical,champneys1996multiplicity}, forming two additional centers at $(p_2,p_3)$, while tranforming itself into a saddle point. The resulting phase space further features homoclinic orbits connecting the saddle point $p_1$ to itself, representing localized buckling modes. The homoclinic orbits or separatrices delineates the shorter closed loop trajectories encircling each center $p_{2,3}$ individually and the longer closed loop trajectories enclosing all three critical points. Also, the non-trivial centers ($\theta_c\neq 0$) correspond to helical buckling modes.

\subsubsection{Soft ferromagnetic rod} \label{subsec:hamiltonian-phase-portrait-soft-ferromagnetic}
\begin{figure}[ht!]
    \centering
    \includegraphics[width=0.8\linewidth]{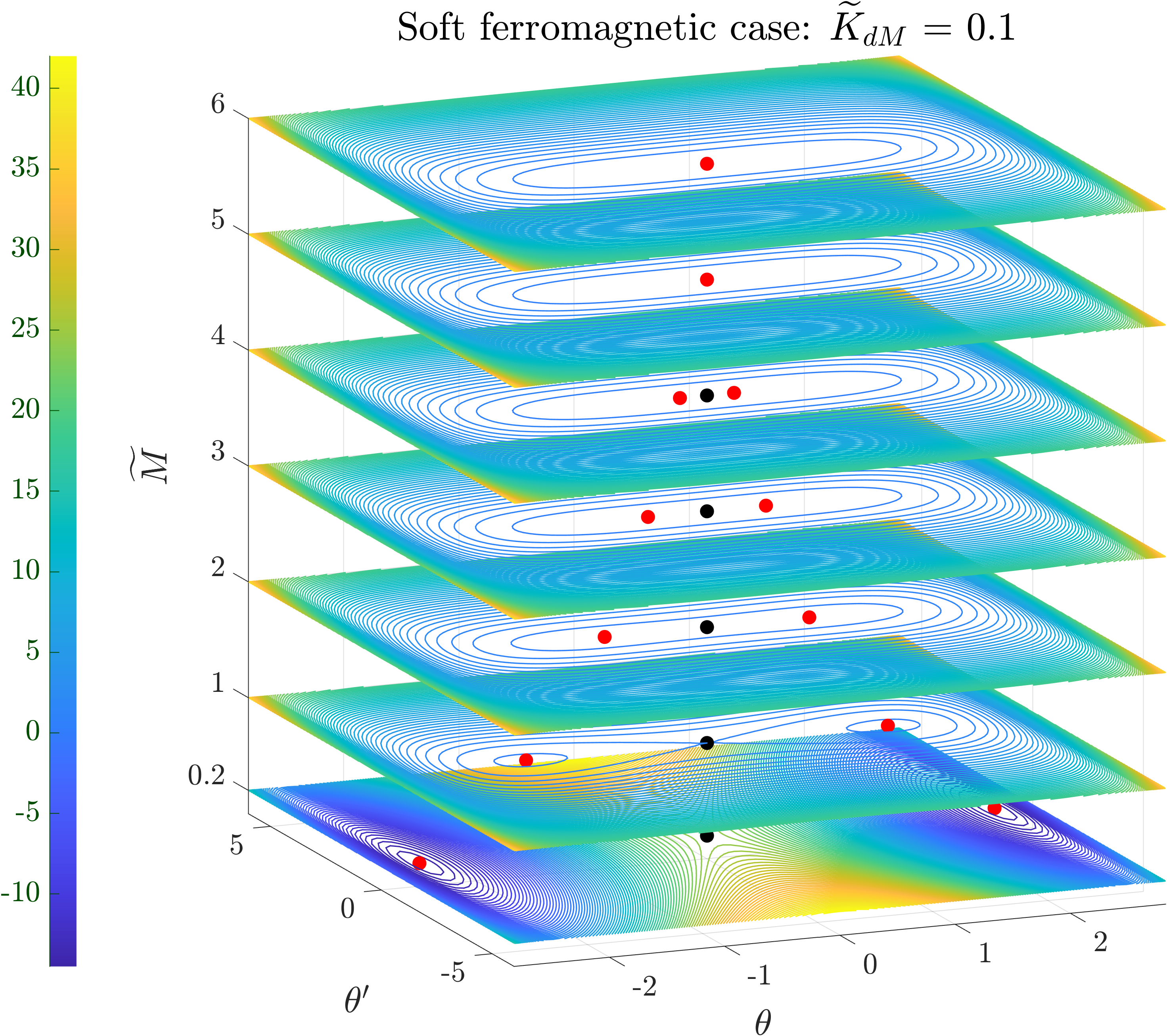}
    \caption{Soft ferromagnetic case: Evolution of phase portraits as $\tilde{M}$ varies for $\tilde{K}_{dM} = 0.1$.}
    \label{fig:phase-portrait-evolution-soft-ferromagnetic-KdM-0.1}
\end{figure}

The evolution of Hamiltonian phase portraits of a soft ferromagnetic rod for a fixed value of $\tilde{K}_{dM} = 0.1$ is illustrated in Fig. \ref{fig:phase-portrait-evolution-soft-ferromagnetic-KdM-0.1}. The supercritical Hamiltonian Hopf pitchfork bifurcation originates at $\tilde{M} = \frac{1}{1 - 8\tilde{K}_{dM}} = \sqrt{20} \approx 4.47$. Interestingly, non-trivial branches emerge in a cusp-like manner at the bifurcation point, representing a distinct departure from the behavior observed in the purely elastic case.

\begin{figure}[ht!]
    \centering
    \includegraphics[width=0.8\linewidth]{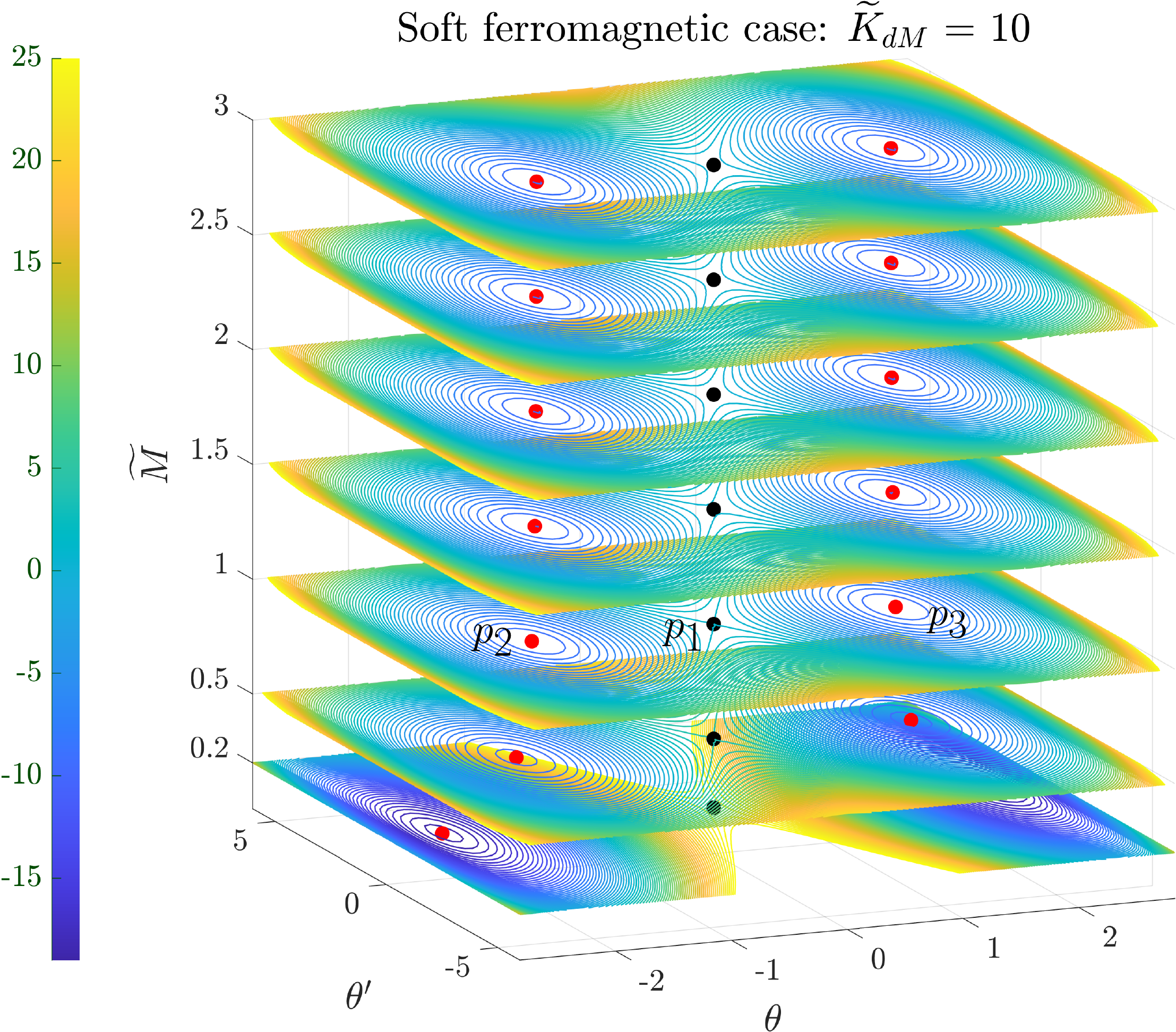}
    \caption{Soft ferromagnetic case: Evolution of phase portraits as $\tilde{M}$ varies for $\tilde{K}_{dM} = 10$.}
    \label{fig:phase-portrait-evolution-soft-ferromagnetic-KdM-10}
\end{figure}

For $\tilde{K}_{dM} > \frac{1}{8}$, three critical points persist for all $\tilde{M}$, as shown in Fig. \ref{fig:phase-portrait-evolution-soft-ferromagnetic-KdM-10}. It is noteworthy that higher values of $\tilde{K}_{dM}$ suppress any bifurcation in the Hamiltonian. The critical points gradually $p_2$ and $p_3$ drifts towards each other with increasing $\tilde{M}$, asymptotically approaching the limits $\mp \theta_{c,\text{asymp}}$, which are determined from the roots of
\begin{equation}
    \frac{1}{(1+\cos\theta)^2} - 2\tilde{K}_{dM}\cos\theta \bigg\rvert_{\theta=\theta_{c,\text{asymp}}} = 0. \label{eqn:threeD-critical-points-asymptotes}
\end{equation}

\clearpage

\subsubsection{Hard ferromagnetic rod}

The Hamiltonian phase portraits of the hard ferromagnetic elastic rod closely resemble those of the purely elastic rod (Fig. \ref{fig:phase-portrait-evolution-purely-elastic}). This resemblance is a direct consequence of the structural equivalence established in Section \ref{subsec:structural-equivalence-Hamiltonian}: the hard ferromagnetic Hamiltonian (Eqn. \ref{eqn:threeD-hard-ferromagnetic-longitudinal-Hamiltonian}) retains the same mathematical form as its purely elastic counterpart, with the bending stiffness and loading parameter appropriately renormalized by the magnetic contributions.

In summary, the phase portraits illustrate a diverse post-buckling landscape. The centers in the phase portraits denote helical deformation mode while the saddle points signify the presence of localized buckling modes. This characterization provides the foundation for the detailed analysis of helical buckling presented in the subsequent section.

\section{Post-buckling into a local helix} \label{sec:threeD-local-helix}
\begin{figure}[ht!]
    \centering
    \includegraphics[width=0.9\linewidth]{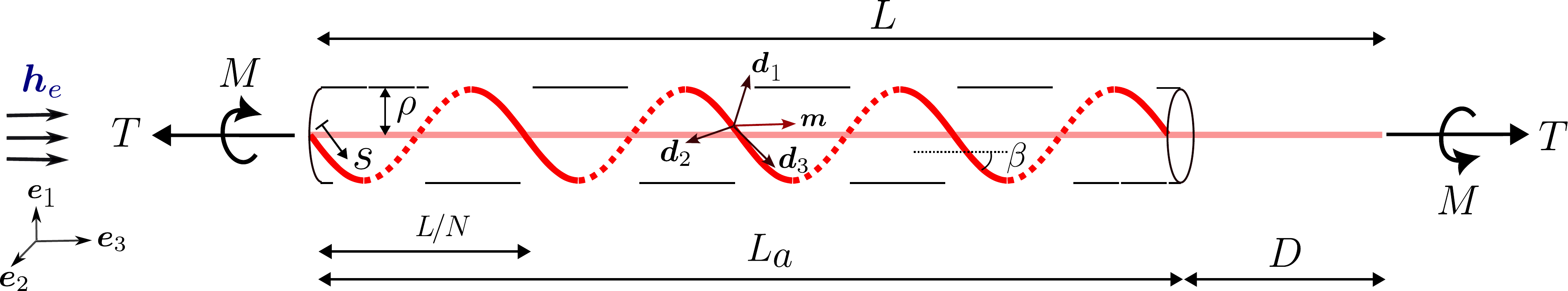}
    \caption{Schematic of a helically deformed ferromagnetic rod.}
    \label{fig:schematic-helix}
\end{figure}

We now study the mechanical response of the helically deformed ferromagnetic rod subjected to the combined effect of twisting moment $M$, tension $T$, and magnetic loading, as sketched in Fig. \ref{fig:schematic-helix}.

Let us briefly describe the geometry of the helically deformed rod. The centerline of the helical rod of length $L$ is defined by two geometrical parameters: radius $\rho$ and pitch or helix angle $\beta$; $\beta = \theta_c$ is obtained from Eqn. \ref{eqn:threeD-critical-points}. The incorporation of the cross sectional effects of the rod requires an additional geometric (kinematic) parameter called internal twist $\tau_i$ as discussed in equations \ref{eqn:tau1} and \ref{eqn:helix-end-rotation-expression}. The projected length of the helix along its axis is $L_a = L \cos\beta$, and the end-shortening length is $D$. The helix geometry composed of $N$ repeated units satisfies
\begin{equation}
\begin{split}
    &\tan\beta = \frac{2\pi\rho}{L_a} \implies L_a = \frac{2\pi\rho}{\tan\beta}, \\
    &\sin\beta = \frac{2\pi\rho}{L/N} \implies N = \frac{L\sin\beta}{2\pi\rho}, \\
    &D = L(1 - \cos\beta) \implies d = \frac{D}{L} = 1 - \cos\beta.
\end{split}
\label{eqn:threeD-helix-geometry}
\end{equation}
From the theory of space curves, the arc length parametrization of the helix is \cite{Pressley2010book}
\begin{equation}
  \begin{split}
      \vb*{r}(s) &= (x(s),y(s),z(s)),   \\
       \text{ where }  & x(s) = \rho\cos(\frac{s\sin\beta}{\rho}), \quad y(s) = \rho\sin(\frac{s\sin\beta}{\rho}), \quad z(s) = s \cos\beta.
  \end{split}
\end{equation}
The corresponding orthonormal Frenet-Seret frame ($\vb*{t},\vb*{n},\vb*{b}$) is defined as
\begin{equation}
    \vb*{t}(s) = \vb*{r}'(s), \quad \vb*{n}(s) = \frac{\vb*{t}'(s)}{\abs{\vb*{t}'(s)}}, \quad \vb*{b}(s) = \vb*{t}(s)\cross\vb*{n}(s),
\end{equation}
from which the geometric or kinematic torsion $\tau_s$ and the curvature $\kappa$, both constants, are derived as
\begin{equation}
    \tau_s = \abs{\vb*{b}'(s)} = \frac{\sin\beta\cos\beta}{\rho},~\kappa = \abs{\vb*{t}'(s)} = \frac{\sin^2\beta}{\rho}. \label{eqn:helix-curvatures}
\end{equation}
According to Love \cite{love1944treatise}, the total torsion of a helix $\tau$ is decomposed as
\begin{equation}
    \tau = \tau_i + \tau_s,
\label{eqn:tau1}
\end{equation}
where the internal twist $\tau_i$ is the twist when the helical curve is unbent.
The end rotation $R$ generated due to the moment $M$ is \cite{thompson1996helix}
\begin{equation}
    r = \frac{R}{L} = \tau_i + \frac{2\pi}{L/N} = \tau_i + \frac{\sin\beta}{\rho}. \label{eqn:helix-end-rotation-expression}
\end{equation}
\subsection{Energy analysis of the post-buckled state}
Following the energy analysis in \cite{thompson1996helix}, the work potentials due to $T$ and $M$ with the corresponding deflections $-D$ (the negative sign is due to the displacement happening in the opposite direction to $T$) and $R$ are:
\begin{equation}
    \begin{split}
        W_T = T(-D) = -TL(1 - \cos\beta), \\
        W_M = MR = ML\left(\tau_i + \frac{\sin\beta}{\rho}\right).
    \end{split}
\end{equation}
The strain energies due to bending $\mathcal{E}_b$ and due to torsion $\mathcal{E}_t$ are
\begin{equation}
\begin{split}
    &\mathcal{E}_b = \frac{1}{2}EI\kappa^2 L = \frac{1}{2}EIL\frac{\sin^4\beta}{\rho^2} \\
    &\mathcal{E}_t = \frac{1}{2}GJ\tau^2 L = \frac{1}{2} GJL\left(\tau_i + \frac{\sin\beta\cos\beta}{\rho}\right)^2.
\end{split}
\end{equation}
For soft ferromagnetic rod with longitudinal $\vb*{m} = \vb*{e}_3$, the magnetic energy is
\begin{equation}
    \begin{split}
        \mathcal{E}_{\text{mag}} &= \frac{K_d\pi a^4}{4} \int_{0}^{L} \left[(\vb*{m}\cdot\vb*{d}_1(s))^2 + (\vb*{m}\cdot\vb*{d}_2(s))^2\right] ds \\
        &= \frac{K_d\pi a^4}{4} \int_{0}^{L} \left[1 - (\vb*{m}\cdot\vb*{d}_3(s))^2\right] ds = \frac{K_d\pi a^4}{4} \int_{0}^{L} \left[1 - \cos^2\beta\right] ds \\
       \implies  \mathcal{E}_{\text{mag}}  &= \frac{K_d\pi a^4 L}{4} \sin^2\beta.
    \end{split}
\end{equation}
The total potential energy is thus
\begin{equation}
    \begin{split}
        \mathcal{E} &= \mathcal{E}_b + \mathcal{E}_t + \mathcal{E}_\text{mag} - W_T - W_M \\
        &= \frac{1}{2}EIL\frac{\sin^4\beta}{\rho^2} + \frac{1}{2} GJL\left(\tau_i + \frac{\sin\beta\cos\beta}{\rho}\right)^2 + \frac{K_d\pi a^4 L}{4} \sin^2\beta \\
        &\qquad + TL(1 - \cos\beta) - ML\left(\tau_i + \frac{\sin\beta}{\rho}\right)
    \end{split}
\end{equation}
For equilibrium, the derivatives of $\mathcal{E}$ with respect to $(\rho,\beta,\tau_i)$ must evaluate to zero. Therefore,
\begin{equation}
    \begin{split}
        \pdv{\mathcal{E}}{\rho} = 0 & \implies M = GJ\tau\cos\beta + \frac{EI\sin^3\beta}{\rho}, \\
        \pdv{\mathcal{E}}{\beta} = 0 & \implies T = GJ\tau\frac{\sin\beta}{\rho} - EI\frac{\cos\beta\sin^2\beta}{\rho^2} - \frac{K_d\pi a^2}{2}\cos\beta, \\
        \pdv{\mathcal{E}}{\tau_i} = 0 & \implies M  = GJ \tau,
    \end{split}
    \label{eqn:threeD-equilibrium-helix}
\end{equation}
The first two equations relate the applied loads $(T,M)$ to the kinematic variables of deformation $(\rho,\beta,\tau)$ while the third equation acts as a constraint and it is identical to Eqn. \ref{eqn:threeD-m-tau-relation}. The constraint enables us to determine the deformation for a prescribed loading.

\subsection{Analysis of the equilibrium equations}
With the helical geometry and energy functional established, we now solve the equilibrium equations to relate loads to deformation. We eliminate $\tau = \frac{M}{GJ}$ from the first two equations in Eqn. \ref{eqn:threeD-equilibrium-helix} to express the loads $(T,M)$ in terms of helix coordinates $(\rho,\beta)$ and the parameter $K_d$ as
\begin{equation}
    \begin{split}
        &T = \frac{M\sin\beta}{\rho} - \frac{EI\cos\beta\sin^2\beta}{\rho^2} - \frac{K_d\pi a^2\cos\beta}{2}, \\
        &M (1 - \cos\beta) = \frac{EI\sin^3\beta}{\rho} \implies M = \frac{EI}{\rho}\frac{\sin^3\beta}{(1-\cos\beta)}.
    \end{split}
    \label{eqn:threeD-loads-rho-beta}
\end{equation}
Substituting the expression of $M$ in the equation for $T$ results in
\begin{equation}
    T = \frac{EI}{\rho^2}\sin^2\beta - \frac{K_d\pi a^2}{2} \cos\beta. \label{eqn:threeD-T-rho-beta}
\end{equation}
Fully eliminating $\rho$ using the equation for $M$ in Eqn. \ref{eqn:threeD-loads-rho-beta} gives the load relation
\begin{equation}
    T + \frac{K_d\pi a^2}{2}\cos\beta = \frac{M^2}{EI(1 + \cos\beta)^2}. \label{eqn:threeD-T-M-Kd-relation}
\end{equation}
We now derive the expression of the average rotation per unit length $r = \frac{R}{L}$. Now, $r = R/L$ follows from $\tau = \tau_i + \tau_s$ and geometry, (see equations \ref{eqn:helix-curvatures} and \ref{eqn:helix-end-rotation-expression}):
\begin{equation}
    \begin{split}
        &r = \tau_i + \frac{2\pi N}{L} = \tau - \tau_s + \frac{\sin\beta}{\rho} \\
        \implies &r = \frac{M}{GJ} - \frac{\sin\beta\cos\beta}{\rho} + \frac{\sin\beta}{\rho} = \frac{M}{GJ} + (1 - \cos\beta)\frac{\sin\beta}{\rho}.
    \end{split}
\end{equation}
From Eqn. \ref{eqn:threeD-T-rho-beta}, $\frac{\sin\beta}{\rho} = \sqrt{\frac{1}{EI} \left(T + \frac{K_d\pi a^2}{2}\cos\beta\right)}$, so
\begin{equation}
    r = \frac{M}{GJ} + \frac{(1 - \cos\beta)}{EI}\sqrt{EI\left(T + \frac{K_d\pi a^2}{2}\cos\beta\right)}. \label{eqn:threeD-r-M-beta}
\end{equation}
Let us derive $r$ in terms of end-shortening parameter $d = \frac{D}{L}$. Rearranging the third equation in Eqns. \ref{eqn:threeD-helix-geometry} gives
\begin{equation}
    d = 1 - \cos\beta \implies \cos\beta = 1 - d \label{eqn:threeD-d-cos-beta}
\end{equation}
which upon substitution in Eqn. \ref{eqn:threeD-T-M-Kd-relation} produces
\begin{equation}
   \begin{split}
       T + \frac{K_d\pi a^2}{2}(1 - d) &= \frac{M^2}{EI(2- d)^2} \\
    \implies M^2 &= EIT(2 -d)^2 + \frac{K_d\pi a^2}{2}EI (1-d)(2-d)^2
   \end{split}
\end{equation}
Therefore,
\begin{equation}
    r = \left(\frac{(2-d)}{GJ} + \frac{d}{EI}\right) \sqrt{EI \left(T + \frac{K_d\pi a^2}{2} (1-d)\right)}. \label{eqn:threeD-r-d-final}
\end{equation}
Having obtained explicit relations between the applied loads and the kinematic variables of the helical configuration, we now examine the rod's equilibrium response under distinct loading sequences.

\subsection{Behaviour under various loading sequences}

We conduct experiments on a helically deformed ferromagnetic rod to determine its equilibrium configurations under different quasi-static loading sequences. As one or more control parameters vary slowly, the rod remains in a given equilibrium state until reaching a critical threshold, where it undergoes a dynamic transition to a distinct equilibrium state, that is qualitatively distinct from the preceding equilibrium state.

Before the detailed analysis, it is instructive to revisit the classification of control parameters as outlined by Thompson and Champneys \cite{thompson1996helix}. A \textit{fixed} parameter denotes a quantity that remains constant throughout the experiment, for example, material properties, geometric characteristics, or a pre-fixed tension or moment. In contrast, a \textit{controlled} parameter is one that is varied quasi-statically during the course of the experiment, thereby serving as the primary driver of equilibrium response. Finally, a \textit{passive} quantity refers to a variable that is neither fixed nor actively controlled, but instead evolves as a consequence of the equilibrium condition. The applied loads are categorized into two classes. The first class, dead loading, comprises the tension $T$ and twisting moment $M$, in which the force and moment resultants are prescribed. The second class, rigid loading, comprises the end rotation $R$ and end-shortening $D$, in which the corresponding kinematic quantities are prescribed.

The three loading sequences analyzed are the following:
\begin{itemize}
    \item Scenario 1: Purely dead loading (tension $T$ is fixed, twisting moment $M$ is varied, and $M$ is fixed, $T$ is varied).
    \item Scenario 2: Mixed loading where $T$ is pre-fixed and end rotation $R$ is varied.
    \item Scenario 3: Mixed loading where end-rotation $R$ is pre-fixed while $T$ is varied.
\end{itemize}
We now analyze these loading scenarios, beginning with Scenario 1.

\subsubsection*{Scenario 1: Purely dead loading }
\begin{figure}[h!]
\centering
\begin{subfigure}{0.49\textwidth}
\centering
\includegraphics[width=\linewidth]{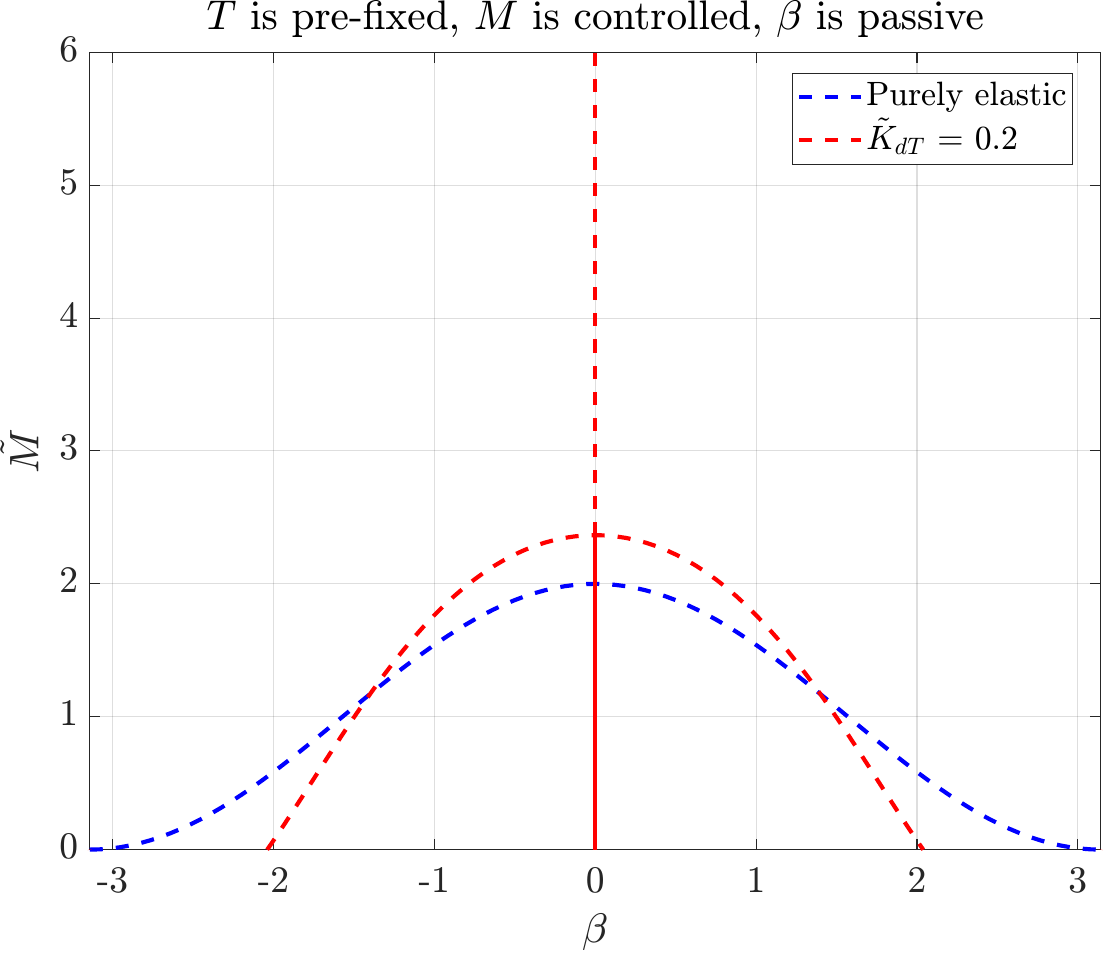}
\caption{T fixed, M controlled: $\tilde{K}_{dT} = 0.2$}
\end{subfigure}
\hfill
\begin{subfigure}{0.49\textwidth}
\centering
\includegraphics[width=\linewidth]{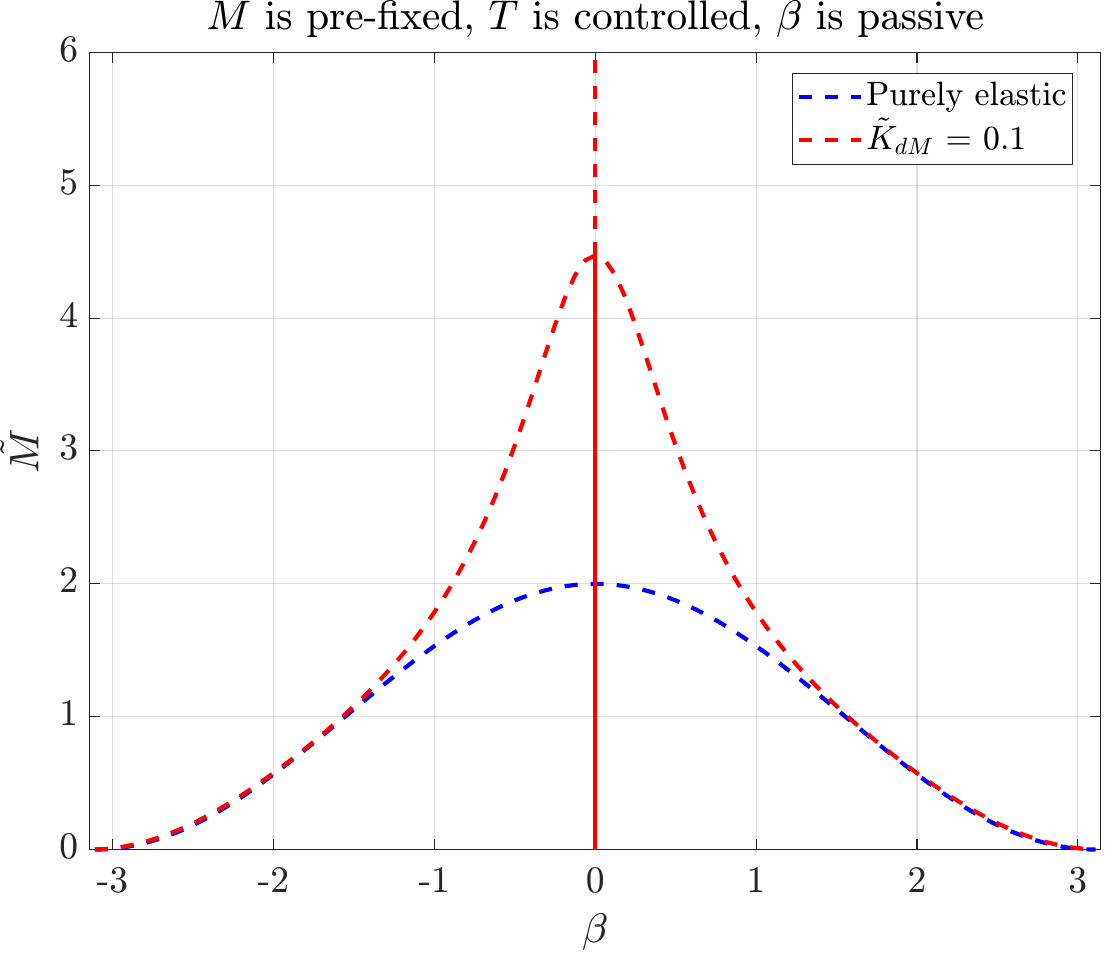}
\caption{M fixed, T controlled: $\tilde{K}_{dM} = 0.1$}
\end{subfigure}
\caption{Totally dead loading: Evolution of $\tilde{M}$ with $\beta$ for purely elastic and soft ferromagnetic rods.}
\label{fig:threeD-helix-totally-dead-loading}
\end{figure}

\textit{$T$ is fixed, twisting moment $M$ is controlled, $\beta$ is passive:} We prescribe a dead tension load $T$ while controlling the twisting moment $M$. Eqn.~\eqref{eqn:threeD-T-M-Kd-relation} connects the quantities $K_d$, $M$, $T$ and $\beta$:
\begin{equation}
    T + \frac{K_d\pi a^2}{2}\cos\beta = \frac{M^2}{EI(1 + \cos\beta)^2}.
\end{equation}
Normalizing by $T$ and introducing the non-dimensional parameters $\tilde{K}_{dT} = \frac{K_d\pi a^2}{4T}$, $\tilde{M} = \frac{M}{\sqrt{EIT}}$ gives
\begin{equation}
    1 + 2\tilde{K}_{dT} \cos\beta = \frac{\tilde{M}^2}{(1 + \cos\beta)^2}.
\end{equation}
At the onset of bifurcation from a straight configuration $(\beta \to 0)$ of the rod into a helical state , the condition reduces to
\begin{equation}
    T + \frac{K_d\pi a^2}{2} = \frac{M^2}{4EI} \Leftrightarrow  1 + 2\tilde{K}_{dT} = \frac{\tilde{M}^2}{4} \implies \tilde{M} = 2\sqrt{1 + 2\tilde{K}_{dT}}.
\end{equation}
Fig. \ref{fig:threeD-helix-totally-dead-loading}(a) illustrates the variation of $\tilde{M}$ with $\beta$ for fixed $T$, comparing purely elastic rods with soft ferromagnetic rods at $\tilde{K}_{dT} = 0.2$. The equilibrium curve for the soft ferromagnetic rod is shifted relative to the purely elastic case, exhibiting a higher bifurcation threshold at $\beta = 0$ and a higher peak value of $\tilde{M}$ attained over the range of admissible helix angles.

\noindent \textit{$M$ is fixed, tension $T$ is controlled, $\beta$ is passive:}

For prescribed $M$, we normalize Eqn.~\eqref{eqn:threeD-T-M-Kd-relation} by $\frac{M^2}{EI}$ and introducing the non-dimensional parameter $\tilde{K}_{dM} = \frac{K_d \pi a^2 EI}{4M^2}$, yielding
\begin{equation}
    \frac{1}{\tilde{M}^2} + 2\tilde{K}_{dM}\cos\beta = \frac{1}{(1 + \cos\beta)^2} \Leftrightarrow \tilde{M}^2 = \frac{1}{\frac{1}{(1 + \cos\beta)^2}  - 2\tilde{K}_{dM}\cos\beta}
    \label{eqn:S1-bifurcation-fixed-M}
\end{equation}
The corresponding $\tilde{M}-\beta$ variation for fixed $M$ is shown in Fig. \ref{fig:threeD-helix-totally-dead-loading}(b) for purely elastic $\tilde{K}_{dM}=0$ and soft ferromagnetic rods, considering $\tilde{K}_{dM} = 0.1$. In this regime, a supercritical Hamiltonian Hopf bifurcation occurs since $\tilde{K}_{dM} < \nicefrac{1}{8}$, see section \ref{sec:bifurcation-points-Hamiltonian}. For the soft ferromagnetic rod, the bifurcation from the straight
configuration into a helical state occurs at a higher value of the
composite load parameter $\tilde{M}$ relative to the purely elastic
case, as is evident from the elevated bifurcation threshold
$\tilde{M}^2 = \frac{1}{\frac{1}{4} - 2\tilde{K}_{dM}}$ established in
Eqn.~\ref{eqn:S1-bifurcation-fixed-M}.

\subsubsection*{Scenario 2: Mixed loading where $T$ is pre-fixed and end rotation $R$ is varied}
\begin{figure}[h!]
\centering
\begin{subfigure}{0.49\textwidth}
\centering
\includegraphics[width=\linewidth]{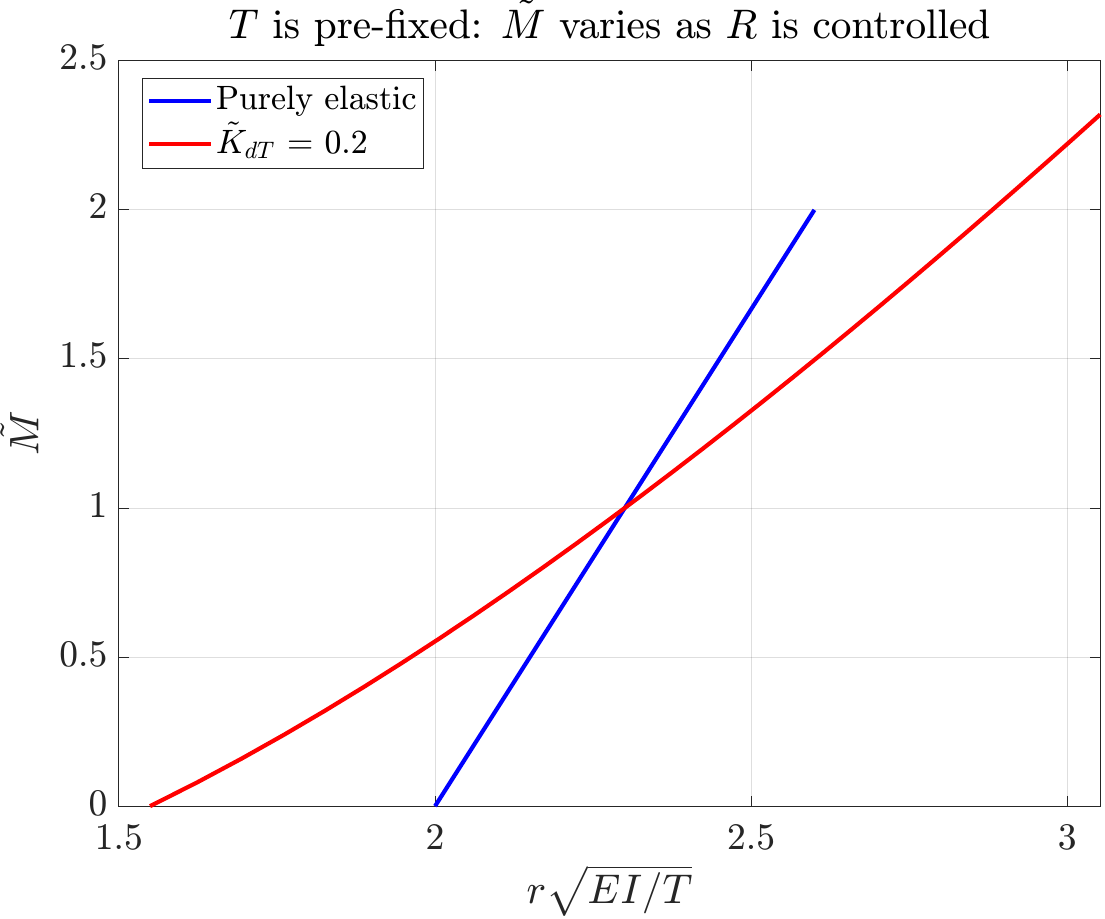}
\caption{$r$ varies with $\tilde{M}$ for $\tilde{K}_{dT} = 0.2$}
\end{subfigure}
\hfill
\begin{subfigure}{0.49\textwidth}
\centering
\includegraphics[width=\linewidth]{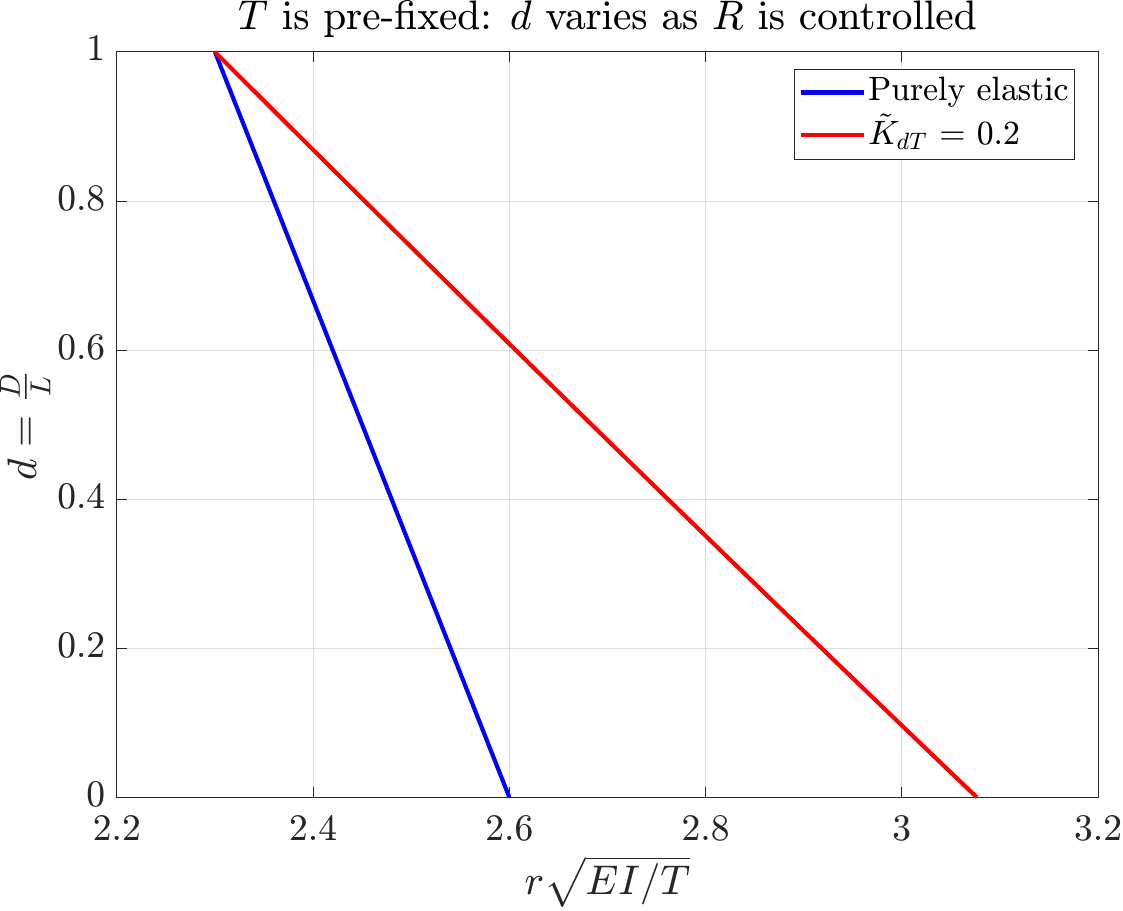}
\caption{$r$ varies with $d$ for $\tilde{K}_{dT} = 0.2$}
\end{subfigure}
\caption{Mixed dead/rigid loading: Variation of (a) $\tilde{M}$ and (b) $d$ with $r\sqrt{EI/T}$ for a pre-fixed $T$ for purely elastic and soft ferromagnetic rods.}
\label{fig:threeD-helix-mixed-dead-and-rigid-loading}
\end{figure}

We now fix dead tension load $T$ and treat $R \, (=rL)$ as the control parameter. From Eqn. \ref{eqn:threeD-r-M-beta}, we have
\begin{equation}
    r = \frac{M}{GJ} + \frac{(1-\cos\beta)}{EI} \sqrt{EI\left(T + \frac{K_d\pi a^2}{2}\cos\beta\right).}
\end{equation}
Normalizing by $\sqrt{EIT}$, we get
\begin{equation}
    \frac{r}{\sqrt{EIT}} = \frac{\tilde{M}}{GJ} + \frac{(1-\cos\beta)}{EI}\sqrt{1 + 2\tilde{K}_{dT}\cos\beta};
\end{equation}
Since $GJ = \frac{E}{2(1+\nu)}\cdot 2I = \frac{EI}{1+\nu}$, this reduces to
\begin{equation}
    \frac{r}{\sqrt{EIT}} = \frac{1}{EI}\left[(1+\nu)\tilde{M} + (1-\cos\beta)\sqrt{1 + 2\tilde{K}_{dT}\cos\beta}\right].
\end{equation}
To obtain $\cos\beta$ in terms of $\tilde{M}$, we divide Eqn. \ref{eqn:threeD-T-M-Kd-relation} throughout by $T$, obtaining
\begin{equation}
    1 + \frac{K_d\pi a^2}{2T}\cos\beta = \frac{M^2}{EIT(1+\cos\beta)^2}.
\end{equation}
Rearranging gives the cubic equation
\begin{equation}
    (\cos\beta + 1)^2 + 2\tilde{K}_{dT} \cos\beta (\cos\beta + 1)^2 = \tilde{M}^2. \label{eqn:threeD-cubic-cos-beta}
\end{equation}
We solve the cubic equation Eqn. \ref{eqn:threeD-cubic-cos-beta} numerically for $\cos\beta$ at fixed $\tilde{K}_{dT}$ with $\tilde{M}$ as the variable, and represent the solution as
\begin{equation}
    \cos\beta = f(\tilde{M}; \tilde{K}_{dT}).
\end{equation}
Substituting this functional form, we get
\begin{equation}
    \frac{r}{\sqrt{EIT}} = \frac{1}{EI}\left[(1+\nu)\tilde{M} + \left(1-f(\tilde{M}; \tilde{K}_{dT})\right)\sqrt{1 + 2\tilde{K}_{dT}f(\tilde{M}; \tilde{K}_{dT})}\right]
\end{equation}
Alternatively, in terms of end-shortening $d = 1 - \cos \beta$ (Eqn.~\eqref{eqn:threeD-d-cos-beta}), the above relation becomes
\begin{equation}
    \frac{r}{\sqrt{EIT}} = \left[(1+\nu)(2-d) + d\right] \frac{\sqrt{1 + 2\tilde{K}_{dT}(1-d)}}{EI}.
\end{equation}
Figures ~\ref{fig:threeD-helix-mixed-dead-and-rigid-loading}(a) and ~\ref{fig:threeD-helix-mixed-dead-and-rigid-loading}(b) present the variation of $\tilde{M}$ and $d$ with $r \sqrt{EI/T}$ for both purely elastic and soft ferromagnetic rods ($\tilde{K}_{dT} = 0.2$) respectively.  In the soft ferromagnetic case, the $\tilde{M}-r\sqrt{\nicefrac{EI}{T}}$ response exhibits a noticeably reduced slope compared to the purely elastic case, indicating a diminished sensitivity of $r$ to changes in $\tilde{M}$. Moreover, the $d-r\sqrt{\nicefrac{EI}{T}}$ response curves show that a larger end rotation $r$ is required in the soft ferromagnetic rod to achieve the same end-shortening $d$ as in the purely elastic case.

\subsubsection*{Scenario 3: Mixed loading where end-rotation $R$ is pre-fixed while $T$ is varied}
\begin{figure}[ht!]
    \centering
    \includegraphics[width=0.5\linewidth]{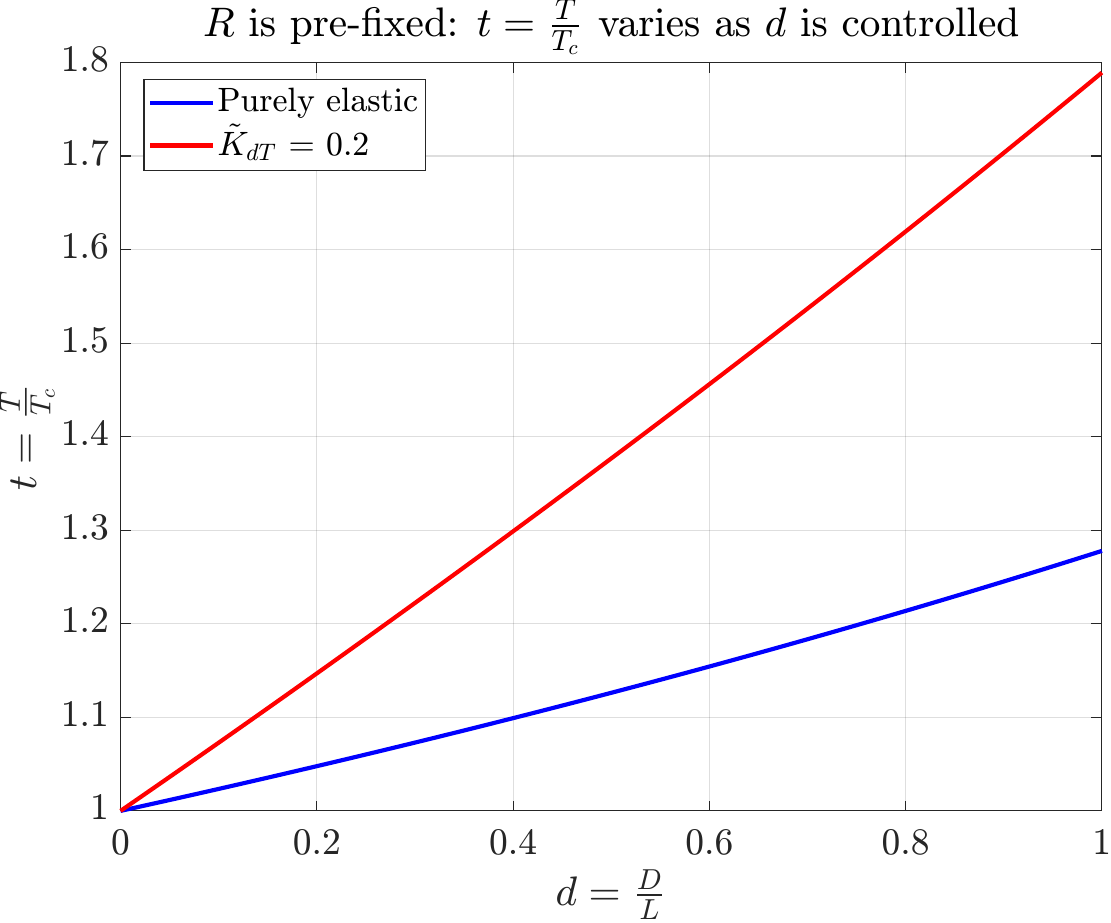}
    \caption{End-rotation $r$ is pre-fixed while $t = \frac{T}{T_c}$ is varied.}
    \label{fig:threeD-helix-r-fixed-T-varied}
\end{figure}

In this scenario, the rigid end-rotation $R$ is prescribed, with the initial pre-twist given by $\tau_0 = \frac{R}{L} = r$.
From Eqn. \ref{eqn:threeD-r-d-final}, the governing relation is
\begin{equation}
     r = \left(\frac{(2-d)}{GJ} + \frac{d}{EI}\right) \sqrt{EI \left(T + \frac{K_d\pi a^2}{2} (1-d)\right)}. \label{eqn:threeD-r-d-final-1}
\end{equation}
The Hamiltonian-Hopf supercritical pitchfork bifurcation occurs at $d = 0$ and $T = T_c$, satisfying the relation $M_0 = GJ\tau_0$, which implies
\begin{equation}
\begin{split}
    & T_c + \frac{K_d\pi a^2}{2} = \frac{M_0^2}{4EI} = \frac{G^2J^2\tau_0^2}{4EI}, \\
    \implies &  r = \tau_0 = \frac{1}{GJ} \sqrt{4EIT_c + 2K_d\pi a^2 EI}.
\end{split} \label{eqn:threeD-r-tau0}
\end{equation}
Equating Eqns. \ref{eqn:threeD-r-d-final-1} and \ref{eqn:threeD-r-tau0}, and simplifying, we have
\begin{equation}
     \left((2-d) + \frac{GJ}{EI}d\right) \sqrt{T + \frac{K_d\pi a^2}{2} (1-d)} = \sqrt{4T_c + 2K_d\pi a^2}.
\end{equation}
Dividing throughout by $\sqrt{T_c}$ and introducing the non-dimensional parameters $t = \frac{T}{T_c}$ and $\tilde{K}_{dT} = \frac{K_d\pi a^2}{4T_c}$ gives
\begin{equation}
\left((2-d) + \frac{d}{(1+\nu)}\right) \sqrt{t +2\tilde{K}_{dT} (1-d)} = 2\sqrt{1 + 2\tilde{K}_{dT}}.
\end{equation}
Fig. \ref{fig:threeD-helix-r-fixed-T-varied} illustrates the variation of $t$ with $d$ both purely elastic and soft ferromagnetic rods at $\tilde{K}_{dT} = 0.2$. The results indicate that, for a given end-shortening $d$, the soft ferromagnetic rod requires a higher applied tension $T$ compared to the purely elastic case.

In summary, across the loading scenarios considered, the soft ferromagnetic rod consistently requires higher values of tension $T$, moment $M$, or end rotation $R$ to attain deformation states comparable to those of the purely elastic rod. This reveals that the magnetoelastic interaction systematically stiffens the soft ferromagnetic rod relative to its purely elastic counterpart, requiring larger applied loads to achieve comparable deformation states. The helical configurations examined in this section, however, represent only the initial stage of the post-buckling response. As established in the Hamiltonian phase portrait analysis of Section \ref{sec:threeD-analysis-Hamiltonian}, homoclinic orbits coexist with the periodic helical orbits in the phase space of the rod equations. These homoclinic orbits correspond to spatially localized deformation modes, in which high-curvature regions are confined to a finite portion of the rod, with the remaining segments approaching a straight configuration. The analysis of such localized buckling modes, is presented in Section \ref{sec:threeD-localized-buckling}.

We now proceed to analyze the localized buckling of the rods.

\section{Localized buckling of rods} \label{sec:threeD-localized-buckling}

The onset of localized buckling in the rod occurs subsequent to the formation of a helical instability, as the composite load parameter $\tilde{M}$ is reduced below the bifurcation threshold. The initial helical deformation evolves into a sinusoidal helical state with reduced applied twisting moment $M$ or increased applied tension $T$. Upon further reduction in $\tilde{M}$, the strain energy gradually becomes localized along the rod, marking the transition toward a localized buckled configuration.

In the purely elastic case, a closed-form solution for the localized deformation can be derived from the associated Hamiltonian, as demonstrated by Coyne \cite{Coyne1990}. In contrast, such an analytical formulation is not readily available for the soft ferromagnetic case due to the additional nonlinear magnetic interactions. Therefore, a numerical approach must be employed to determine the localized buckling modes in these systems. The numerical procedure adopted for constructing the deformed shapes is presented in the following section.

\begin{figure}[h!]
\centering
\begin{subfigure}{0.49\textwidth}
    \centering
    \includegraphics[width=\linewidth]{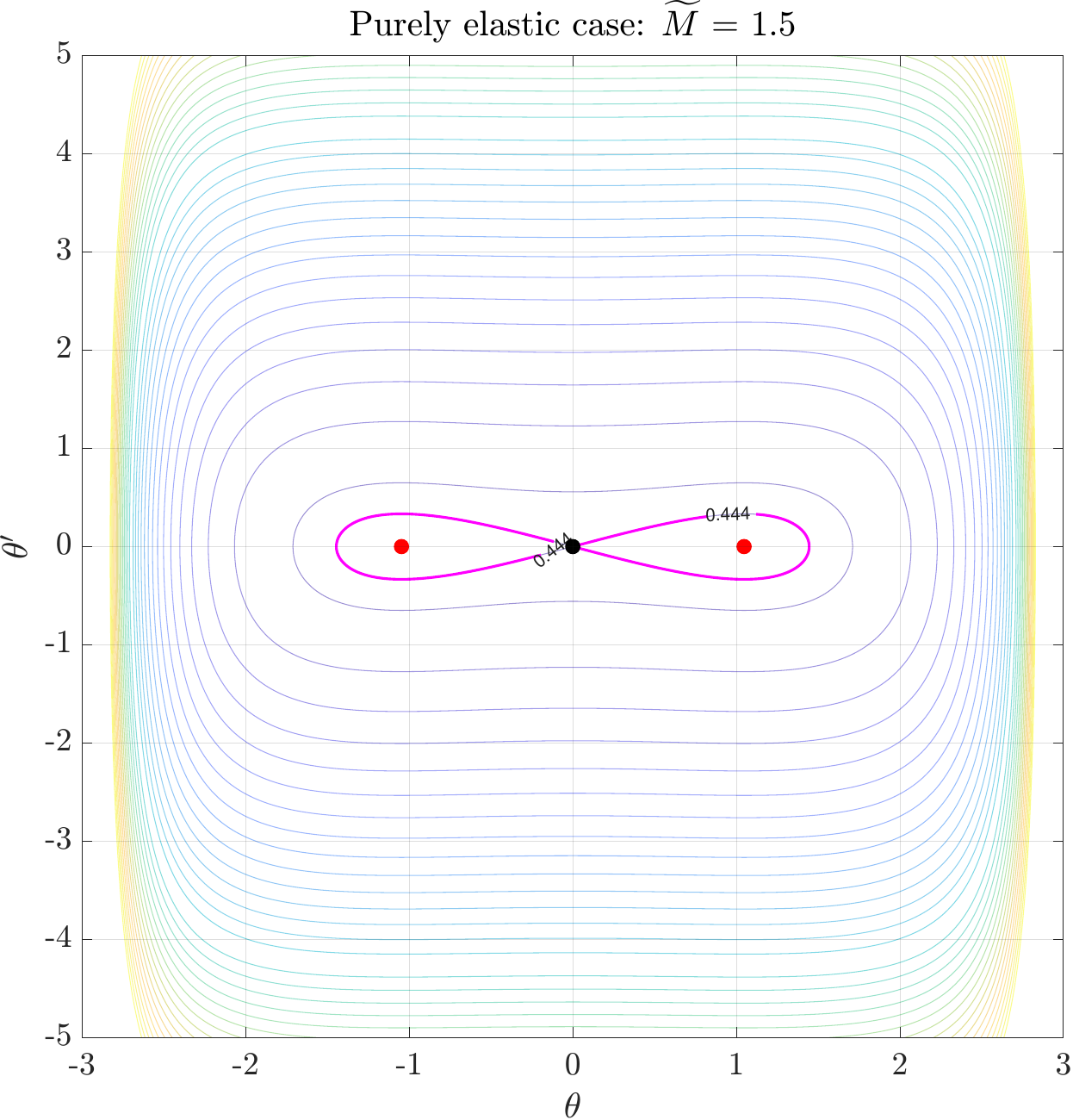}
   \label{fig:sub1}
\end{subfigure}
\hfill
\begin{subfigure}{0.49\textwidth}
    \centering
    \includegraphics[width=\linewidth]{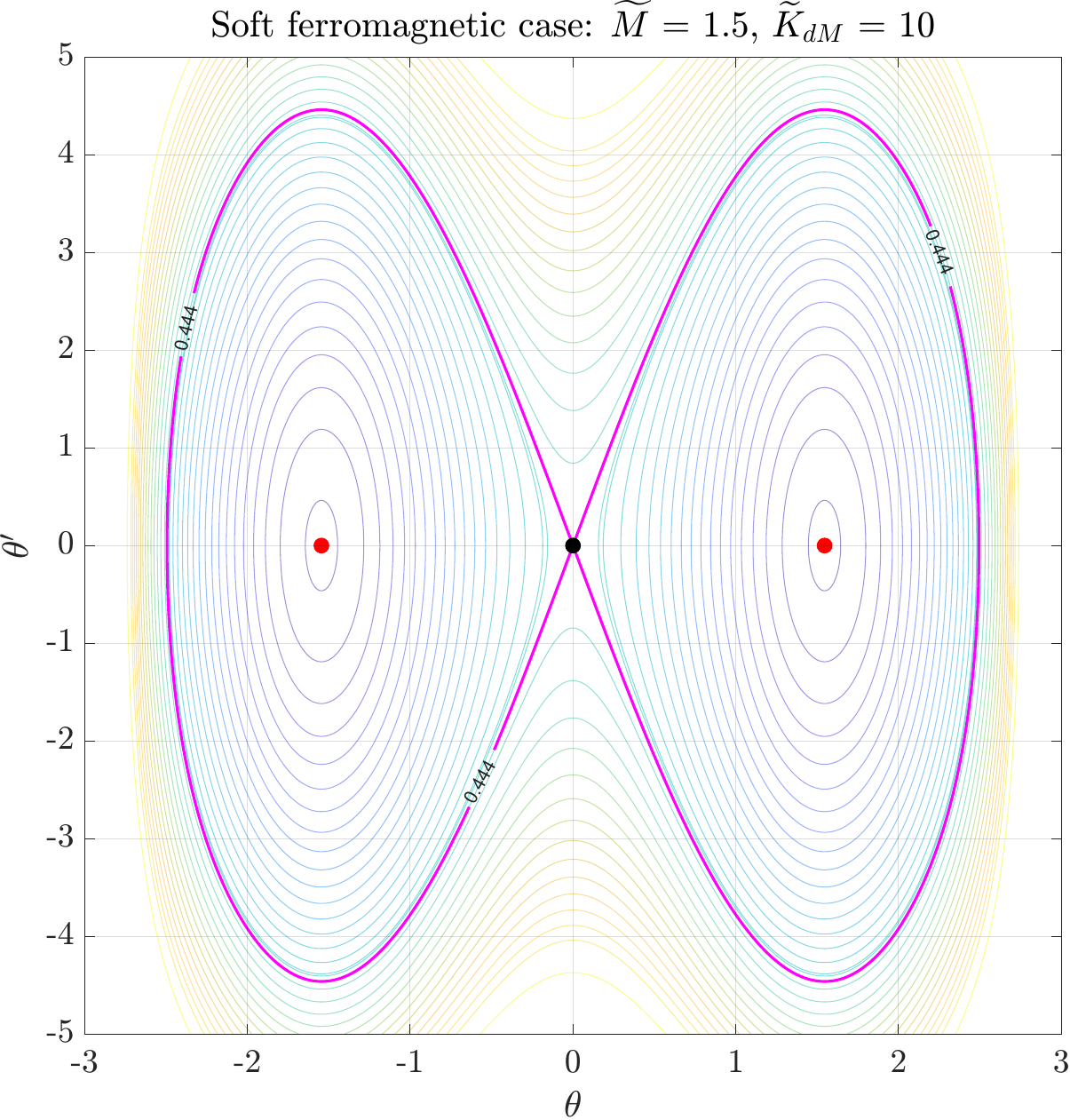}
    \label{fig:phase-portrait-soft-ferromagnetic}
\end{subfigure}
\caption{Phase portraits for $\tilde{M} = 1.5$: (left) purely elastic case, and (right) soft ferromagnetic case with $\tilde{K}_{dM} = 10$.}
\label{fig:phase-portraits-localised-buckling}
\end{figure}

To determine the deformed buckling configurations, we employ  the numerical methodology detailed in \cite[Section 4]{AvatarDabade2025}. Fixed values of the parameters $\tilde{M}$ and $\tilde{K}_{dM}$ are prescribed, and a trajectory is selected from the corresponding phase portrait (e.g., Fig.~\ref{fig:phase-portraits-localised-buckling}) along which the Hamiltonian (Eqn.~\ref{eqn:threeD-soft-ferromagnetic-longitudinal-Hamiltonian-non-dimensionalized-M}) assumes a constant value $C$. This Hamiltonian equation is rearranged to relate the differentials $d\tilde{s}$ and $d\theta$, which is then used to modify the Euler angle rate equations (Eqns.~\ref{eqn:euler-angle-rates-non-dimensionalized-M}) linking $d\psi$ and $d\phi$ to $d\theta$. A point $\theta_0$ on the $\theta$-axis of the phase portrait is chosen as the starting point for integration. The equations are then integrated concurrently with the initial conditions $\tilde{s}(\theta_0) = 0$, $\psi(\theta_0) = \psi_0$ and $\phi(\theta_0) = \phi_0$ (typically, $\psi_0 = \phi_0 = 0$):
\begin{equation}
    \begin{split}
      \tilde{s} = \int_{0}^{\tilde{s}} d\tilde{s} &= \int_{\theta_0}^{\theta}\frac{d\theta}{\sqrt{2} \sqrt{C - \frac{1}{2}\frac{(1 - \cos\theta)}{(1 + \cos\theta)} + \tilde{K}_{dM}\sin^2\theta - \frac{\cos\theta}{\tilde{M}^2}}}, \\
      \int_{\psi_0}^{\psi} d\psi &= \int_{\theta_0}^{\theta} \frac{d\theta}{\sqrt{2} (1 + \cos\theta)\sqrt{C - \frac{1}{2}\frac{(1 - \cos\theta)}{(1 + \cos\theta)} + \tilde{K}_{dM}\sin^2\theta - \frac{\cos\theta}{\tilde{M}^2}}},\\
      \int_{\phi_0}^{\phi} d\phi &=  \int_{\theta_0}^{\theta} \left(\nu + \frac{1}{1 + \cos\theta}\right) \frac{d\theta}{\sqrt{2} \sqrt{C - \frac{1}{2}\frac{(1 - \cos\theta)}{(1 + \cos\theta)} + \tilde{K}_{dM}\sin^2\theta - \frac{\cos\theta}{\tilde{M}^2}}} .
    \end{split}
\end{equation}
This procedure yields the numerical solution $(\tilde{s}(\theta),\psi(\theta),\phi(\theta))$, which is subsequently mapped into $(\theta(\tilde{s}),\psi(\tilde{s}),\phi(\tilde{s}))$. The centerline coordinates $\left(x(\tilde{s}),y(\tilde{s}),z(\tilde{s})\right)$ are then reconstructed by integrating Eqns.~\ref{eqn:centerline-description} using trapezoidal quadrature scheme.

\begin{figure}[ht!]
    \centering
    \includegraphics[width=0.85\linewidth]{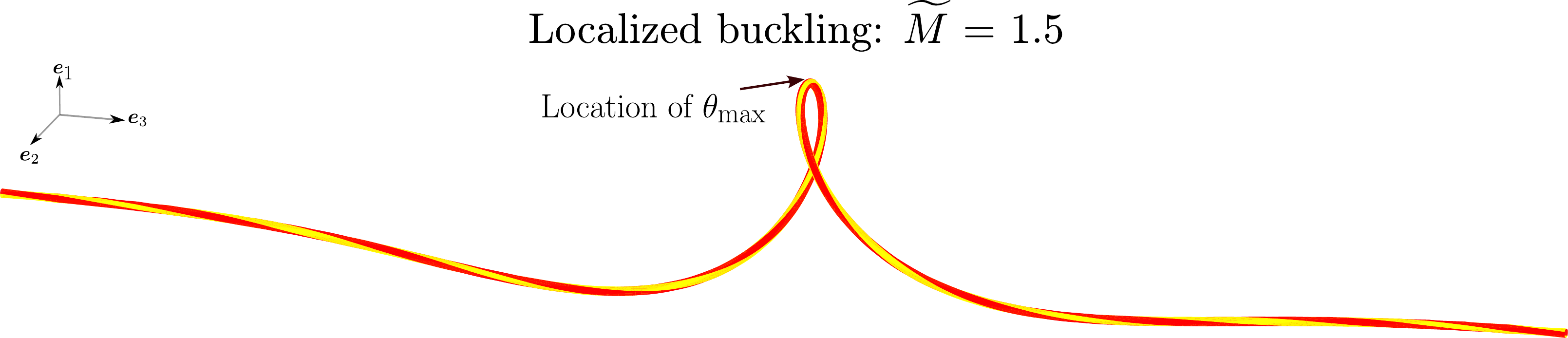}
    \caption{Localized buckling configuration of a purely elastic rod for $\tilde{M} = 1.5$.}
    \label{fig:localized-buckling-purely-elastic-m-1.5}
\end{figure}

\begin{figure}[ht!]
    \centering
    \includegraphics[width=0.85\linewidth]{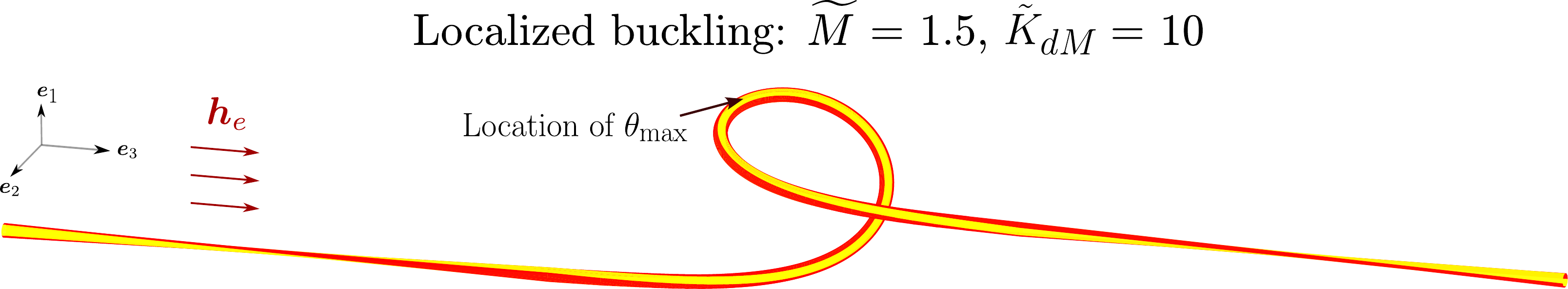}
    \caption{Localized buckling configuration of a soft ferromagnetic elastic rod for $\tilde{M} = 1.5$ and $\tilde{K}_{dM} = 10$.}
    \label{fig:localized-buckling-soft-ferromagnetic-m-1.5}
\end{figure}

The localized buckling deformations are determined by integrating along the homoclinic orbits present in the Hamiltonian phase portraits, highlighted in pink color in Fig. \ref{fig:phase-portraits-localised-buckling}. For a purely elastic rod, such homoclinic orbits exist only when $\tilde{M} < 2$. In contrast, for a soft ferromagnetic rod, they occur for all $\tilde{M}>0$ provided $\tilde{K}_{dM} > \frac{1}{8}$. As an illustration, Fig.~\ref{fig:localized-buckling-purely-elastic-m-1.5} shows the localized deformation of a purely elastic rod at $\tilde{M} = 1.5$. The configuration is characterized by a looped shape in which the extended straight segments remain collinear. By comparison, the corresponding localized shape for the soft ferromagnetic rod under the same loading parameter $\tilde{M} = 1.5$ with the value $\tilde{K}_{dM} = 10$ (see Fig.~\ref{fig:localized-buckling-soft-ferromagnetic-m-1.5}), exhibits a shifted configuration, with extended straight segments that are no longer collinear. This contrast highlights the distinct mechanical response imparted by magnetic coupling.

\subsection{Analysis of localized solution}
We now derive an expression for the maximum value $\theta_{\max}$  of the Euler angle $\theta$, the angle subtended between the rod's local tangent vector $\vb*{d}_3(s)$ and the fixed reference axis $\vb*{e}_3$ as a function of the composite load parameter $\tilde{M}$, for a fixed value of the magnetoelastic parameter $\tilde{K}_{dM}$. From Eqn. \ref{eqn:threeD-soft-ferromagnetic-longitudinal-Hamiltonian-non-dimensionalized-M}, we have
\begin{equation}
    \tilde{\mathcal{H}}_\text{soft} = \frac{1}{2}\theta'^2 + \frac{1}{2}\frac{(1 - \cos\theta)}{(1 + \cos\theta)} - \tilde{K}_{dM}\sin^2\theta + \frac{\cos\theta}{\tilde{M}^2}.
\end{equation}

The Hamiltonian $\tilde{\mathcal{H}}_\text{soft}$, being a conserved quantity along the rod, takes the constant value
\begin{equation}
    \tilde{\mathcal{H}}_\text{soft} = \frac{1}{\tilde{M}^2}.
\end{equation}

Therefore,
\begin{equation}
    \frac{1}{2}\theta'^2 + \frac{1}{2}\frac{(1 - \cos\theta)}{(1 + \cos\theta)} - \tilde{K}_{dM}\sin^2\theta + \frac{\cos\theta}{\tilde{M}^2} = \frac{1}{\tilde{M}^2}.
\end{equation}
At the location on the centerline where $\theta$ attains a maximum $\theta_\text{max}$, ($\implies \theta' = 0$), and so
\begin{equation}
\begin{split}
   \frac{1}{2}\frac{(1 - \cos\theta)}{(1 + \cos\theta)} - \tilde{K}_{dM} (1 - \cos^2\theta)  - \frac{1}{\tilde{M}^2} (1 - \cos\theta)\bigg\rvert_{\theta=\theta_\text{max}} &= 0 \\
   \implies \frac{1}{2}\frac{(1 - \cos\theta)}{(1 + \cos\theta)} - \tilde{K}_{dM} (1 - \cos\theta) (1 + \cos\theta)  - \frac{1}{\tilde{M}^2} (1 - \cos\theta)\bigg\rvert_{\theta=\theta_\text{max}} &= 0.
\end{split}
\end{equation}
Factoring out $(1 - \cos\theta_\text{max})$, we have
\begin{equation}
    \begin{split}
        \frac{1}{2}\frac{1}{(1 + \cos\theta_{\text{max}})} - \tilde{K}_{dM} (1 + \cos\theta_{\text{max}})  - \frac{1}{\tilde{M}^2} &= 0 \\
        \implies 2 \tilde{K}_{dM} \tilde{M}^2 \cos^2\theta_{\text{max}}  + \left(4\tilde{K}_{dM}\tilde{M}^2 + 2\right)\cos\theta_{\text{max}} + 2\tilde{K}_{dM}\tilde{M}^2 + 2 - \tilde{M}^2 &= 0.
    \end{split}
\end{equation}

\begin{figure}[ht!]
    \centering
    \includegraphics[width=0.75\linewidth]{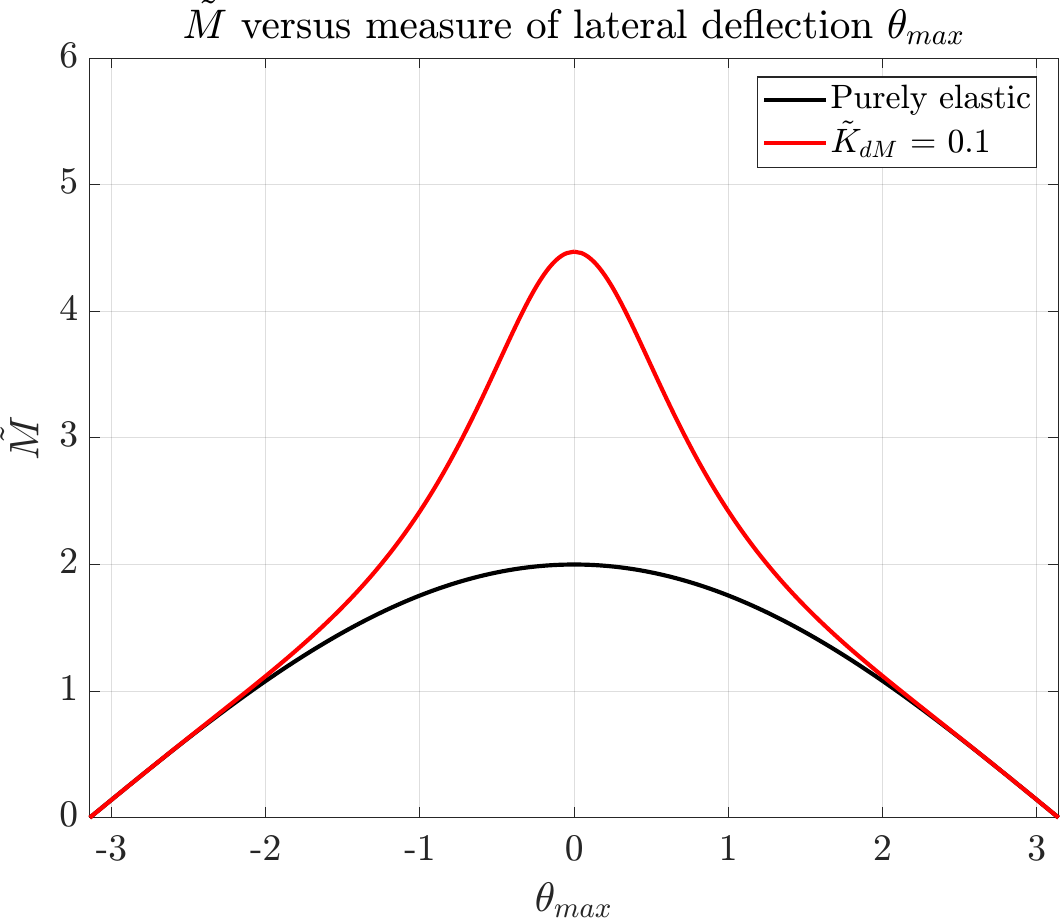}
    \caption{Localized buckling: Variation of $\tilde{M}$ with a measure of lateral deflection $\theta_\text{max}$ for purely elastic and soft ferromagnetic rods.}
    \label{fig:localized-buckling-soft-ferromagnetic-m-versus-theta-max}
\end{figure}

The solution to the above quadratic expression gives the value of $\theta_\text{max}$ for a given $(\tilde{M},\tilde{K}_{dM})$. Figure \ref{fig:localized-buckling-soft-ferromagnetic-m-versus-theta-max} presents a comparative illustration of the variation of the composite load parameter $\tilde{M}$ with the maximum lateral deflection $\theta_\text{max}$ for purely elastic and soft ferromagnetic rods, highlighting the distinct influence of magnetoelastic coupling on the extent of localized deformation.

\subsection{Displacement of the end terminals}
Inspired by Coyne \cite{Coyne1990}, we now proceed to obtain the expression relating the end displacement to the composite load factor $\tilde{M}$ and $\tilde{K}_{dM}$.

In the straight segments of the rod, where $\theta = \theta' = 0$, the Hamiltonian (Eqn. \ref{eqn:threeD-soft-ferromagnetic-longitudinal-Hamiltonian-non-dimensionalized-M}) evaluates to $\tilde{H}_{\mathrm{soft}} = \frac{1}{\tilde{M}^2}$. Consequently, the first integral along the localized solution satisfies the following:
\begin{equation}
    \frac{1}{2}\theta'^2 + \frac{1}{2}\frac{(1 - \cos\theta)}{(1 + \cos\theta)} - \tilde{K}_{dM}\sin^2\theta + \frac{\cos\theta}{\tilde{M}^2} = \frac{1}{\tilde{M}^2}.
\end{equation}
Making appropriate trigonometric substitutions, $$1 - \cos\theta = 2\sin^2\frac{\theta}{2}, ~~1 + \cos\theta = 2\cos^2\frac{\theta}{2},~~\sin\theta = 2\sin\frac{\theta}{2}\cos\frac{\theta}{2},$$
we now have
\begin{equation}
  \begin{split}
    &\frac{1}{2}\left(\dv{\theta}{\tilde{s}}\right)^2 + \frac{1}{2}\frac{2\sin^2\frac{\theta}{2}}{2\cos^2\frac{\theta}{2}} - \tilde{K}_{dM} \left(2\sin\frac{\theta}{2}\cos\frac{\theta}{2}\right)^2  - \frac{2\sin^2\frac{\theta}{2}}{\tilde{M}^2} = 0 \\
    \implies & \left(\dv{\theta}{\tilde{s}}\right)^2 + \tan^2\frac{\theta}{2} - 8\tilde{K}_{dM} \sin^2\frac{\theta}{2}\cos^2\frac{\theta}{2} - \frac{4\sin^2\frac{\theta}{2}}{\tilde{M}^2} = 0 \\
    \implies & \left(\dv{\theta}{\tilde{s}}\right)^2 = 8\tilde{K}_{dM} \tan^2\frac{\theta}{2}\cos^4\frac{\theta}{2} + \frac{4}{\tilde{M}^2} \tan^2\frac{\theta}{2}\cos^2\frac{\theta}{2} -\tan^2\frac{\theta}{2} \\
    \implies & \dv{\theta}{\tilde{s}} = \tan\frac{\theta}{2} \sqrt{8\tilde{K}_{dM} \cos^4\frac{\theta}{2} + \frac{4}{\tilde{M}^2} \cos^2\frac{\theta}{2} - 1} \\
    \implies & d\tilde{s} = \frac{\cos\frac{\theta}{2} ~d\theta}{\sin\frac{\theta}{2} \sqrt{8\tilde{K}_{dM} \cos^4\frac{\theta}{2} + \frac{4}{\tilde{M}^2} \cos^2\frac{\theta}{2} - 1}}
  \end{split}
\end{equation}
Putting $t = \sin\frac{\theta}{2}$ and $dt = \cos\frac{\theta}{2} \frac{d\theta}{2}$, we have
\begin{equation}
    d\tilde{s} = \frac{2~dt}{t \sqrt{8\tilde{K}_{dM} (1 - t^2)^2 + \frac{4}{\tilde{M}^2} (1 - t^2) - 1}}
\end{equation}
On solving this using \textit{Mathematica},
\begin{equation}
    \tilde{s} = - \frac{2\tilde{M}}{\sqrt{(1 - 8\tilde{K}_{dM})\tilde{M}^2 - 4}} \tan^{-1}\left( \frac{\tilde{M}\left(\sqrt{8\tilde{K}_{dM}} t^2 - \sqrt{8\tilde{K}_{dM} (1 - t^2)^2 + \frac{4}{\tilde{M}^2} (1 - t^2) - 1}\right)}{\sqrt{(1 - 8\tilde{K}_{dM})\tilde{M}^2 - 4}} \right)
\end{equation}
or,
\begin{equation}
  \tan \left(-\frac{\tilde{s} \sqrt{(1 - 8\tilde{K}_{dM})\tilde{M}^2 - 4}}{2\tilde{M}}\right) = \frac{\tilde{M}\left(\sqrt{8\tilde{K}_{dM}} t^2 - \sqrt{8\tilde{K}_{dM} (1 - t^2)^2 + \frac{4}{\tilde{M}^2} (1 - t^2) - 1}\right)}{\sqrt{(1 - 8\tilde{K}_{dM})\tilde{M}^2 - 4}}.
\end{equation}
For the range $0 < \tilde{K}_{dM} < \frac{1}{8}$, the equation assumes the form
\begin{equation}
  \tanh\left(\frac{\tilde{s} \sqrt{4 - (1 - 8\tilde{K}_{dM})\tilde{M}^2}}{2\tilde{M}}\right) = \frac{\tilde{M}\left(\sqrt{8\tilde{K}_{dM}} t^2 - \sqrt{8\tilde{K}_{dM} (1 - t^2)^2 + \frac{4}{\tilde{M}^2} (1 - t^2) - 1}\right)}{\sqrt{4 - (1 - 8\tilde{K}_{dM})\tilde{M}^2}}.
\end{equation}
We can express $t^2 = \sin^2\frac{\theta}{2}$ explicitly as
\begin{equation}
    \sin^2\frac{\theta}{2} = \frac{Q^2 \left(\sqrt{2\tilde{K}_{dM}}\tilde{M}\sqrt{4 - (1 - 8\tilde{K}_{dM})\tilde{M}^2}\tanh\left(\frac{\tilde{s}Q}{2\tilde{M}}\right) + 4\tilde{K}_{dM} \tilde{M}^2 + 1\right)}{4\tilde{K}_{dM} \tilde{M}^2 ((16\tilde{K}_{dM} -1)\tilde{M}^2 + 8) + (4\tilde{K}_{dM} \tilde{M}^2 + 2)  \cosh\left(\frac{\tilde{s}Q}{2\tilde{M}}\right) + 2}. \label{eqn:threeD-sin-2-theta/2}
\end{equation}
where $Q = \sqrt{4 - (1 - 8\tilde{K}_{dM})\tilde{M}^2}$.

The displacement of a material point $\tilde{s}$ of the rod along the $\vb*{e}_3$-axis is $\tilde{s}-\tilde{z}(\tilde{s})$, where $\tilde{z}(\tilde{s})$ is the non-dimensional $z$-coordinate of centerline coordinates as given in Eqn. \ref{eqn:centerline-coordinates}. The $\tilde{s}$-derivative of the axial displacement is
\begin{equation}
    \dv{}{\tilde{s}} ~(\tilde{s} - \tilde{z}(\tilde{s})) = 1 - \cos\theta = 2\sin^2\frac{\theta}{2}.
\end{equation}

\begin{figure}[ht!]
    \centering
    \includegraphics[width=0.7\linewidth]{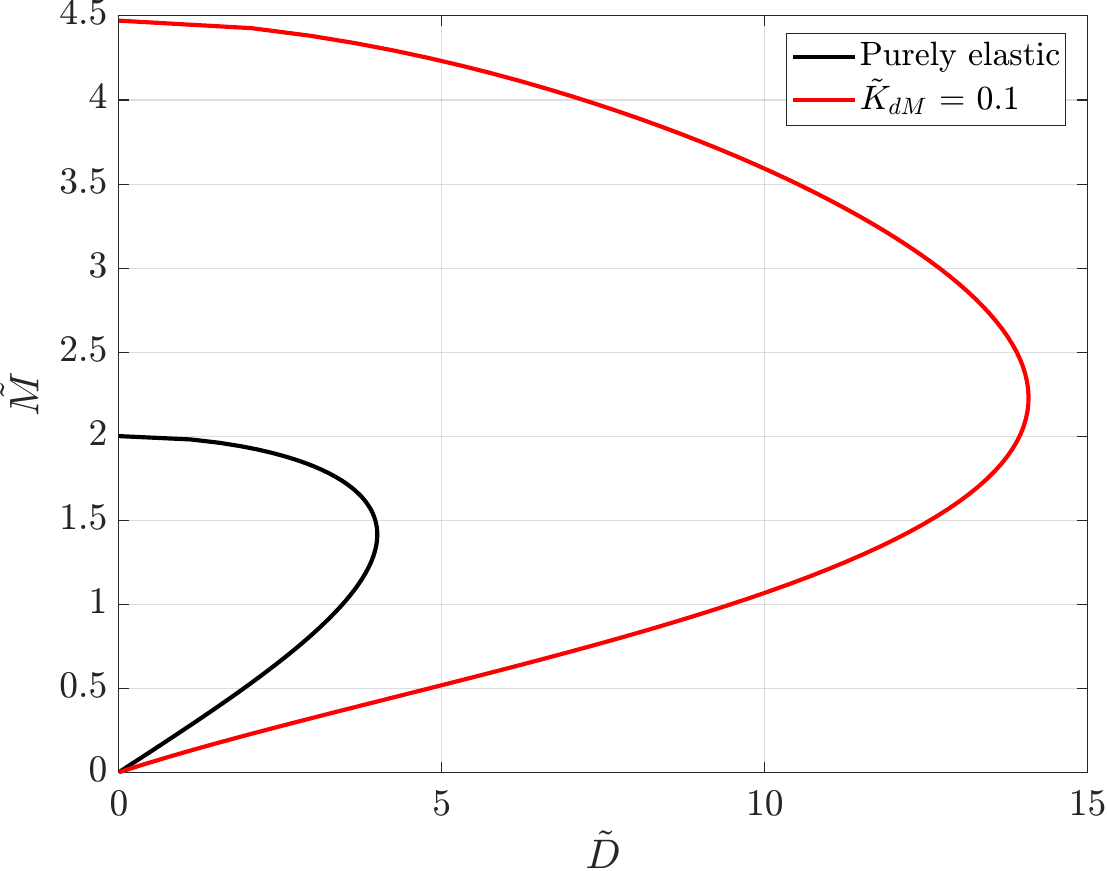}
    \caption{Localized buckling: Variation of end displacement $\tilde{D}$ with $\tilde{M}$ for purely elastic and soft ferromagnetic rods as $l \rightarrow \infty$.}
    \label{fig:localized-buckling-soft-ferromagnetic-d-versus-m}
\end{figure}

Centering the origin of the axis at $\tilde{s}=0$ where $\theta$ attains its maximum for reasons of symmetry, we have $\tilde{z}(\tilde{s}=0) = 0$. Further, $\tilde{s}$ varies from $-\frac{l}{2}$ to $\frac{l}{2}$. At the rod terminals $\tilde{s} = \pm \frac{l}{2}$, $\tilde{z} = \pm \frac{\tilde{D}}{2}$, where $\tilde{D} = \frac{DM}{EI}$ is the non-dimensional end displacement of the terminals. Therefore, integrating the above equation and exploiting the symmetry $\tilde{D}(-\frac{l}{2}) = -\tilde{D}(\frac{l}{2})$ leads to
\begin{equation}
    \tilde{D} = \int_{-l/2}^{l/2} d\left(\tilde{s} - \tilde{z}(\tilde{s})\right) = 4\int_{0}^{l/2} \sin^2\frac{\theta}{2} d\tilde{s}.
\label{eqn:beyound Coyne}
\end{equation}
The expression for $\sin^2\frac{\theta}{2}$ is substituted from Eqn. \ref{eqn:threeD-sin-2-theta/2} to obtain $\tilde{D}$. In the limiting case of an infinitely long rod ($l \rightarrow \infty$) composed of a purely elastic material ($\tilde{K}_{dM} = 0$), the relation reduces to
\begin{equation}
\left(\frac{\tilde{D}}{4\tilde{M}}\right)^2 + \left(\frac{\tilde{M}}{2}\right)^2 = 1,
\label{eqn:Coyne ellipse}
\end{equation}
which, upon appropriate non-dimensionalization, recovers Equation 32 of Coyne \cite{Coyne1990}. Figure \ref{fig:localized-buckling-soft-ferromagnetic-d-versus-m} presents the variation of the total terminal displacement $\tilde{D}$ with the composite load parameter $\tilde{M}$ for both purely elastic and soft ferromagnetic rods in this infinite-length limit. This correspondence establishes the present work  as a proper analytical extension of Coyne's loop-formation analysis to the magnetoelastic setting. Whereas Coyne's result describes the  end-displacement of a purely elastic rod under combined tension and  torsion, Eqn.~\ref{eqn:beyound Coyne} generalizes this to a soft  ferromagnetic rod by introducing the magnetoelastic parameter  $\tilde{K}_{dM}$, which continuously deforms the ellipse  (Eqn.~\ref{eqn:Coyne ellipse}) into the family of curves shown in  Fig.~\ref{fig:localized-buckling-soft-ferromagnetic-d-versus-m}.

\section{Conclusions} \label{sec:conclusions-threeD}

The total energy functional of a three-dimensional ferromagnetic elastic rod, incorporating both elastic strain energy and micromagnetic energy contributions, was rigorously derived. The Hamiltonian was constructed via the Legendre transform of the energy density and, exploiting the conserved Casimir invariants arising from the circular cross-sectional symmetry and the applied loading, was reduced to a single-degree-of-freedom system in the primary Euler angle. This reduction was carried out for purely elastic rods and for both soft and hard ferromagnetic rods subjected to a longitudinal magnetic field. Analysis of the resulting Hamiltonian phase portraits revealed that purely elastic and hard ferromagnetic rods exhibit a supercritical Hamiltonian Hopf pitchfork bifurcation. In contrast, soft ferromagnetic rods exhibit this bifurcation only within the restricted magnetoelastic parameter regime $0<\tilde{K}_{dM}< \nicefrac{1}{8}$; for $\tilde{K}_{dM} \geq \nicefrac{1}{8}$, the number of critical points of the Hamiltonian remains unchanged as the composite mechanical load parameter is varied, and no bifurcation from the straight configuration is observed.

The analysis of helical and localised post-buckling behaviour closely follows the framework established by Thompson and Champneys \cite{thompson1996helix} and Coyne \cite{Coyne1990} for purely elastic rods, extended here to the magnetoelastic setting. A systematic investigation of helical post-buckling under three distinct loading sequences revealed significant differences between purely elastic and soft ferromagnetic rods. In particular, the magneto-mechanical response of the helically deformed soft ferromagnetic rod depends qualitatively on the loading sequence: when the twisting moment $M$ is varied at fixed tension $T$, the response differs markedly from the case in which $T$ is varied at fixed $M$. Closed-form expressions were derived relating the maximum lateral deflection and the end displacement of the rod to the composite load parameter, extending the loop-formation analysis of Coyne \cite{Coyne1990} to the soft ferromagnetic case. The localised buckled configurations of the soft ferromagnetic rod exhibit a geometrically distinctive feature absent in the purely elastic case: the extended straight segments lying on either side of the localised deformation zone are mutually non-collinear, a direct consequence of the magnetoelastic coupling.

Several directions remain open for future investigation. The force-displacement relationship for a soft ferromagnetic rod subjected to a prescribed twisting moment, specifically, the relation between the applied tension $T$ and the axial end-displacement has not been derived in the present work and constitutes an immediate next step. More broadly, future work will be directed towards the formulation and solution of the full Euler–Lagrange equations characterising three-dimensional deformations of ferromagnetic elastic rods under combined mechanical and magnetic loading, encompassing a range of boundary conditions and loading scenarios relevant to experimental settings. The present Hamiltonian phase-space analysis identifies equilibrium configurations but does not address their stability, a systematic stability analysis of the obtained equilibria is needed to determine critical bifurcation thresholds rigorously and to characterise post-buckling behaviour in the presence of perturbations. These investigations aim to provide a deeper theoretical foundation for the mechanics of ferromagnetic elastic rods and to inform the design of magnetically actuated structures.

\backmatter
\bmhead{Supplementary information} Nil.

\bmhead{Acknowledgements}
The authors acknowledge research funding through Prime Minister Research Fellowship (PMRF ID: 0201857).

\bmhead{Declarations}
The authors have no competing interests to declare that are relevant to the content of this article.

\begin{appendices}
\input{appendix}

\end{appendices}

\bibliography{ActaMechanica_Manuscript}

\end{document}

%% file: appendix.tex
\section{Magnetostatic energy of a soft ferromagnetic rod}\label{sec:app-slastikov-derivation}
\begin{figure}
    \centering
    \includegraphics[width=0.4\linewidth]{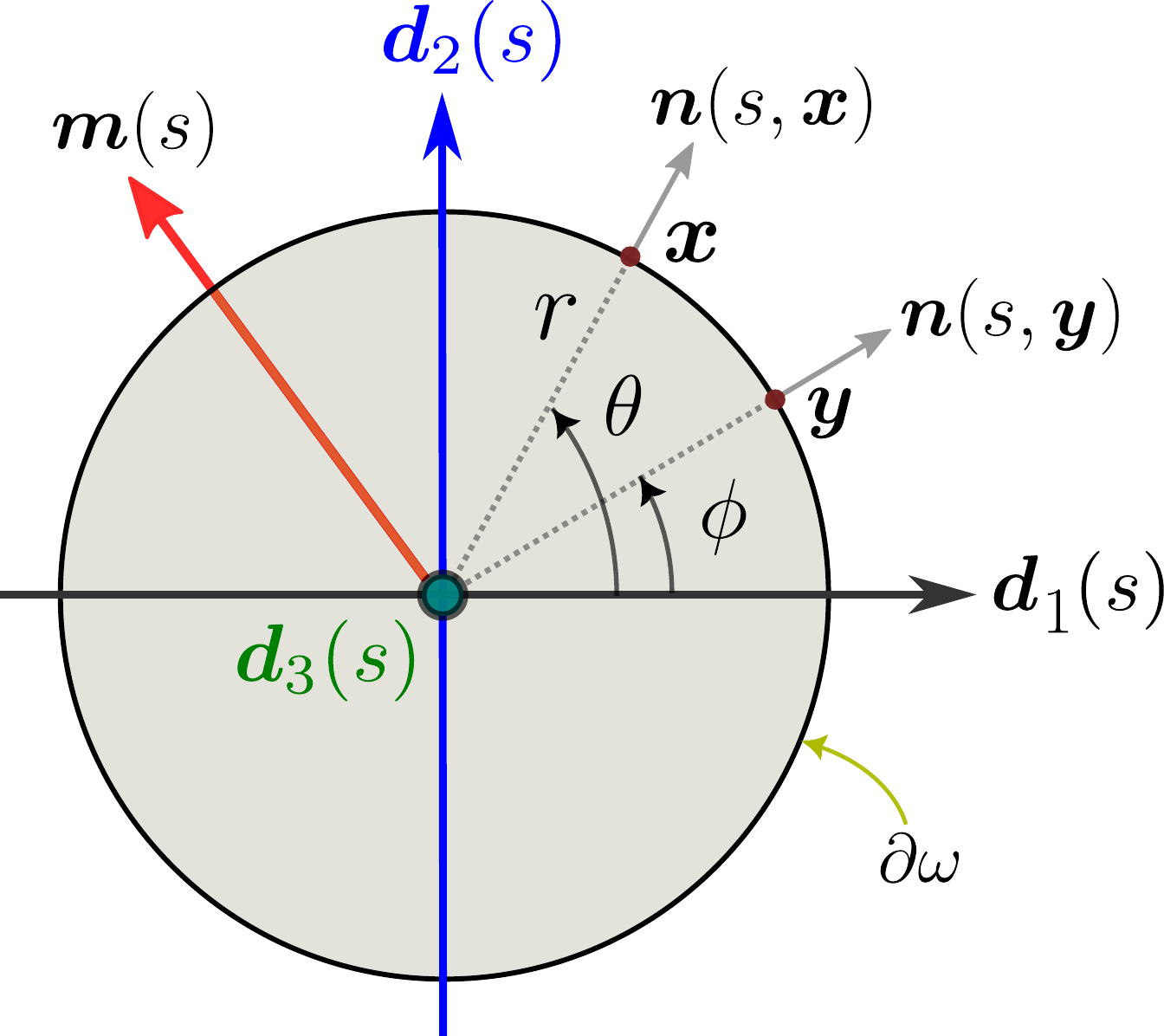}
    \caption{Cross-section of the circular rod at a location $s$ indicating the director basis, the points $\vb*{x}$ and $\vb*{y}$ on the periphery.}
    \label{fig:slastikov-derivation-schematic}
\end{figure}

We present a brief derivation of the magnetic energy of a circular rod, inspired by \cite{Slastikov2011}.
For a soft ferromagnetic rod composed of uniaxial ferromagnet, the micromagnetic energy is given as:
\begin{equation}
    \mathcal{E}_{\text{mag,soft}}(\vb*{m}) = A\int_{-L}^{L} \abs{\vb*{m}'(s)}^2~ds + K_d\int_{-L}^{L} \left(\vb*{Mm}(s),\vb*{m}(s)\right)~ds,
\end{equation}
{\color{black} where $(\cdot,\cdot)$ in the second integral denotes inner product} and the constant symmetric matrix $\vb*{M}$ is defined as
\begin{equation}
    \vb*{M} = -\frac{1}{2\pi} \int_{\partial \omega} \int_{\partial \omega} \vb*{n}(s,\vb*{x})\otimes \vb*{n}(s,\vb*{y}) \ln|\vb*{x} - \vb*{y}| d\vb*{x}d\vb*{y},
\end{equation}
{\color{black} where $\otimes$ denotes tensor (dyadic) product of two vectors \cite[Section 1.2]{holzapfel2000nonlinear} and} $\vb*{n}(s,\vb*{x})$ is a normal vector to $\partial \omega$; $\partial \omega$ denotes the boundary of the cross section (see Fig. \ref{fig:slastikov-derivation-schematic}). Thus,
\begin{equation}
\begin{split}
    \mathcal{E}_{\text{mag,soft}}(\vb*{m}) &= A\int_{-L}^{L} \abs{\vb*{m}'(s)}^2~ds - \frac{K_d}{2\pi}\int_{-L}^{L}\int_{\partial \omega} \int_{\partial \omega} (\vb*{m}(s)\cdot\vb*{n}(s,\vb*{x}))(\vb*{m}(s)\cdot\vb*{n}(s,\vb*{y})) \ln|\vb*{x}-\vb*{y}| d\vb*{x}d\vb*{y} ds \\
     &= A\int_{-L}^{L} \abs{\vb*{m}'(s)}^2~ds - \frac{K_d}{2\pi}\int_{-L}^{L} f(s)~ds.
\end{split}
\end{equation}
We ought to determine
\begin{equation}
    f(s) = \int_{\partial \omega} \int_{\partial \omega} (\vb*{m}(s)\cdot\vb*{n}(s,\vb*{x}))(\vb*{m}(s)\cdot\vb*{n}(s,\vb*{y})) \ln|\vb*{x}-\vb*{y}| d\vb*{x}d\vb*{y}.
\end{equation}
We consider a rod of circular cross section of diameter $a = 2r$ ($r$ is the radius). We parametrize any two arbitrary points $\vb*{x}$ and $\vb*{y}$ lying on the circular boundary with respect to the director basis $\mathcal{D} = (\vb*{d}_1,\vb*{d}_2,\vb*{d}_3)$ (see Fig. \ref{fig:slastikov-derivation-schematic}) along with the corresponding outward normal vectors $\vb*{n}(s,\vb*{x})$ and $\vb*{n}(s,\vb*{y})$ as
\begin{equation}
    \begin{split}
        \vb*{x} = (r\cos\theta,r\sin\theta,0), &\quad \vb*{y} = (r\cos\phi,r\sin\phi,0), \\
        \vb*{n}(s,\vb*{x}) = (\cos\theta,\sin\theta,0), &\quad \vb*{n}(s,\vb*{y}) = (\cos\phi,\sin\phi,0),
    \end{split}
\end{equation}
where, $\theta, \phi \in [0,2\pi)$. Likewise, the components of the magnetization vector $\vb*{m}(s)$ in terms of the director basis are
\begin{equation}
    \vb*{m}(s) = (m_1(s), m_2(s), m_3(s)).
\end{equation}
Now, a few expansions:
\begin{equation}
\begin{split}
    \vb*{m} \cdot \vb*{n}(s,\vb*{x}) &= m_1(s)\cos\theta + m_2(s)\sin\theta, \\
    \vb*{m} \cdot \vb*{n}(s,\vb*{y}) &= m_1(s)\cos\phi + m_2(s)\sin\phi, \\
    \ln|\vb*{x} - \vb*{y}| &= \ln(r\sqrt{(\cos\theta-\cos\phi)^2 + (\sin\theta - \sin\phi)^2}) \\
                           &= \ln r + \frac{1}{2}\ln[2(1 - (\cos\theta\cos\phi + \sin\theta\sin\phi))].
\end{split}
\end{equation}
Along the boundary, $d\vb*{x} = rd\theta,~d\vb*{y} = rd\phi$. Now,
\begin{multline}
     f(s) = \int_{\partial \omega} \int_{\partial \omega} (\vb*{m}(s)\cdot\vb*{n}(s,\vb*{x}))(\vb*{m}(s)\cdot\vb*{n}(s,\vb*{y})) \ln|\vb*{x}-\vb*{y}| d\vb*{x}d\vb*{y} \\
     = \int_{\theta=0}^{2\pi}\int_{\phi=0}^{2\pi} (m_1\cos\theta + m_2\sin\theta)(m_1(s)\cos\phi + m_2(s)\sin\phi) \\
        \left[\ln r + \frac{1}{2}\ln[2(1 - (\cos\theta\cos\phi + \sin\theta\sin\phi))]\right]d\theta d\phi.
\end{multline}
On expanding the terms,
\begin{equation}
\begin{split}
   f(s) &= r^2 \underbrace{\int_{\theta=0}^{2\pi}\int_{\phi=0}^{2\pi} [m_1^2 \cos\theta\cos\phi + m_2^2\sin\theta\sin\phi + m_1 m_2 (\cos\theta\sin\phi + \sin\theta\cos\phi)] \ln r d\theta d\phi}_{=0}\\
    &\qquad  +  \frac{r^2}{2} \int_{\theta=0}^{2\pi}\int_{\phi=0}^{2\pi} [m_1^2 \cos\theta\cos\phi + m_2^2\sin\theta\sin\phi + m_1 m_2 (\cos\theta\sin\phi + \sin\theta\cos\phi)] \\
    &\qquad \qquad \qquad \qquad \qquad \ln[2(1 - (\cos\theta\cos\phi + \sin\theta\sin\phi))]d\theta d\phi,
\end{split}
\end{equation}
and further simplifying, we get
\begin{equation}
    f(s) = -\pi^2 r^2 [m_1^2(s) + m_2^2(s)] = -\pi^2 r^2 [(\vb*{m}\cdot\vb*{d}_1)^2 + (\vb*{m}\cdot\vb*{d}_2)^2]
\end{equation}
Substituting the expression for $f(s)$ in the micromagnetic energy functional, we have
\begin{equation}
\begin{split}
   \mathcal{E}_{\text{mag,soft}}(\vb*{m}) &= A\int_{-L}^{L} \abs{\vb*{m}'(s)}^2 + \frac{K_d\pi r^2}{2}\int_{-L}^{L} [(\vb*{m}\cdot\vb*{d}_1)^2 + (\vb*{m}\cdot\vb*{d}_2)^2] ds,  \\
   \implies \mathcal{E}_{\text{mag,soft}}(\vb*{m}) &= 2A\int_{0}^{L} \abs{\vb*{m}'(s)}^2 + \frac{K_d\pi a^2}{4}\int_{0}^{L} [(\vb*{m}\cdot\vb*{d}_1)^2 + (\vb*{m}\cdot\vb*{d}_2)^2] ds.
\end{split}
\end{equation}
We now consider the case of a soft ferromagnetic rod subjected to a high external magnetic field $\vb*{h}_e$. The magnetization of the rod $\vb*{m}(s)$ becomes saturated and aligns with $\vb*{h}_e$, such that $\vb*{m}'(s) = \vb*{0}$. Thus, the micromagnetic energy of a soft ferromagnetic rod takes the form
\begin{equation}
    \mathcal{E}_{\text{mag,soft}}(\vb*{m}) = \frac{K_d\pi a^2}{4}\int_{0}^{L} [(\vb*{m}\cdot\vb*{d}_1)^2 + (\vb*{m}\cdot\vb*{d}_2)^2] ds. \label{app:soft-magnetic-energy-rod}
\end{equation}
Let us derive the expression of the distributed magnetic couple by taking the first variation of the above expression:
\begin{equation}
    \delta \mathcal{E}_{\text{mag,soft}}(\vb*{m}) = \frac{K_d\pi a^2}{2}\int_{0}^{L} [(\vb*{m}\cdot\vb*{d}_1)(\vb*{m}\cdot\delta\vb*{d}_1) + (\vb*{m}\cdot\vb*{d}_2)(\vb*{m}\cdot\delta\vb*{d}_2)] ds. \label{eqn:appendix-slastikov-derivation-soft-first-variation-1}
\end{equation}
As presented in \cite[Section 3.6]{Audoly2010elasticity}, the perturbations to the director basis vectors are expressed as
\begin{equation}
    \delta\vb*{d}_i = \delta\vb*{\phi}\times\vb*{d}_i, ~i=1,2,3,
\end{equation}
where the infinitesimal rotation vector $\delta\vb*{\phi}(s)$ is
\begin{equation}
    \delta\vb*{\phi}(s) = \delta\phi_1(s)\vb*{d}_1(s) + \delta\phi_2(s)\vb*{d}_2(s) + \delta\phi_3(s)\vb*{d}_3
\end{equation}
which denotes the perturbations to the orientation of the director basis. Now,
\begin{equation}
    \begin{split}
        \vb*{m}\cdot\delta\vb*{d}_1 &= (\vb*{m}\cdot\vb*{d}_2)\delta\phi_3 - (\vb*{m}\cdot\vb*{d}_3)\delta\phi_2 \\
        \vb*{m}\cdot\delta\vb*{d}_2 &= -(\vb*{m}\cdot\vb*{d}_1)\delta\phi_3 + (\vb*{m}\cdot\vb*{d}_3)\delta\phi_1.
    \end{split}
\end{equation}
Inserting the expressions in the first variation,
\begin{multline}
    \delta \mathcal{E}_{\text{mag,soft}}(\vb*{m}) = \frac{K_d\pi a^2}{2}\int_{0}^{L} [(\vb*{m}\cdot\vb*{d}_2)(\vb*{m}\cdot\vb*{d}_3)\delta\phi_1 - (\vb*{m}\cdot\vb*{d}_1)(\vb*{m}\cdot\vb*{d}_3)\delta\phi_2] ds \\
    = \frac{K_d\pi a^2}{2}\int_{0}^{L}[(\vb*{m}(s)\cdot\vb*{d}_3(s))(\vb*{m}(s)\times\vb*{d}_3(s))]\cdot\delta\vb*{\phi} ds. \label{eqn:appendix-slastikov-derivation-soft-first-variation-2}
\end{multline}
It can be inferred from the above that the distributed magnetic couple due to magnetostatic energy is
\begin{equation}
    \vb*{q}_{\text{mag,soft}}(s) = \frac{K_d\pi a^2}{2}(\vb*{m}(s)\cdot\vb*{d}_3(s))(\vb*{m}(s)\times\vb*{d}_3(s)). \label{app:soft-magnetic-couple-rod}
\end{equation}

\section{Exchange and anisotropy energies of a hard ferromagnetic rod}\label{sec:app-exchange-energy-hard-magnetic}
We now proceed to deriving expressions for the exchange and anisotropy energies for hard ferromagnetic rods inspired by \cite{Sheka2015}. The energy formulation considers a general magnetization distribution along the centerline of the rod $\vb*{m}(s)$.

\begin{figure}
    \centering
    \includegraphics[width=0.7\linewidth]{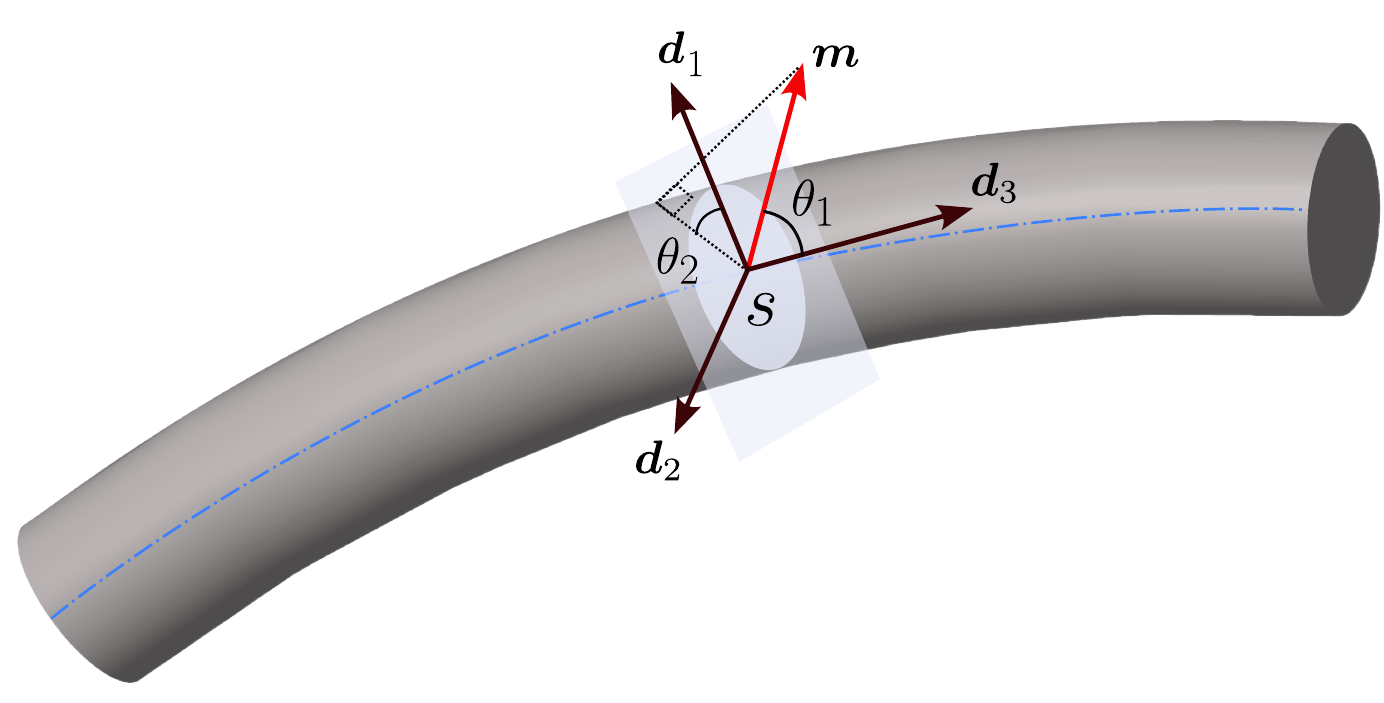}
    \caption{A segment of a circular ferromagnetic rod depicting the magnetization $\vb*{m}$ and its angular position with respect to the director basis $(\vb*{d}_1,\vb*{d}_2,\vb*{d}_3)$ described the coordinates $(\theta_1,\theta_2)$ at a location $s$.}
    \label{fig:hard-magnetic-energy-derivation-schematic}
\end{figure}
In the reference configuration, the magnetization distribution is
\begin{equation}
    \vb*{m}_0(s) = \sin\theta_1(s)\cos\theta_2(s)\vb*{e}_1 + \sin\theta_1(s)\sin\theta_2(s)\vb*{e}_2 + \cos\theta_1(s)\vb*{e}_3,
\end{equation}
where $\theta_1(s)\in[0,\pi]$ and $\theta_2(s)\in [0,2\pi)$. The magnetization distribution of the rod in the deformed configuration, as illustrated in Fig. \ref{fig:hard-magnetic-energy-derivation-schematic}, is
\begin{equation}
    \vb*{m}(s) = \sin\theta_1(s)\cos\theta_2(s)\vb*{d}_1(s) + \sin\theta_1(s)\sin\theta_2(s)\vb*{d}_2(s) + \cos\theta_1(s)\vb*{d}_3(s).
\end{equation}
We determine the micromagnetic energy functional of the hard ferromagnetic rod:
\begin{equation}
    \mathcal{E}_{\text{mag,hard}} = \frac{A\pi a^2}{4}\int_{0}^{L} \mathbb{E}_{\text{ex}} ds + \frac{K_a\pi a^2}{4}\int_{0}^{L}(\vb*{m}\cdot\vb*{p})^2ds - \frac{2K_d\pi a^2}{4}\int_{0}^{L}\vb*{m}\cdot\vb*{h}_e ds,
\end{equation}
where $\mathbb{E}_{\text{ex}}$ is the exchange energy density and $\vb*{p}$ is the anisotropy easy axis. We derive the exchange energy density in the curvilinear rod frame. Recall that the centerline-based parametrization of the rod is:
\begin{equation}
    \vb*{x}(s,a_1,a_2) = \vb*{r}(s) + a_1\vb*{d}_1 + a_2\vb*{d}_2,~\tilde{\vb*{a}}=(a_1,a_2)
\end{equation}
such that $\abs{\tilde{\vb*{a}}} \leq \nicefrac{a}{2}$. Also, the director basis vectors evolve as
\begin{equation}
    \vb*{d}_i'(s) = \vb*{\kappa}(s)\cross\vb*{d}_i(s) = \vb*{K}(s)\cdot\vb*{d}_i(s),
\end{equation}
where $\vb*{\kappa} = \kappa_1\vb*{d}_1 + \kappa_2\vb*{d}_2 + \kappa_3\vb*{d}_3$ is the strain gradient vector, while $\vb*{K}$ denotes the corresponding skew-symmetric matrix composed of its components as
\begin{equation}
    \vb*{K} = \begin{bmatrix}
        0 & -\kappa_3 & \kappa_2 \\ \kappa_3 & 0 & -\kappa_1 \\ -\kappa_2 & \kappa_1 & 0
    \end{bmatrix}.
\end{equation}
The exchange energy density is given in terms of Cartesian frame of reference as:
\begin{equation}
    \mathbb{E}_{\text{ex}} = (\grad m_i)\cdot (\grad m_i),~i=1,2,3,
\end{equation}
where the Cartesian components of the magnetization vector, $m_i$, in terms of the curvilinear components $m_\alpha$ as follows:
\begin{equation}
    m_i = m_\alpha (\vb*{d}_\alpha\cdot\vb*{e}_i).
\end{equation}
We substitute this expression into $\mathbb{E}_{\text{ex}}$ and apply a \textit{del} operator in its curvilinear form, $\grad \equiv \vb*{e}_3\pdv{}{s}$. Finally, the energy density in the curvilinear frame is expressed as:
\begin{equation}
    \begin{split}
        \mathbb{E}_{\text{ex}} &= (m_\alpha \vb*{d}_\alpha)'\cdot(m_\beta\vb*{d}_\beta)' \\
        &=  (m_\alpha' \vb*{d}_\alpha + m_\alpha \vb*{d}_\alpha')\cdot(m_\beta'\vb*{d}_\beta + m_\beta\vb*{d}_\beta') \\
        &= m_\alpha'm_\beta'\vb*{d}_\alpha\cdot\vb*{d}_\beta + (m_\alpha'm_\beta \vb*{d}_\alpha\cdot\vb*{d}_\beta' + m_\alpha m_\beta'\vb*{d}_\alpha'\cdot\vb*{d}_beta) + m_\alpha m_\beta \vb*{d}_\alpha'\cdot\vb*{d}_\beta' \\
        &= \underbrace{m_\alpha'm_\alpha'}_{\mathbb{E}_{\text{ex}}^o} + \underbrace{m_\alpha m_\beta \vb*{d}_\alpha'\cdot\vb*{d}_\beta'}_{\mathbb{E}_{\text{ex}}^A} + \underbrace{(m_\alpha'm_\beta \vb*{d}_\alpha\cdot\vb*{d}_\beta' + m_\alpha m_\beta'\vb*{d}_\alpha'\cdot\vb*{d}_\beta)}_{\mathbb{E}_{\text{ex}}^D}.
    \end{split}
\end{equation}
We further expand each of these terms. Now,
\begin{equation}
    \mathbb{E}_{\text{ex}}^o = m_\alpha'm_\alpha' = \abs{\vb*{m}'(s)}^2.
\end{equation}
The $\mathbb{E}_{\text{ex}}^A$-component is
\begin{equation}
    \begin{split}
        \mathbb{E}_{\text{ex}}^A &= m_\alpha m_\beta \vb*{d}_\alpha'\cdot\vb*{d}_\beta' = m_\alpha m_\beta (\vb*{K}\vb*{d}_\alpha)\cdot(\vb*{K}\vb*{d}_\beta) = m_\alpha m_\beta \vb*{d}_\alpha\cdot(\vb*{K}^T\vb*{K})\vb*{d}_\beta = m_\alpha m_\beta \vb*{d}_\alpha\cdot\vb*{B}\vb*{d}_\beta \\
        &= B_{\alpha\beta}m_\alpha m_\beta,
    \end{split}
\end{equation}
where
\begin{equation}
    \vb*{B} = \vb*{K}^T\vb*{K} = -\vb*{K}^2 = \begin{bmatrix}
        \kappa_2^2 + \kappa_3^2 & -\kappa_1\kappa_2 & -\kappa_1\kappa_3 \\ -\kappa_1\kappa_2 & \kappa_1^2 + \kappa_3^2 & -\kappa_2\kappa_3 \\ -\kappa_1\kappa_3 & -\kappa_2\kappa_3 & \kappa_1^2 + \kappa_2^2
    \end{bmatrix}.
\end{equation}
The Dzyaloshinskii interaction component is
\begin{equation}
   \begin{split}
       \mathbb{E}_{\text{ex}}^D &= m_\alpha'm_\beta \vb*{d}_\alpha\cdot\vb*{d}_\beta' + m_\alpha m_\beta'\vb*{d}_\alpha'\cdot\vb*{d}_\beta \\
       &= m_\alpha'm_\beta \vb*{d}_\alpha\cdot\vb*{K}\vb*{d}_\beta + m_\alpha m_\beta'\vb*{K}\vb*{d}_\alpha\cdot\vb*{d}_\beta \\
       &= m_\alpha'm_\beta \vb*{d}_\alpha\cdot\vb*{K}\vb*{d}_\beta + m_\alpha m_\beta'\vb*{d}_\alpha\cdot\vb*{K}^T\vb*{d}_\beta \\
       &= (m_\alpha'm_\beta - m_\alpha m_\beta')K_{\alpha\beta}.
   \end{split}
\end{equation}
It can be further simplified as
\begin{equation}
    \begin{split}
        \mathbb{E}_{\text{ex}}^D &= (m_\alpha'm_\beta - m_\alpha m_\beta')K_{\alpha\beta} \\
        &= m_\alpha'K_{\alpha\beta} m_\beta - m_\alpha K_{\alpha\beta}m_\beta' \\
        &= m_\alpha'K_{\alpha\beta} m_\beta + m_\alpha K_{\beta\alpha}m_\beta' \qquad (K_{\beta\alpha} = -K_{\alpha\beta})  \\
        &= m_\alpha'K_{\alpha\beta} m_\beta + m_\beta' K_{\beta\alpha}m_\alpha' \\
        &= 2m_\alpha'K_{\alpha\beta} m_\beta.
    \end{split}
\end{equation}
Given that
\begin{equation}
    \vb*{m}(s) = \underbrace{\sin\theta_1(s)\cos\theta_2(s)}_{m_1}\vb*{d}_1(s) + \underbrace{\sin\theta_1(s)\sin\theta_2(s)}_{m_2}\vb*{d}_2(s) + \underbrace{\cos\theta_1(s)}_{m_3}\vb*{d}_3(s).
\end{equation}
Now,
\begin{multline}
    \mathbb{E}_{\text{ex}}^o = m_\alpha'm_\alpha' = (\cos\theta_1\cos\theta_2\theta_1' - \sin\theta_1\sin\theta_2\theta_2')^2 + (\cos\theta_1\sin\theta_2\theta_1' + \sin\theta_1\cos\theta_2\theta_2')^2 \\ + (-\sin\theta_1\theta_1')^2
    = \cos^2\theta_1\theta_1'^2 + \sin^2\theta_1\theta_2'^2 + \sin^2\theta_1\theta_1'^2
\end{multline}
Therefore,
\begin{equation}
    \mathbb{E}_{\text{ex}}^o = \theta_1'^2 + \sin^2\theta_1\theta_2'^2.
\end{equation}
\begin{equation}
   \begin{split}
        \mathbb{E}_{\text{ex}}^A &= \begin{bmatrix}
            \sin\theta_1\cos\theta_2 \\ \sin\theta_1\sin\theta_2 \\ \cos\theta_1
        \end{bmatrix} \cdot
        \begin{bmatrix}
        \kappa_2^2 + \kappa_3^2 & -\kappa_1\kappa_2 & -\kappa_1\kappa_3 \\ -\kappa_1\kappa_2 & \kappa_1^2 + \kappa_3^2 & -\kappa_2\kappa_3 \\ -\kappa_1\kappa_3 & -\kappa_2\kappa_3 & \kappa_1^2 + \kappa_2^2
    \end{bmatrix} \begin{bmatrix}
            \sin\theta_1\cos\theta_2 \\ \sin\theta_1\sin\theta_2 \\ \cos\theta_1
        \end{bmatrix}
   \end{split}
\end{equation}
\begin{multline}
    \mathbb{E}_{\text{ex}}^A = (\kappa_1^2 + \kappa_2^2)\cos^2\theta_1 - \kappa_3\sin(2\theta_1)(\kappa_1\cos\theta_2 + \kappa_2\sin\theta_2)  \\ + \frac{1}{2}\sin^2\theta_1(\kappa_1^2 + \kappa_2^2 + 2\kappa_3^2 + (\kappa_2^2 - \kappa_1^2)\cos(2\theta_2) - 2\kappa_1\kappa_2\sin(2\theta_2)).
\end{multline}
Now,
\begin{equation}
    \mathbb{E}_{\text{ex}}^D  =  2 \begin{bmatrix}
        \cos\theta_1\cos\theta_2\theta_1' - \sin\theta_1\sin\theta_2\theta_2' \\ \cos\theta_1\sin\theta_2\theta_1' + \sin\theta_1\cos\theta_2\theta_2' \\ -\sin\theta_1\theta_1'
    \end{bmatrix} \cdot
    \begin{bmatrix}
        0 & -\kappa_3 & \kappa_2 \\
        \kappa_3 & 0 & -\kappa_1 \\
        -\kappa_2 & \kappa_1 & 0
    \end{bmatrix}
     \begin{bmatrix}
        \sin\theta_1\cos\theta_2 \\ \sin\theta_1\sin\theta_2 \\ \cos\theta_1
        \end{bmatrix}
\end{equation}
Therefore,
\begin{equation}
     \mathbb{E}_{\text{ex}}^D  = 2(\kappa_2\cos\theta_2 - \kappa_1\sin\theta_2)\theta_1' + (2\kappa_3\sin^2\theta_1 - \sin (2\theta_1)(\kappa_1\cos\theta_2 + \kappa_2\sin\theta_2))\theta_2'.
\end{equation}
Gathering and summing $\mathbb{E}_{\text{ex}}^o$, $\mathbb{E}_{\text{ex}}^A$ and $\mathbb{E}_{\text{ex}}^D$, we have
\begin{multline}
    \mathbb{E}_{\text{ex}} = \theta_1'^2 + \sin^2\theta_1\theta_2'^2 +(\kappa_1^2 + \kappa_2^2)\cos^2\theta_1 - \kappa_3\sin(2\theta_1)(\kappa_1\cos\theta_2 + \kappa_2\sin\theta_2)  \\ + \frac{1}{2}\sin^2\theta_1(\kappa_1^2 + \kappa_2^2 + 2\kappa_3^2 + (\kappa_2^2 - \kappa_1^2)\cos(2\theta_2) - 2\kappa_1\kappa_2\sin(2\theta_2)) \\
    + 2(\kappa_2\cos\theta_2 - \kappa_1\sin\theta_2)\theta_1' + (2\kappa_3\sin^2\theta_1 - \sin (2\theta_1)(\kappa_1\cos\theta_2 + \kappa_2\sin\theta_2))\theta_2'
\end{multline}
We consider that the magnetization vector $\vb*{m}$ is uniformly distributed in the reference (straight, untwisted) configuration, that is, $\theta_1'(s) = \theta_2'(s) = 0$. We now analyze the resulting special cases of $\vb*{m}$ which assumes the following forms in the deformed configuration:
\begin{itemize}
    \item $\vb*{m} = \vb*{d}_3(s) \implies \theta_1 = 0$: $\mathbb{E}_{\text{ex}} = \kappa_1^2 + \kappa_2^2$

    \item $\vb*{m} \perp \vb*{d}_3(s) \implies \theta_1 = \frac{\pi}{2}$: $\mathbb{E}_{\text{ex}} = \frac{1}{2}(\kappa_1^2 + \kappa_2^2 + 2\kappa_3^2 + (\kappa_2^2 - \kappa_1^2)\cos(2\theta_2) - 2\kappa_1\kappa_2\sin(2\theta_2))$; the following are the two special sub-cases:
    \begin{itemize}
        \item $\vb*{m} = \vb*{d}_1~(\theta_2 = 0)$: $\mathbb{E}_{\text{ex}} =\kappa_2^2 + \kappa_3^2$
        \item $\vb*{m} = \vb*{d}_2~(\theta_2 = \frac{\pi}{2})$: $\mathbb{E}_{\text{ex}} =\kappa_1^2 + \kappa_3^2$
    \end{itemize}
\end{itemize}
For a hard ferromagnetic circular rod composed of a uniaxial ferromagnet with tangential magnetization distribution $\vb*{m}(s) = \vb*{d}_3(s)$ and easy axis $\vb*{p} = \vb*{d}_3(s)$, the magnetic energy is given as
\begin{equation}
    \mathcal{E}_{\text{mag,hard}}(\vb*{m}) = \frac{A\pi a^2}{4}\int_{0}^{L} (\kappa_1^2 + \kappa_2^2) ds + \underbrace{\frac{K_a\pi a^2 L}{4}}_\text{constant} - \frac{K_d\pi a^2}{2}\int_{0}^{L}\vb*{d}_3\cdot\vb*{h}_e ds. \label{app:hard-magnetic-energy-rod}
\end{equation}
Repeating the steps as followed in Eqns. \ref{eqn:appendix-slastikov-derivation-soft-first-variation-1}-\ref{eqn:appendix-slastikov-derivation-soft-first-variation-2}, the corresponding distributed magnetic couple is
\begin{equation}
    \vb*{q}_{\text{mag,hard}}(s) = \frac{A\pi a^2}{2}\left(\kappa_1(s)\vb*{d}_1(s) + \kappa_2(s)\vb*{d}_2(s)\right)' + \frac{K_d\pi a^2}{2}\vb*{d}_3(s)\cross\vb*{h}_e.  \label{app:hard-magnetic-rod-moment}
\end{equation}